\documentclass[12pt, draftclsnofoot, onecolumn]{IEEEtran}
\usepackage{geometry}
\ifCLASSINFOpdf

\else

\fi
\usepackage[cmex10]{amsmath}
\usepackage{amssymb}
\usepackage{cite}
\usepackage{graphicx}
\usepackage{array,color}
\usepackage{amsmath}
\usepackage{stfloats}
\usepackage{graphicx}
\usepackage{subfigure}
\usepackage{tabularx}
\usepackage{epsfig,epsf,color,balance,cite}
\usepackage{algorithmic}
\usepackage{algorithm}
\usepackage{bm}
\usepackage{textcomp}

\usepackage{epstopdf}
\usepackage{graphicx}
\usepackage{epstopdf}
\usepackage{color}
\definecolor{skyblue}{rgb}{0,0.5,1}

\begin{document}

\title{Artificial-Noise-Aided Secure MIMO Wireless Communications via Intelligent Reflecting Surface}
\author{ Sheng Hong, Cunhua Pan, Hong Ren, Kezhi Wang, and Arumugam Nallanathan, \IEEEmembership{Fellow, IEEE}
\thanks{This work was supported by the National Natural Science Foundation of China (61661032), the Young Natural Science Foundation of Jiangxi Province (20181BAB202002), the China Postdoctoral Science Foundation (2017M622102), the Foundation from China Scholarship Council (201906825071).}
\thanks{S. Hong is with Information Engineering School of Nanchang University, Nanchang 330031, China. (email: shenghong@ncu.edu.cn). C. Pan, H. Ren, and A. Nallanathan are with the School of Electronic Engineering and Computer Science at Queen Mary University of London, London E1 4NS, U.K. (e-mail:\{c.pan, h.ren, a.nallanathan\}@qmul.ac.uk). K. Wang is with Department of Computer and Information Sciences, Northumbria University, UK. (email: kezhi.wang@northumbria.ac.uk).
}
}

\maketitle
\vspace{-1.9cm}
\begin{abstract}
This paper considers a MIMO secure wireless communication system aided by the physical layer security technique of sending artificial noise (AN). To further enhance the system security performance, the advanced intelligent reflecting surface (IRS) is invoked in the AN-aided communication system, where the base station (BS), legitimate information receiver (IR) and eavesdropper (Eve) are equipped with multiple antennas. With the aim for maximizing the secrecy rate (SR), the transmit precoding (TPC) matrix at the BS, covariance matrix of AN and phase shifts at the IRS are jointly optimized subject to constrains of transmit power limit and unit modulus of IRS phase shifts. Then, the secrecy rate maximization (SRM) problem is formulated, which is a non-convex problem with multiple coupled variables. To tackle it, we propose to utilize the block coordinate descent (BCD) algorithm to alternately update the TPC matrix, AN covariance matrix, and phase shifts while keeping SR non-decreasing. Specifically, the optimal TPC matrix and AN covariance matrix are derived by Lagrangian multiplier method, and the optimal phase shifts are obtained by Majorization-Minimization (MM) algorithm. Since all variables can be calculated in closed form, the proposed algorithm is very efficient. We also extend the SRM problem to the more
general multiple-IRs scenario and propose a BCD algorithm to solve it. Finally, simulation results validate
the effectiveness of system security enhancement via an IRS.
\end{abstract}

\IEEEpeerreviewmaketitle
\section{Introduction}
The next-generation (i.e, 6G) communication is expected to be a sustainable green, cost-effective and secure communication system \cite{saad2019vision}. In particular, secure communication is crucially important in 6G communication networks since communication environment becomes increasingly complicated and the security of
private information is imperative \cite{wang2019energy}. The information security using crytographic encryption (in the network layer) is a conventional secure communication technique, which suffers from the vulnerabilities, such as secret key distribution, protection and management \cite{liao2010qos}. Unlike this network layer security approach, the physical layer security can guarantee good security performance bypassing the relevant manipulations on the secret key, thus is more attractive for the academia and industry \cite{wu2018survey}. There are various physical-layer secrecy scenarios. The first one is the classical physical-layer secrecy setting where there is one legitimate information receiver (IR) and one eavesdropper (Eve) operating over a single-input-single-output (SISO) channel (i.e., the so-called three-terminal SISO Gaussian wiretap channel) \cite{wyner1975wire,csiszar1978broadcast}. The second one considers the physical-layer secrecy with an IR and Eve operating over a multiple-input-single-output (MISO) channel, which is called as three-terminal MISO Gaussian wiretap channel. The third one is a renewed and timely scenario with one IR and one Eve operating over a multiple-input-multiple-output (MIMO) channel, which is named as three-terminal MIMO Gaussian wiretap channel\cite{khisti2010secure,oggier2011secrecy} and is the focus of this paper. For MIMO systems, a novel idea in physical-layer security is to transmit  artificial noise (AN) from the base station (BS) to contaminate the Eve's received signal \cite{mukherjee2009fixed,swindlehurst2009fixed,goel2008guaranteeing}. For these AN-aided methods, a portion of transmit power is assigned to the artificially generated noise to interfere the Eve, which should be carefully designed. For AN-aided secrecy systems, while most of the existing AN-aided design papers focused on the MISO wiretap channel and null-space AN \cite{khisti2010secure,zhou2010secure}, designing the transmit precoding (TPC) matrix together with AN covariance matrix for the MIMO wiretap channel is more challenging \cite{li2013transmit}.

In general, the secrecy rate (SR) achieved by the mutual information difference between the legitimate IR and the Eve is limited by the channel difference between the BS-IR link and the BS-Eve link. The AN-aided method can further improve the SR, but it consumes the transmit power destined for the legitimate IR. When the transmit power is confined, the performance bottleneck always exists for the AN-aided secure communication. To conquer the dilemma, the recently proposed intelligent reflecting surface (IRS) technique can be exploited. Since higher SR can be achieved by enhancing the channel quality in the BS-IR link and degrading the channel condition in the BS-Eve link, the IRS can serve as a powerful complement to AN-aided secure communication due to its capability of reconfiguring the wireless propagation environment.

The IRS technique has been regarded as a revolutionary technique to control and reconfigure the wireless environment \cite{di2019smart,qingqing2019towards},\cite{huang2019holographic}. An IRS comprises an array of reflecting elements, which can reflect the incident electromagnetic (EM) wave passively, and the complex reflection coefficient contains the phase shift and amplitude. In practical applications, the phase shifts of the reflection coefficients are discrete due to the manufacturing cost\cite{wu2019beamforming}. However, many works on IRS aided wireless communications are based on the assumption of continuous phase shifts \cite{huang2019reconfigurable},\cite{huang2020reconfigurable}. To investigate the potential effect of IRS on the secure communication, we also assume continuous phase shifts to simplify the problem. We evaluate its impact on the
system performance in the simulation section. Theoretically, the reflection amplitude of each IRS element can be adjusted for different purpose \cite{wu2019towards}. However, considering the hardware cost, the reflection amplitude is usually assumed to be 1 for simplicity. Hence, by smartly tuning the phase shifts with a preprogrammed controller, the direct signals from the BS and the reflected signals from the IRS can be combined constructively or destructively according to different requirements. In comparison to the existing related techniques which the IRS resembles, such as active intelligent surface \cite{hu2018beyond}, traditional reflecting surfaces\cite{ford1984electromagnetic}, backscatter communication \cite{yang2017modulation} and amplify-and-forward (AF) relay \cite{zhang2009optimal}, the IRSs have the advantages of flexible reconfiguration on the phase shifts in real time, minor additional power consumption, easy installation with many reflecting elements, etc. Furthermore, due to the light weight and compact size, the IRS can be integrated into the traditional communication systems with minor modifications \cite{pan2019multicell}. Because of these appealing virtues, IRS has  introduced into various wireless communication systems, including the single-user case \cite{yu2019miso,yang2019intelligent}, the downlink  multiuser case \cite{wu2019intelligent,huang2019reconfigurable,guo2019weighted,nadeem2019large,zhou2019intelligent}, mobile edge computing \cite{bai2019latency}, wireless information and power transfer design \cite{pan2019intelligent}, and the physical layer security design \cite{yu2019enabling,cui2019secure,shen2019secrecy,chen2019intelligent}.

IRS is  promising to strengthen the system security of wireless communication. In \cite{yu2019enabling,shen2019secrecy,feng2019physical}, the authors investigated the problem of maximizing the achievable SR in a secure MISO communication system aided by IRS, where both the legitimate user and eavesdropper are equipped with a  single antenna. The TPC matrix at the BS and the phase shifts at the IRS were optimized by an alternate optimization (AO) strategy. To handle the nonconvex unit modulus constraint, the semidefinite relaxation (SDR) \cite{ma2010semidefinite}, majorization-minimization (MM) \cite{huang2019reconfigurable,sun2016majorization}, complex circle manifold (CCM) \cite{absil2009optimization} techniques were proposed to optimize phase shifts. An IRS-assisted MISO secure communication with a single IR and single Eve was also considered in \cite{cui2019secure}, but it was limited to a special scenario, where the Eve has a stronger channel than the IR, and the two channels from BS to Eve and IR are highly correlated. Under this assumption, the transmit beamforming and the IRS reflection beamforming are jointly optimized to improve the SR. Similarly, a secure IRS-assisted downlink MISO broadcast system was considered in \cite{chen2019intelligent}, and it assumes that multiple legitimate IRs and multiple Eves are in the same directions to the BS, which implies that the IR channels are highly correlated with the Eve channels. \cite{feng2019secure} considered the transmission design for an IRS-aided secure MISO communication with a single IR and single Eve, in which the system energy consumption is minimized under two assumptions that the channels of access point (AP)-IRS links are rank-one and full-rank. An IRS-assisted MISO network with cooperative jamming was investigated in \cite{wang2019energy}. The physical layer security in a simultaneous wireless information and power transfer (SWIPT) system was considered with the aid of IRS \cite{shi2019enhanced}. However, there are a paucity of papers considering the IRS-assisted secure communication with AN. A secure MISO communication system aided by the transmit jamming and AN was considered in \cite{guan2019intelligent}, where a large number of Eves exist, and the AN beamforming vector and jamming vector were optimized to reap the additional degrees of freedom (DoF) brought by the IRS. \cite{xu2019resource} investigated the resource allocation problem in an IRS-assisted MISO communication by jointly optimizing the beamforming vectors, the phase shifts of the IRS, and AN covariance matrix for secrecy rate maximization (SRM), but the direct BS-IRs links and direct BS-Eves link are assumed to be blocked.

Although a few papers have studied security enhancement for an AN-aided system through the IRS, the existing papers related to this topic either only studied the MISO scenario or assumed special settings to the channels. The investigation on the MIMO scenario with general channel settings is absent in the existing literature. Hence, we investigate this problem in this paper by employing an IRS in an AN-aided MIMO communication system for the physical layer security enhancement. Specifically, by carefully designing the phase shifts of the IRS, the reflected signals are combined with the direct signals constructively for enhancing the data rate at the IR and destructively for decreasing the rate at the Eve. As a result, the TPC matrix and AN covariance matrix at the BS can be designed flexibly with a higher DoF than the case without IRS. In this work, the TPC matrix, AN covariance matrix and the phase shift matrix are jointly optimized. Since these optimization variables are highly coupled, an efficient algorithm based on the block coordinate descent (BCD) and MM techniques for solving the problem is proposed.

We summarize our main contributions as follows:
\begin{enumerate}
  \item This is the first research on exploiting an IRS to enhance security in AN-aided MIMO communication systems. Specifically, an SRM problem is formulated by jointly optimizing the TPC matrix and AN covariance matrix at the BS, together with the phase shifts of the IRS subject to maximum transmit power limit and the unit modulus constraint of the phase shifters. The objective function (OF) of this problem is the difference of two Shannon capacity expressions, thus is not jointly concave over the three highly-coupled variables. To handle it, the popular minimum mean-square error (MMSE) algorithm is used to reformulate the SRM problem.
  \item The BCD algorithm is exploited to optimize the variables alternately. Firstly, given the phase shifts of IRS, the optimal TPC matrix and AN covariance matrix are obtained in closed form by utilizing the Lagrangian multiplier method. Then, given the TPC matrix and AN covariance matrix, the optimization problem for IRS phase shifts is transformed by sophisticated matrix manipulations into a quadratically constrained quadratic program (QCQP) problem subject to unit modulus constraints. To solve it, the MM algorithm is utilized, where the phase shifts are derived in closed form iteratively. Based on the BCD-MM algorithm, the original formulated SRM problem can be solved efficiently.
  \item The SRM problem is also extended to the more general scenario of multiple legitimate IRs. A new BCD algorithm is proposed to solve it, where the optimal TPC matrix and AN covariance matrix are obtained by solving a QCQP problem, and the unit modulus constraint is handled by the penalty convex-concave procedure (CCP) method.
  \item The simulation results confirm that on the one hand, the IRS can greatly enhance the security of an AN-aided MIMO communication system; on the other hand, the phase shifts of IRS should be properly optimized. Simulation results also show that larger IRS element number and more transmit power is beneficial to the security. Moreover, properly-selected IRS location and good channel states of the IRS-related links are important to realize the full potential of IRS.
\end{enumerate}

This paper is organized as follows. Section II provides the signal model of an AN-aided MIMO communication system assisted by an IRS, and the SRM problem formulation. The SRM problem is reformulated in Section III, where the BCD-MM algorithm is proposed to optimize the TPC
matrix, AN covariance matrix and phase shifts of IRS. Section IV extends the SRM problem to a more general scenario of multiple IRs. In Section V, numerical simulations are given to validate the algorithm efficiency and security enhancement. Section VI concludes this paper.

\emph{Notations}: Throughout this paper, boldface lower case, boldface upper case and regular letters are used to denote vectors, matrices, and scalars respectively. ${\bf{X}} \odot {\bf{Y}}$ is the Hadamard product of $\bf X$ and $\bf Y$. ${\rm{Tr}}\left( {\bf{X}} \right)$ and $\left| {\bf{X}} \right|$ denote the trace and determinant of ${\bf{X}}$ respectively. ${{\mathbb{ C}}^{M \times N}}$ denotes the space of $M \times N$ complex matrices. ${\rm{Re}}\{\cdot\}$ and $\arg\{\cdot\}$ denote the real part of a complex value and the extraction of phase information respectively. ${\rm{diag}}\{\cdot\}$ is the operator for diagonalization. ${\cal C}{\cal N}({\bm{\mu}},{\bf{Z}})$ represents a circularly symmetric complex gaussian (CSCG) random vector with mean ${\bm{\mu}}$ and covariance matrix ${\bf{Z}}$. ${\left( \cdot \right)^{\rm{T}}}$, ${\left(  \cdot \right)^{\rm{H}}}$ and ${\left( \cdot \right)^{\rm{\ast}}}$ denote the transpose, Hermitian and conjugate operators respectively. $(\cdot)^{ \star }$ stands for the optimal value, and $(\cdot)^{ \dag }$ means the pseudo-inverse. $[\cdot]^{+}$ is the projection onto the non-negative number, i.e, if $y=[x]^{+}$, then $y=\rm{max}\{0,x\}$.

\section{Signal Model and Problem Formulation}\label{system}
\subsection{Signal Model}

\begin{figure}[h!]
\centering
\includegraphics[width=3.5in]{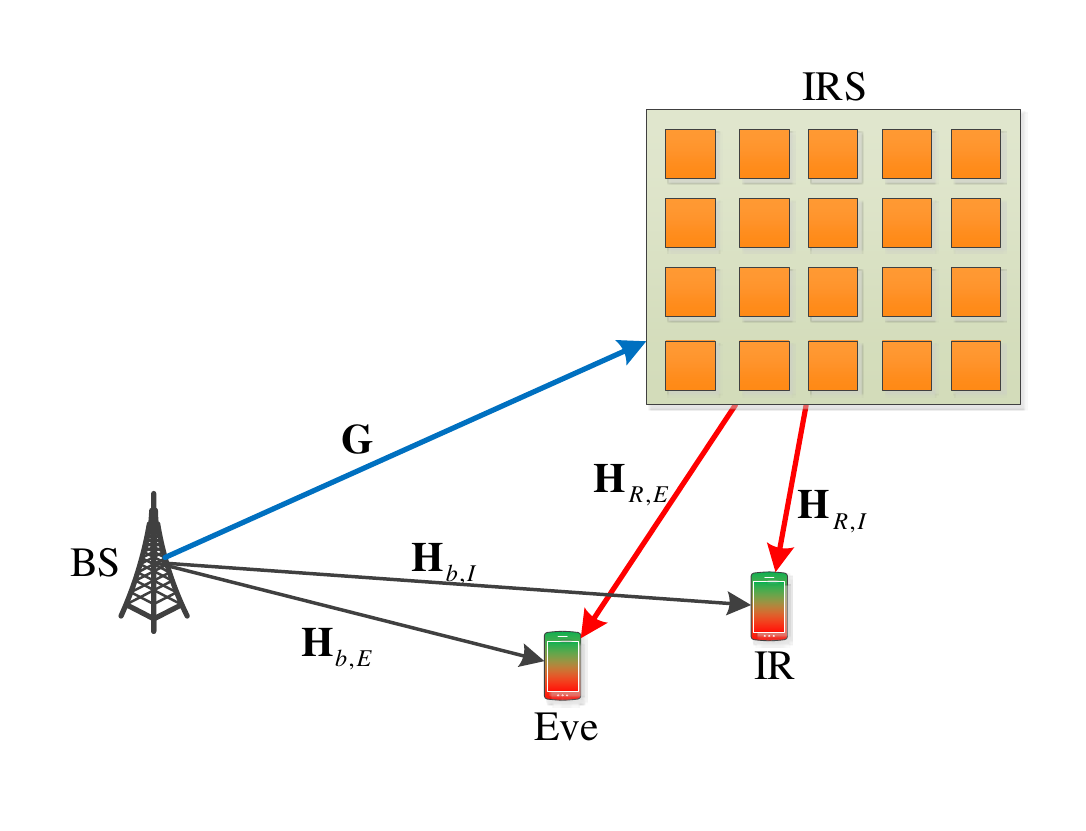}
\caption{An AN-aided MIMO secure communication system with IRS.}\vspace{-0.8cm}
\label{fig1}
\end{figure}

We consider an IRS-aided communication network shown in Fig.~\ref{fig1} that consists of a BS, a legitimate IR and an Eve, all of which are equipped with multiple antennas. The number of transmit antennas at the BS is ${{N}_{T}}\ge 1$, and the numbers of receive antennas at the legitimate IR and Eve are ${{N}_{I}}\ge 1$ and ${{N}_{E}}\ge 1$ respectively. To ensure secure transmission from the BS to the IR, the AN is sent from the BS to interfere the eavesdropper to achieve strong secrecy.

With above assumptions, the BS employed the TPC matrix to transmit data streams with AN. The transmitted signal can be modeled as
\begin{align}
{\bf{x}} = {\bf{Vs}} + {\bf{n}}, \label{eq1t}
\end{align}
where ${\bf{V}}\in {{\mathbb{C}}^{{{N}_{T}}\times d}}$ is the TPC matrix; the number of data streams is $d\le \min ({{N}_{T}},{{N}_{I}})$; the transmitted data towards the IR is $\mathbf{s} \sim \mathcal{C}\mathcal{N}(0,{\mathbf{I}_{d}})$; and $\mathbf{n} \in {\cal C}{\cal N}({\bm{0}},{\bf{Z}})$ represents the AN random vector with zero mean and covariance matrix $\mathbf{Z}$.

Assuming that the wireless signals are propagated in a non-dispersive and narrow-band way, we model the equivalent channels of the BS-IRS link, the BS-IR link, the BS-Eve link, the IRS-IR link, the IRS-Eve link by the matrices $\mathbf{G} \in {{\mathbb{C}}^{M\times {{N}_{T}}}}$, ${{\bf{H}}_{b,I}}\in {{\mathbb{C}}^{{{N}_{I}}\times {{N}_{T}}}}$, ${{\bf{H}}_{b,E}}\in {{\mathbb{C}}^{{{N}_{E}}\times {{N}_{T}}}}$, ${{\bf{H}}_{R,I}}\in {{\mathbb{C}}^{{{N}_{I}}\times M}}$,${{\bf{H}}_{R,E}}\in {{\mathbb{C}}^{{{N}_{E}}\times M}}$, respectively. The phase shift coefficients of IRS are collected in a diagonal matrix defined by ${\bf{\Phi}} =\text{diag }\!\!\{\!\!\text{ }{{\phi }_{1}},\cdots ,{{\phi }_{m}},\cdots ,{{\phi }_{M}}\text{ }\!\!\}\!\!\text{ }$ and ${{\phi }_{m}}={{e}^{j{{\theta }_{m}}}}$, where ${{\theta }_{m}}\in [0,2\pi ]$ denotes the phase shift of the $m$-th reflection element. The multi-path signals that have been reflected by multiple times are considered to be absorbed and diffracted, then the signal received at the legitimate IR is given by
\begin{equation}
{\bf{y}}_I = ({\bf{H}}_{b,I} + {\bf{H}}_{R,I}{\bf{\Phi}} {\bf{G}}) {\bf{x}} + {\bf{n}}_{I}, \label{eq2t}
\end{equation}
where ${{\bf{n}}_{I}}$ is the random noise vector at IR obeying the distribution ${{\bf{n}}_{I}}\sim \mathcal{C}\mathcal{N}({\bf{0}},\sigma _{I}^{2}{\bf{I}}_{{{N}_{I}}})$. The signal received at the Eve is
\begin{align}
{\bf{y}}_E = ({\bf{H}}_{b,E} + {\bf{H}}_{R,E}{\bf{\Phi}} {\bf{G}}){\bf{x}} + {{\bf{n}}_E}, \label{eq3t}
\end{align}
where ${\bf{n}}_{E}$ is the Eve's noise vector following the distribution ${\bf{n}}_{E}\sim \mathcal{C}\mathcal{N}({\bf{0}},\sigma _{E}^{2}{\bf{I}}_{{{N}_{E}}})$.

Assume that the BS has acquired the prior information of all the channel state information (CSI). Then the BS is responsible for optimizing the IRS phase shifts and sending them back to the IRS controller through a separate low-rate communication link such as wireless links \cite{wu2019towards},\cite{wu2019beamforming} or wired lines \cite{tan2016increasing}. The assumption of perfect CSI knowledge is idealistic, since the CSI estimation for IRS networks is challenging. However, the algorithms developed allow us to derive the relevant performance upper bounds for realistic scenarios in the presence of realistic CSI errors. Recently, we have investigated the design of robust and secure transmission in IRS-aided MISO wireless communication systems in \cite{hong2020robust} by considering the statistical CSI error model associated with the cascaded channels for the eavesdropper. Its extension to the MIMO scenario will be studied in our future work.

Upon substituting $\bf{x}$ into \eqref{eq2t}, ${\bf{y}}_{I}$ can be rewirtten as
\begin{align}
{\bf{y}}_I = {\hat {\bf{H}}_I}({\bf{V}}{\bf{s}} + {\bf{n}}) + {{\bf{n}}_I}{\rm{ = }}{\hat {\bf{H}}_I}{\bf{V}}{\bf{s}} + {\hat {\bf{H}}_I}{\bf{n}} + {{\bf{n}}_I}, \label{eq4t}
\end{align}
where ${{\hat{\bf{H}}}_{I}}\overset{\triangle}{=} {{\bf{H}}_{b,I}}+{{\bf{H}}_{R,I}}{\bf{\Phi}} {\bf{G}}$ is defined as the equivalent channel spanning from the BS to the legitimate IR. Then, the data rate (bit/s/Hz) achieved by the legitimate IR is given by
\begin{align}
{R_I}({\bf{V}},{\bf{\Phi}} ,{\bf{Z}}) = {\rm{log}}\left| {{\bf{I}} + {{\hat {\bf{H}}}_I}{\bf{V}}{{\bf{V}}^H}\hat {\bf{H}}_I^H{\bf{J}}_I^{ - 1}} \right|, \label{eq5t}
\end{align}
where ${{\bf{J}}_{I}}$ is the interference-plus-noise covariance matrix given by ${{\bf{J}}_{I}}\overset{\triangle}{=}  {{\hat{\bf{H}}}_{I}}{\bf{Z}}{{\hat{\bf{H}}}_{I}}^{H}+\sigma _{I}^{2}{{\bf{I}}_{{{N}_{I}}}}$.

Upon substituting $\bf{x}$ into \eqref{eq3t}, ${{\bf{y}}_{E}}$ can be rewritten as
\begin{align}
{\bf{y}}_E = {\hat {\bf{H}}_E}({\bf{Vs}} + {\bf{n}}) + {\bf{n}}_E = {\hat {\bf{H}}_E}{\bf{Vs}} + {\hat {\bf{H}}_E}{\bf{n}} + {\bf{n}}_E,
 \label{eq6t}
\end{align}
where ${{\hat{\bf{H}}}_{E}}\overset{\triangle}{=} {{\bf{H}}_{b,E}}+{{\bf{H}}_{R,E}}{\bf{\Phi}} {\bf{G}}$ is defined as the equivalent channel spanning from the BS to the Eve. Then, the data rate (bit/s/Hz) achieved by the Eve is given by
\begin{align}
{R_E}({\bf{V}},{\bf{\Phi}} ,{\bf{Z}}) = {\rm{log}}\left| {{\bf{I}} + {{\hat {\bf{H}}}_E}{\bf{V}}{{\bf{V}}^H}\hat {\bf{H}}_E^H{\bf{J}}_E^{ - 1}} \right|, \label{eq7t}
\end{align}
where ${{\bf{J}}_{E}}$ is the interference-plus-noise covariance matrix given by ${{\bf{J}}_{E}}\overset{\triangle}{=} {{\hat{\bf{H}}}_{E}}{\bf{Z}}{{\hat{\bf{H}}}_{E}}^{H}+\sigma _{E}^{2}{{\bf{I}}_{{{N}_{E}}}}$.
The achievable secrecy rate is given by
\begin{align}\label{CANorign}
{{\rm{C}}_{AN}}{\rm{(}}{\bf{V}},{\bf{\Phi}} ,{\bf{Z}})&{\rm{ = }}[{R_I}({\bf{V}},{\bf{\Phi}} ,{\bf{Z}}) - {R_E}({\bf{V}},{\bf{\Phi}} ,{\bf{Z}}){]^ + }\nonumber \\
&= {\rm{log}}\left| {{\bf{I}} + {{\hat {\bf{H}}}_I}{\bf{V}}{{\bf{V}}^H}\hat {\bf{H}}_I^H{\bf{J}}_I^{ - 1}} \right| - {\rm{log}}\left| {{\bf{I}} + {{\hat {\bf{H}}}_E}{\bf{V}}{{\bf{V}}^H}\hat {\bf{H}}_E^H{\bf{J}}_E^{ - 1}} \right| \nonumber \\
&= {\rm{log}}\left| {{\bf{I}} + {{\hat {\bf{H}}}_I}{\bf{V}}{{\bf{V}}^H}\hat {\bf{H}}_I^H{{({{\hat {\bf{H}}}_I}{\bf{Z}}{{\hat {\bf{H}}}_I}^H + \sigma _I^2{{\bf{I}}_{{N_I}}})}^{ - 1}}} \right| \nonumber \\
&\quad - {\rm{log}}\left| {{\bf{I}} + {{\hat {\bf{H}}}_E}{\bf{V}}{{\bf{V}}^H}\hat {\bf{H}}_E^H{{({{\hat {\bf{H}}}_E}{\bf{Z}}{{\hat {\bf{H}}}_E}^H + \sigma _E^2{{\bf{I}}_{{N_E}}})}^{ - 1}}} \right|.
\end{align}

\subsection{Problem Formulation}

With the aim for maximizing SR, the TPC matrix ${\bf{V}}$ at the BS, the AN covariance matrix ${\bf{Z}}$ at the BS, and the phase shift matrix ${\bf{\Phi}}$ at the IRS should be optimized jointly subject to the constraints of the maximum transmit power and unit modulus of phase shifts. Hence, we formulate the SRM problem as
\begin{subequations} \label{optorig}
\begin{align}
\ &\ \underset{{\bf{V}},{\bf{\Phi}} ,{\bf{Z}}}{\mathop\text{max}} \ \ {{\rm{C}}_{AN}}{\rm{(}}{\bf{V}},{\bf{\Phi}} ,{\bf{Z}}{\rm{)}} \label{eq9ta} \\
& \ \ \text{s.t.} \quad \ {\rm{Tr(}}{\bf{V}}{{\bf{V}}^H}{\rm{ + }}{\bf{Z}}{\rm{)}} \le {P_T},\label{eq9tb} \\
& \quad \quad  \quad {\bf{Z}}\succeq 0, \label{eq9tc}\\
& \quad \quad  \quad \! \left| {{{{\phi}} _m}} \right| = 1,m = 1, \cdots ,M, \label{phaseshifconstrnt}
\end{align}
\end{subequations}
where ${P_T}$ is the maximum transmit power limit. The optimal value of SR in (9) is always non-negative, which can be proved by using contradiction. Assume that the optimal value of SR is negative, then we can simply set the TPC matrix ${\bf{V}}$ to zero matrix, and the resulted SR will be equal to zero, which is larger than a negative SR.

By variable substitution ${\bf{Z}}={{\bf{V}}_{E}}{{\bf{V}}^{H}_{E}}$, where ${{\bf{V}}_{E}}\in {{\mathbb{C}}^{{{N}_{T}}\times {{N}_{T}}}}$, Problem (\ref{optorig}) is equivalent to
\begin{subequations} \label{optorigVE}
\begin{align}
&\ \underset{{\bf{V}} ,{{\bf{V}}_E},{\bf{\Phi}}}{\mathop\text{max}} \ \ {{\rm{C}}_{AN}}{\rm{(}}{\bf{V}},{{\bf{V}}_{E}},{\bf{\Phi}} {\rm{)}} \label{optorigVE_a} \\
& \ \ \text{s.t.} \quad \ {\rm{Tr(}}{\bf{V}}{{\bf{V}}^H}{\rm{ + }}{{\bf{V}}_{E}}{{\bf{V}}^{H}_{E}}{\rm{)}} \le {P_T},\label{optorigVE_b} \\
& \quad \quad  \quad \! \left| {{{{\phi}} _m}} \right| = 1,m = 1, \cdots ,M, \label{optorigVE_c}
\end{align}
\end{subequations}
where the OF of (\ref{optorigVE_a}) is obtained by substituting ${\bf{Z}}={{\bf{V}}_{E}}{{\bf{V}}_{E}}^{H}$ into (\ref{CANorign}). In (\ref{optorigVE_a}), the expression of OF is difficult to tackle, and the variables of ${\bf{V}}$, ${\bf{V}}_{E}$ and ${\bf{\Phi}}$ are coupled with each other, which make Problem (\ref{optorigVE}) difficult to solve. In addition, the unit modulus constraint imposed on the phase shifts in (\ref{optorigVE_c}) aggravates the difficulty. In the following, we provide a low-complexity algorithm to solve this problem.

\vspace{-0.4cm}\section{A Low-Complexity Algorithm of BCD-MM}\label{algo}
Firstly, the OF of Problem (\ref{optorigVE}) is reformulated into a more tractable expression equivalently. Then, the BCD-MM method is proposed for optimizing the TPC matrix ${\bf{V}}$, ${\bf{V}}_{E}$, and the phase shift matrix ${\bf{\Phi}}$ alternately.
\subsection{Reformulation of the Original Problem}

Firstly, the achievable SR ${{\rm{C}}_{AN}}{\rm{(}}{\bf{V}},{{\bf{V}}_E},{\bf{\Phi}}{\rm{)}}$ in (\ref{CANorign}) can be further simplified as
\begin{align} \label{CAN}
{{\rm{C}}_{AN}}\rm{(}{\bf{V}},{{\bf{V}}_E},{\bf{\Phi}}\rm{)}&{\rm{ = log}}\left| {{\bf{I}}_{{N_I}} + {{\hat {\bf{H}}}_I}{\bf{V}}{{\bf{V}}^H}\hat {\bf{H}}_I^H{{({{\hat {\bf{H}}}_I}{\bf{Z}}{{\hat {\bf{H}}}_I}^H + \sigma _I^2{{\bf{I}}_{{N_I}}})}^{ - 1}}} \right|{\rm{ + log}}\left| {{{\hat {\bf{H}}}_E}{\bf{Z}}{{\hat {\bf{H}}}_E}^H + \sigma _E^2{{\bf{I}}_{{N_E}}}} \right| \nonumber \\
&\quad - {\rm{log}}\left| {{{\hat {\bf{H}}}_E}{\bf{Z}}{{\hat {\bf{H}}}_E}^H + \sigma _E^2{{\bf{I}}_{{N_E}}} + {{\hat {\bf{H}}}_E}{\bf{V}}{{\bf{V}}^H}\hat {\bf{H}}_E^H} \right| \nonumber  \\
&=\underbrace {{\rm{log}}\left| {{\bf{I}}_{{N_I}} + {{\hat {\bf{H}}}_I}{\bf{V}}{{\bf{V}}^H}\hat {\bf{H}}_I^H{{({{\hat {\bf{H}}}_I}{{\bf{V}}_E}{{\bf{V}}^H_E}{{\hat {\bf{H}}}_I}^H + \sigma _I^2{{\bf{I}}_{{N_I}}})}^{ - 1}}} \right|}_{{f_1}} \nonumber \\
&\quad {\rm{ + }}\underbrace {{\rm{log}}\left| {{{\bf{I}}_{{N_E}}} + {{\hat {\bf{H}}}_E}{{\bf{V}}_E}{{\bf{V}}^H_E}{{\hat {\bf{H}}}_E}^H(\sigma _E^2{{\bf{I}}_{{N_E}}})^{-1}} \right|}_{{f_2}} \nonumber  \\
&\quad  \underbrace {- {\rm{log}}\left| {{{\bf{I}}_{{N_E}}} + \sigma _E^{-2}{{\hat {\bf{H}}}_E}({\bf{V}}{{\bf{V}}^H}+{{\bf{V}}_E}{{\bf{V}}^H_E})\hat {\bf{H}}_E^H} \right|}_{{f_3}}.
\end{align}
The expression in $f_1$ represents the data rate of the legitimate IR, which can be reformulated by exploiting the relationship between the data rate and the mean-square error (MSE) for the optimal decoding matrix. Specifically, the linear decoding matrix ${{\bf{U}}_{I}}\in {{\mathbb{C}}^{{{N}_{T}}\times {d}}}$ is applied to estimate the signal vector $\hat{{\bf{s}}}$ for the legitimate IR, and the MSE matrix of the legitimate IR is given by
\begin{align} \label{MSE_EI}
{{\bf{E}}_I}({{\bf{U}}_I},{\bf{V}},{{\bf{V}}_E})& \buildrel \Delta \over = {{\mathbb{E}}_{{\bf{s}},{\bf{n}},{{\bf{n}}_I}}}\left[ {(\hat {\bf{s}}-{\bf{s}} ){{(\hat {\bf{s}}-{\bf{s}} )}^H}} \right] \nonumber \\
&{\rm{ = }}({{\bf{U}}_I}^H{{\hat {\bf{H}}}_I}{\bf{V}}-{\bf{I}}_{d}){({{\bf{U}}_I}^H{{\hat {\bf{H}}}_I}{\bf{V}}-{\bf{I}}_{d} )^H} + {{\bf{U}}_I}^H({{\hat {\bf{H}}}_I}{{\bf{V}}_E}{{\bf{V}}_E}^H{{\hat {\bf{H}}}_I}^H{\rm{ + }}\sigma _I^2{{\bf{I}}_{{N_I}}}){{\bf{U}}_I}.
\end{align}
By introducing an auxiliary matrix ${{\bf{W}}_I}\succeq 0$, ${{\bf{W}}_I}\in {{\mathbb{C}}^{{d}\times {d}}}$ and exploiting the fact 3) of Lemma 4.1 in \cite{shi2015secure}, we have
\begin{align} \label{lowboundh1f1}
f_1&{\rm{ = }} \mathop \text{max}\limits_{{{\bf{U}}_I},{{\bf{W}}_I}\succeq 0} h_1({{\bf{U}}_I},{\bf{V}},{{\bf{V}}_E},{{\bf{W}}_I})\nonumber \\
&\buildrel \Delta \over = \mathop \text{max}\limits_{{{\bf{U}}_I},{{\bf{W}}_I}\succeq 0} \log \left| {{{\bf{W}}_I}} \right| - {\rm{Tr}}({{\bf{W}}_I}{{\bf{E}}_I}({{\bf{U}}_I},{\bf{V}},{{\bf{V}}_E})) + d.
\end{align}
$h_1({{\bf{U}}_I},{\bf{V}},{{\bf{V}}_E},{{\bf{W}}_I})$ is concave with respect to (w.r.t.) each matrix of the matrices ${{\bf{U}}_I}$,${\bf{V}}$,${{\bf{V}}_E}$,${{\bf{W}}_I}$ by fixing the other three matrices. According to the facts 1) and 2) of Lemma 4.1 in \cite{shi2015secure}, the optimal ${{\bf{U}}^ {\star}_{I}}$, ${{\bf{W}}^ {\star}_I}$ to achieve the maximum value of $h_1({{\bf{U}}_I},{\bf{V}},{{\bf{V}}_E},{{\bf{W}}_I})$ is given by
\begin{align}\label{optUI}
{{\bf{U}}^ {\star}_I}{\rm{ = }}\text{arg} \mathop \text{max}\limits_{{{\bf{U}}_I}} h_1({{\bf{U}}_I},{\bf{V}},{{\bf{V}}_E},{{\bf{W}}_I}) {\rm{ = }}({\hat {\bf{H}}_I}{{\bf{V}}_E}{{\bf{V}}^H_E}{\hat {\bf{H}}_I}^H{\rm{ + }}\sigma _I^2{{\bf{I}}_{{N_I}}}{\rm{ + }}{\hat {\bf{H}}_I}{\bf{V}}{{\bf{V}}^H}{\hat {\bf{H}}_I}^H)^{-1}{\hat {\bf{H}}_I}{\bf{V}},
\end{align}
\begin{align}\label{optWI}
{{\bf{W}}^ {\star}_I}{\rm{ = }}\text{arg} \mathop \text{max}\limits_{{{\bf{W}}_I}\succeq 0} h_1({{\bf{U}}_I},{\bf{V}},{{\bf{V}}_E},{{\bf{W}}_I}) {\rm{ = [}}{{\bf{E}}^ {\star}_I}({{\bf{U}}^ {\star}_I},{\bf{V}},{{\bf{V}}_E}){]^{ - 1}},
\end{align}
where ${{\bf{E}}^ {\star}_I}$ is obtained by plugging the expression of ${{\bf{U}}^ {\star}_I}$ into ${{\bf{E}}_I}({{\bf{U}}_I},{\bf{V}},{{\bf{V}}_E})$ as
\begin{align}\label{optEI}
{{\bf{E}}^ {\star}_I}({{\bf{U}}^ {\star}_I},{\bf{V}},{{\bf{V}}_E}){\rm{ = }}( {{\bf{U}}^{{\star}H}_I}{\hat {\bf{H}}_I}{\bf{V}}-{\bf{I}}_{d}){({{\bf{U}}^{{\star}H}_I}{\hat {\bf{H}}_I}{\bf{V}}-{\bf{I}}_{d})^H} + {{\bf{U}}^{{\star}H}_I}({\hat {\bf{H}}_I}{{\bf{V}}_E}{{\bf{V}}^H_E}{\hat {\bf{H}}_I}^H{\rm{ + }}\sigma _I^2{{\bf{I}}_{{N_I}}}){{\bf{U}}^{{\star}}_I}{{\rm{}}}.
\end{align}
Similarly, by introducing the auxiliary variables ${{\bf{W}}_E}\succeq 0$, ${{\bf{W}}_E}\in {{\mathbb{C}}^{{{N}_{T}}\times {{N}_{T}}}}$, ${{\bf{U}}_{E}}\in {{\mathbb{C}}^{{{N}_{E}}\times {{N}_{T}}}}$, and exploiting the fact 3) of Lemma 4.1 in \cite{shi2015secure}, we have
\begin{align}  \label{lowboundh2f2}
f_2&{\rm{ = }} \mathop \text{max}\limits_{{{\bf{U}}_E},{{\bf{W}}_E}\succeq 0} h_2({{\bf{U}}_E},{{\bf{V}}_E},{{\bf{W}}_E}) \nonumber \\
&\buildrel \Delta \over = \mathop \text{max}\limits_{{{\bf{U}}_E},{{\bf{W}}_E}\succeq 0} \log \left| {{{\bf{W}}_E}} \right| - {\rm{Tr}}({{\bf{W}}_E}{{\bf{E}}_E}({{\bf{U}}_E},{{\bf{V}}_E})) + {N}_{t},
\end{align}
$h_2({{\bf{U}}_E},{{\bf{V}}_E},{{\bf{W}}_E})$ is concave w.r.t each matrix of the matrices ${{\bf{U}}_E}$,${{\bf{V}}_E}$,${{\bf{W}}_E}$ when the other two matrices are given. According to the facts 1) and 2) of Lemma 4.1 in \cite{shi2015secure}, the optimal ${{\bf{U}}^ {\star}_{E}}$, ${{\bf{W}}^ {\star}_E}$ to achieve the maximum value of $h_2({{\bf{U}}_E},{{\bf{V}}_E},{{\bf{W}}_E})$ is given by
\begin{align} \label{optUE}
{{\bf{U}}^ {\star}_E}{\rm{ = }}\text{arg} \mathop \text{max}\limits_{{{\bf{U}}_E}} h_2({{\bf{U}}_E},{{\bf{V}}_E},{{\bf{W}}_E}) {\rm{ = }}(\sigma _E^2{{\bf{I}}_{{N_E}}}{\rm{ + }}{\hat {\bf{H}}_E}{{\bf{V}}_E}{{{\bf{V}}^H_E}}{\hat {\bf{H}}_E}^H)^{-1}{\hat {\bf{H}}_E}{{\bf{V}}_E},
\end{align}
\begin{align} \label{optWE}
{{\bf{W}}^ {\star}_E}{\rm{ = }}\text{arg} \mathop \text{max}\limits_{{{\bf{W}}_E}\succeq 0} h_2({{\bf{U}}_E},{{\bf{V}}_E},{{\bf{W}}_E}) {\rm{ = [}}{{\bf{E}}^ {\star}_E}({{\bf{U}}^ {\star}_E},{{\bf{V}}_E}){]^{ - 1}},
\end{align}
where ${{\bf{E}}^ {\star}_E}$ is obtained by plugging the expression of ${{\bf{U}}^ {\star}_E}$ into ${{\bf{E}}_E}({{\bf{U}}_E},{{\bf{V}}_E})$ as
\begin{align}\label{optEE}
\begin{array}{l}
{{\bf{E}}^ {\star}_E}({{\bf{U}}^ {\star}_E},{{\bf{V}}_E}) = ({{\bf{U}}_E}^{{\star}H}{{\hat {\bf{H}}}_E}{\bf{V}}_E-{\bf{I}}_{N_{T}}){({{\bf{U}}^{{\star}H}_E}{{\hat {\bf{H}}}_E}{\bf{V}}_E-{\bf{I}}_{N_{T}} )^H} + {{\bf{U}}^{{\star}H}_E}(\sigma _E^2{{\bf{I}}_{{N_E}}}){{\bf{U}}^ {\star}_E}.
\end{array}
\end{align}
By using the Lemma 1 in \cite{li2013transmit}, we have
\begin{align} \label{lowboundh3f3}
f_3&{\rm{ = }} \mathop \text{max}\limits_{{{\bf{W}}_X}\succeq 0} h_3({{\bf{V}}},{{\bf{V}}_E},{{\bf{W}}_X}) \nonumber \\
&{\rm{ = }} \mathop \text{max}\limits_{{{\bf{W}}_X}\succeq 0} \log \left| {{{\bf{W}}_X}} \right| - {\rm{Tr}}({{\bf{W}}_X}{{\bf{E}}_X}({{\bf{V}}},{{\bf{V}}_E})) + {N}_{E},
\end{align}
where ${{\bf{W}}_X}\succeq 0$, ${{\bf{W}}_X}\in {{\mathbb{C}}^{{{N}_{E}}\times {{N}_{E}}}}$ are the introduced auxiliary variable, and
\begin{align} \label{MSE_EX}
\begin{array}{l}
{{\bf{E}}_X}({{\bf{V}}},{{\bf{V}}_E})\buildrel \Delta \over = {{{\bf{I}}_{{N_E}}} + \sigma _E^{-2}{{\hat {\bf{H}}}_E}({\bf{V}}{{\bf{V}}^H}+{{\bf{V}}_E}{{\bf{V}}^H_E})\hat {\bf{H}}_E^H}.
\end{array}
\end{align}
$h_3({{\bf{V}}},{{\bf{V}}_E},{{\bf{W}}_X})$ is concave w.r.t each matrix of the matrices ${{\bf{V}}},{{\bf{V}}_E},{{\bf{W}}_X}$ when the other two matrices are given. The optimal ${{\bf{W}}^ {\star}_{X}}$ to achieve the maximum value of $h_3({{\bf{V}}},{{\bf{V}}_E},{{\bf{W}}_X})$ is
\begin{align} \label{optWX}
{{\bf{W}}^ {\star}_X}{\rm{ = }}\text{arg} \mathop \text{max}\limits_{{{\bf{W}}_X}\succeq 0} h_3({{\bf{V}}},{{\bf{V}}_E},{{\bf{W}}_X}) {\rm{ = [}}{{\bf{E}}_X}({{\bf{V}}},{{\bf{V}}_E}){]^{ - 1}}.
\end{align}

By substituting (\ref{lowboundh1f1}), (\ref{lowboundh2f2}), (\ref{lowboundh3f3}) into (\ref{CAN}), we have
\begin{align}\label{lowboundCAN}
 {{\rm{C}}_{AN}}{\rm{(}}{\bf{V}},{{\bf{V}}_E}{\rm{)}}&=\mathop \text{arg max}\limits_{{{\bf{U}}_I},{{\bf{W}}_I},{{\bf{U}}_E},{{\bf{W}}_E},{{\bf{W}}_X}} {{\rm{C}}^{l}_{AN}}({{\bf{U}}_I},{{\bf{W}}_I},{{\bf{U}}_E},{{\bf{W}}_E},{{\bf{W}}_X},{\bf{V}},{{\bf{V}}_E}),
\end{align}
where
\begin{align} \label{lowboundCANlow}
{{\rm{C}}^{l}_{AN}}({{\bf{U}}_I},{{\bf{W}}_I},{{\bf{U}}_E},{{\bf{W}}_E},{{\bf{W}}_X},{\bf{V}},{{\bf{V}}_E})\buildrel \Delta \over =&h_1({{\bf{U}}_I},{\bf{V}},{{\bf{V}}_E},{{\bf{W}}_I})+h_2({{\bf{U}}_E},{{\bf{V}}_E},{{\bf{W}}_E}) \nonumber \\
&+h_3({{\bf{V}}},{{\bf{V}}_E},{{\bf{W}}_X}).
\end{align}

Obviously, ${{\rm{C}}^{l}_{AN}}({{\bf{U}}_I},{{\bf{W}}_I},{{\bf{U}}_E},{{\bf{W}}_E},{{\bf{W}}_X},{\bf{V}},{{\bf{V}}_E})$ is a concave function for each of the matrices ${{\bf{U}}_I}$,${{\bf{W}}_I}$,${{\bf{U}}_E}$,${{\bf{W}}_E}$,${{\bf{W}}_X}$,${\bf{V}}$,${{\bf{V}}_E}$ when the other six matrices are given. By substituting (\ref{lowboundCAN}) into Problem (\ref{optorigVE}), we have the following equivalent problem:
\begin{subequations}\label{optorigVElowerbnd}
\begin{align}
\ &\ \underset{{{\bf{U}}_I},{{\bf{W}}_I}\succeq 0,{{\bf{U}}_E},{{\bf{W}}_E}\succeq 0,{{\bf{W}}_X}\succeq 0,{\bf{V}} ,{{\bf{V}}_E},{\bf{\Phi}}}{\mathop\text{max}} \ \ {{\rm{C}}^{l}_{AN}}({{\bf{U}}_I},{{\bf{W}}_I},{{\bf{U}}_E},{{\bf{W}}_E},{{\bf{W}}_X},{\bf{V}},{{\bf{V}}_E},{\bf{\Phi}})  \label{optorigVElowerbnda} \\
& \quad\quad\quad\quad\quad\quad \text{s.t.} \quad \ {\rm{Tr(}}{\bf{V}}{{\bf{V}}^H}{\rm{ + }}{{\bf{V}}_{E}}{{\bf{V}}_{E}}^{H}{\rm{)}} \le {P_{T}},\label{optorigVElowerbndb} \\
& \quad\quad\quad\quad\quad\quad \quad\quad \ \ \left| {{{{\phi}} _m}} \right| = 1,m = 1, \cdots ,M. \label{optorigVElowerbndc}
\end{align}
\end{subequations}
To solve Problem \eqref{optorigVElowerbnd}, we apply the BCD method, each iteration of which consists the following two sub-iterations. Firstly, with given ${\bf{V}} ,{{\bf{V}}_E},{\bf{\Phi}}$, update ${{\bf{U}}_I},{{\bf{W}}_I},{{\bf{U}}_E},{{\bf{W}}_E},{{\bf{W}}_X}$ by using \eqref{optUI}, \eqref{optWI}, \eqref{optUE}, \eqref{optWE}, \eqref{optWX} respectively. Secondly, with given ${{\bf{U}}_I},{{\bf{W}}_I},{{\bf{U}}_E},{{\bf{W}}_E},{{\bf{W}}_X}$, update ${\bf{V}} ,{{\bf{V}}_E},{\bf{\Phi}}$ by solving the following subproblem:
\begin{subequations} \label{optorigVElowerbndSmp}
\begin{align}
&\ \underset{{\bf{V}} ,{{\bf{V}}_E},{\bf{\Phi}}}{\mathop\text{min}} \ \ -\text{Tr}({{\mathbf{W}}_{I}}{{\mathbf{V}}^{H}}{{{\mathbf{\hat{H}}}}_{I}}^{H}{{\mathbf{U}}_{I}})-\text{Tr}({{\mathbf{W}}_{I}}{{\mathbf{U}}^{H}_{I}}{{{\mathbf{\hat{H}}}}_{I}}\mathbf{V})+\text{Tr}({{\mathbf{V}}^{H}}{{\mathbf{H}}_{V}}\mathbf{V}) \nonumber  \\
 &\quad \quad \quad \quad -\text{Tr}({{\mathbf{W}}_{E}}{{\mathbf{V}}^{H}_{E}}{{{\mathbf{\hat{H}}}}_{E}}^{H}{{\mathbf{U}}_{E}})-\text{Tr}({{\mathbf{W}}_{E}}{{\mathbf{U}}^{H}_{E}}{{{\mathbf{\hat{H}}}}_{E}}{{\mathbf{V}}_{E}})+\text{Tr}({{\mathbf{V}}^{H}_{E}}{{\mathbf{H}}_{VE}}{{\mathbf{V}}_{E}}) \label{optorigVElowerbndSmpa} \\
& \ \ \text{s.t.} \quad {\rm{Tr(}}{\bf{V}}{{\bf{V}}^H}{\rm{ + }}{{\bf{V}}_{E}}{{\bf{V}}^{H}_{E}}{\rm{)}} \le P_{T},\label{optorigVElowerbndSmpb} \\
& \quad \quad  \quad \! \left| {{{{\phi}} _m}} \right| = 1,m = 1, \cdots ,M, \label{optorigVElowerbndSmpc}
\end{align}
\end{subequations}
where
\begin{align} \label{HV}
{{\mathbf{H}}_{V}}={{\mathbf{\hat{H}}}_{I}}^{H}{{\mathbf{U}}_{I}}{{\mathbf{W}}_{I}}{{\mathbf{U}}^{H}_{I}}{{\mathbf{\hat{H}}}_{I}}+\sigma _E^{-2}\mathbf{\hat{H}}_{E}^{H}{{\mathbf{W}}_{X}}{{\mathbf{\hat{H}}}_{E}},
\end{align}
\begin{align} \label{HVE}
{{\mathbf{H}}_{VE}}={{\mathbf{\hat{H}}}_{I}}^{H}{{\mathbf{U}}_{I}}{{\mathbf{W}}_{I}}{{\mathbf{U}}^{H}_{I}}{{\mathbf{\hat{H}}}_{I}}+{{\mathbf{\hat{H}}}_{E}}^{H}{{\mathbf{U}}_{E}}{{\mathbf{W}}_{E}}{{\mathbf{U}}^{H}_{E}}{{\mathbf{\hat{H}}}_{E}}+\sigma _E^{-2}{{\mathbf{\hat{H}}}_{E}}^{H}{{\mathbf{W}}_{X}}{{\mathbf{\hat{H}}}_{E}}.
\end{align}

Problem (\ref{optorigVElowerbndSmp}) is obtained from Problem \eqref{optorigVElowerbnd} by taking the ${{\bf{U}}_I},{{\bf{W}}_I},{{\bf{U}}_E},{{\bf{W}}_E},{{\bf{W}}_X}$ as constant values, and the specific derivations are given in Appendix \ref{firstconstrntDeriv}.

It is obvious that Problem (\ref{optorigVElowerbndSmp}) is much easier to tackle than Problem (\ref{optorigVE}) due to the convex quadratic OF in (\ref{optorigVElowerbndSmpa}). Now, we devote to solve Problem (\ref{optorigVElowerbndSmp}) equivalently instead of Problem (\ref{optorigVE}), and the matrices ${\bf{V}}$, ${\bf{V}}_{E}$, and phase shift matrix $\mathbf{\Phi }$ will be optimized.

\vspace{-0.4cm}\subsection{Optimizing the Matrices ${\bf{V}}$ and ${\bf{V}}_{E}$}\label{kodsijcosakpdc}
In this subsection, the TPC matrix ${\bf{V}}$ and matrix ${\bf{V}}_{E}$ are optimized by fixing $\mathbf{\Phi }$. Specifically, the unit modulus constraint on the phase shifts $\mathbf{\Phi }$ is removed, and the updated optimization problem reduced from Problem (\ref{optorigVElowerbndSmp}) is given by
\begin{subequations} \label{optorigVElowerbndSmpNOfai}
\begin{align}
& \ \ \underset{{\bf{V}} ,{{\bf{V}}_E}}{\mathop\text{min}} \ \ -\text{Tr}({{\mathbf{W}}_{I}}{{\mathbf{V}}^{H}}{{{\mathbf{\hat{H}}}}_{I}}^{H}{{\mathbf{U}}_{I}})-\text{Tr}({{\mathbf{W}}_{I}}{{\mathbf{U}}^{H}_{I}}{{{\mathbf{\hat{H}}}}_{I}}\mathbf{V})+\text{Tr}({{\mathbf{V}}^{H}}{{\mathbf{H}}_{V}}\mathbf{V}) \nonumber \\
 &\quad \quad \quad \quad -\text{Tr}({{\mathbf{W}}_{E}}{{\mathbf{V}}^{H}_{E}}{{{\mathbf{\hat{H}}}}_{E}}^{H}{{\mathbf{U}}_{E}})-\text{Tr}({{\mathbf{W}}_{E}}{{\mathbf{U}}^{H}_{E}}{{{\mathbf{\hat{H}}}}_{E}}{{\mathbf{V}}_{E}})+\text{Tr}({{\mathbf{V}}^{H}_{E}}{{\mathbf{H}}_{VE}}{{\mathbf{V}}_{E}})  \label{eq10ta} \\
& \ \ \text{s.t.} \quad {\rm{Tr(}}{\bf{V}}{{\bf{V}}^H}{\rm{ + }}{{\bf{V}}_{E}}{{\bf{V}}^{H}_{E}}{\rm{)}} \le P_{T}.\label{eq10tb}
\end{align} \label{eq10t}
\end{subequations}
The above problem is a convex QCQP problem, and the standard optimization packages, such as CVX \cite{grant2014cvx} can be exploited to solve it. However, the calculation burden is heavy. To reduce the complexity, the near-optimal closed form expressions of the TPC matrix and AN covariance matrix are provided by applying the Lagrangian multiplier method.

Since Problem \eqref{optorigVElowerbndSmpNOfai} is a convex problem, the Slater's condition is satisfied, where the duality gap between Problem \eqref{optorigVElowerbndSmpNOfai} and its dual problem is zero. Thus, Problem \eqref{optorigVElowerbndSmpNOfai} can be solved by addressing its dual problem if the dual problem is easier. For this purpose, by introducing Lagrange multiplier $\lambda$ to combine the the constraint and OF of Problem \eqref{optorigVElowerbndSmpNOfai}, the Lagrangian function of Problem \eqref{optorigVElowerbndSmpNOfai} is obtained as
\begin{align} \label{LagrgnforVVE}
\mathcal{L}\left( {{\bf{V}},{{\bf{V}}_E},\lambda } \right) &\!\buildrel \Delta \over = \!- {\rm{Tr}}\left( {{\bf{W}}_I}{{\bf{V}}^H}{\bf{\hat H}}_I^H{{\bf{U}}_I} \right)\! - \!{\rm{Tr}}\left( {{{\bf{W}}_I}{\bf{U}}_I^H{{{\bf{\hat H}}}_I}{\bf{V}}} \right)\! +\! {\rm{Tr}}\left( {{{\bf{V}}^H}{{\bf{H}}_V}{\bf{V}}} \right) \!-\! {\rm{Tr}}\left( {{{\bf{W}}_E}{\bf{V}}_E^H{\bf{\hat H}}_E^H{{\bf{U}}_E}} \right)\nonumber \\
& \quad- {\rm{Tr}}\left( {{{\bf{W}}_E}{\bf{U}}_E^H{{{\bf{\hat H}}}_E}{{\bf{V}}_E}} \right) + {\rm{Tr}}\left( {{\bf{V}}_E^H{{\bf{H}}_{VE}}{{\bf{V}}_E}} \right)+ \lambda[ {{\rm{Tr}}\left( {{\bf{V}}{{\bf{V}}^H} + {{\bf{V}}_E}{\bf{V}}_E^H} \right)} -P_{T}] \nonumber \\
&=  - {\rm{Tr}}\left( {{{\bf{W}}_I}{{\bf{V}}^H}{\bf{\hat H}}_I^H{{\bf{U}}_I}} \right) - {\rm{Tr}}\left( {{{\bf{W}}_I}{\bf{U}}_I^H{{{\bf{\hat H}}}_I}{\bf{V}}} \right) + {\rm{Tr}}\left[ {{{\bf{V}}^H}\left( {{{\bf{H}}_V} + \lambda{\bf{I}}} \right){\bf{V}}} \right] \nonumber \\
&\quad- {\rm{Tr}}\left( {{{\bf{W}}_E}{\bf{V}}_E^H{\bf{\hat H}}_E^H{{\bf{U}}_E}} \right)\! - \!{\rm{Tr}}\left( {{{\bf{W}}_E}{\bf{U}}_E^H{{{\bf{\hat H}}}_E}{{\bf{V}}_E}} \right)\! +\! {\rm{Tr}}\left[ {{\bf{V}}_E^H\left( {{{\bf{H}}_{VE}} \!+ \!\lambda {\bf{I}}} \right){{\bf{V}}_E}} \right] \!-\! \lambda {P_T}.
\end{align}

Then the dual problem of Problem \eqref{optorigVElowerbndSmpNOfai} is
\begin{subequations} \label{DualoptforVVE}
\begin{align}
\mathop {\max }\limits_\lambda \quad \quad  {\rm{  }}h\left( \lambda  \right)
\label{DualforVVEa} \\
\text{s.t.}\quad \quad {\rm{  }}\lambda  \ge 0,
 \label{DualforVVEb}
\end{align}
\end{subequations}
where $h\left( \lambda  \right)$ is the dual function given by
\begin{align}\label{DualobjforVVE}
h\left( \lambda  \right) \buildrel \Delta \over = \mathop {\min }\limits_{{\bf{V}},{{\bf{V}}_E}} {\rm{  }}\mathcal{L}\left( {{\bf{V}},{{\bf{V}}_E},\lambda } \right).
\end{align}
Note that Problem \eqref{DualobjforVVE} is a convex quadratic optimization problem with no constraint, which can be solved in closed form. The optimal solution ${\bf{V}^{\star}},{{\bf{V}^{{\star}}}_E}$ for Problem \eqref{DualobjforVVE} is
\begin{align}\label{optVVE}
[{\bf{V}^{\star}},{{\bf{V}^{{\star}}}_E}]=\text{arg}\mathop {\min }\limits_{{\bf{V}},{{\bf{V}}_E}} {\rm{  }}\mathcal{L}\left( {{\bf{V}},{{\bf{V}}_E},\lambda } \right).
\end{align}
By setting the first-order derivative of $\mathcal{L}\left( {{\bf{V}},{{\bf{V}}_E},\lambda } \right)$ w.r.t. ${{{\bf{V}}}}$ to zero matrix, we can obtain the optimal solution of ${\bf{V}}$ as follows:
\begin{subequations} \label{DerivLagrgnforVVE1L1L2}
\begin{align}
\frac{\partial{\mathcal{L}\left( {{\bf{V}},{{\bf{V}}_E},\lambda } \right)}}{{\partial {\bf{V}}}}=\bf{0}, \label{DerivLagrgnforVVE1L1L2a} \\
\frac{\partial{\mathcal{L}\left( {{\bf{V}},{{\bf{V}}_E},\lambda } \right)}}{{\partial {{\bf{V}}_E}}}=\bf{0}. \label{DerivLagrgnforVVE1L1L2b}
\end{align}
\end{subequations}
The left hand side of Equation \eqref{DerivLagrgnforVVE1L1L2a} can be expanded as
\begin{align} \label{DerivLagrgnforVVE1L1L2_expand}
\frac{\partial{\mathcal{L}\left( {{\bf{V}},{{\bf{V}}_E},\lambda } \right)}}{{\partial {\bf{V}}}}&=\frac{{\partial {\rm{Tr}}\left[ {{{\bf{V}}^H}\left( {{{\bf{H}}_V} + \lambda {\bf{I}}} \right){\bf{V}}} \right]}}{{\partial {\bf{V}}}}-\left( {{{\bf{W}}_I}{\bf{U}}_I^H{{{\bf{\hat H}}}_I}} \right)^H-\left( {{\bf{\hat H}}_I^H{{\bf{U}}_I}{{\bf{W}}_I}} \right) \nonumber \\
&=2\left( {{{\bf{H}}_V} + \lambda {\bf{I}}} \right){\bf{V}}-2\left( {{\bf{\hat H}}_I^H{{\bf{U}}_I}{{\bf{W}}_I}} \right).
\end{align}
The equation \eqref{DerivLagrgnforVVE1L1L2a} becomes
\begin{align} \label{DerivLagrgnforVVE1L1L2Zero}
\left( {{{\bf{H}}_V} + \lambda {\bf{I}}} \right){\bf{V}}=\left( {{\bf{\hat H}}_I^H{{\bf{U}}_I}{{\bf{W}}_I}} \right).
\end{align}
Then the optimal solution ${\bf{V}^{\star}}$ for Problem \eqref{optVVE} is
\begin{align} \label{optimalV}
{{\bf{V}}^{\star}}&=\left( {{{\bf{H}}_V} + \lambda {\bf{I}}} \right)^{ \dag }\left( {{\bf{\hat H}}_I^H{{\bf{U}}_I}{{\bf{W}}_I}} \right) \nonumber \\
&\buildrel \Delta \over= {{\bf{\Theta }}_V}\left( \lambda  \right)\left( {{\bf{\hat H}}_I^H{{\bf{U}}_I}{{\bf{W}}_I}} \right).
\end{align}
Similarly, we solve Problem \eqref{optVVE} by setting the first-order derivative of $\mathcal{L}\left( {{\bf{V}},{{\bf{V}}_E},\lambda } \right)$ w.r.t. ${{{\bf{V}}_E}}$ to zero matrix, which becomes
\begin{align} \label{DerivLagrgnforVE1Zero1}
2\left( {{{\bf{H}}_{VE}} + \lambda {\bf{I}}} \right){{\bf{V}}_E} - 2{\bf{\hat H}}_E^H{{\bf{U}}_E}{\bf{W}}_E^H = \bf{0}.
\end{align}
Then the optimal solution ${\bf{V}}_E^{\star}$ for Problem \eqref{optVVE} is
\begin{align} \label{optimalVE}
{\bf{V}}_E^{\star} &= \left( {{{\bf{H}}_{VE}} + \lambda {\bf{I}}} \right)^{ \dag }{\bf{\hat H}}_E^H{{\bf{U}}_E}{\bf{W}}_E^H \nonumber \\
&\buildrel \Delta \over ={{\bf{\Theta }}_{VE}}\left( \lambda  \right){\bf{\hat H}}_E^H{{\bf{U}}_E}{\bf{W}}_E^H.
\end{align}
Once the optimal solution ${\lambda}^{\star}$ for Problem \eqref{DualoptforVVE} is found, the final optimal ${\bf{V}^{\star}},{\bf{V}}_E^{\star}$ can be obtained. The value of ${\lambda}^{\star}$ should be chosen in order to guarantee the complementary slackness condition as
\begin{align} \label{Compslack}
\lambda[ {\rm{Tr(}}{\bf{V}^{\star}}{{\bf{V}}^{{\star}H}}{\rm{ + }}{{\bf{V}}_{E}^{{\star}}}{{\bf{V}}^{{\star}H}_{E}}{\rm{)}}-P_{T}]=0.
\end{align}
We define
\begin{align} \label{Plambda}
P(\lambda)&\buildrel \Delta \over={\rm{Tr(}}{\bf{V}^{\star}}{{\bf{V}}^{{\star}H}}{\rm{ + }}{{\bf{V}}_{E}^{{\star}}}{{\bf{V}}^{{\star}H}_{E}}{\rm{)}} ={\rm{Tr(}}{\bf{V}^{\star}}{{\bf{V}}^{{\star}H}}{\rm{)}}+{\rm{Tr(}}{{\bf{V}}_{E}^{{\star}}}{{\bf{V}}^{{\star}H}_{E}}{\rm{)}},
\end{align}
where
\begin{align}
{\rm{Tr}}\left( {{\bf{V}}^{\star}{{\bf{V}}^{{\star}H}}} \right)&={\rm{Tr}}\left( {{{\bf{\Theta }}_V}\left( \lambda  \right)( {{\bf{\hat H}}_I^H{{\bf{U}}_I}{\bf{W}}_I^H} )}( {{\bf{\hat H}}_I^H{{\bf{U}}_I}{\bf{W}}_I^H} )^H{{\bf{\Theta }}^H_V}\left( \lambda  \right) \right) \nonumber \\
&={\rm{Tr}}\left( {{{\bf{\Theta }}^H_V}\left( \lambda  \right){{\bf{\Theta }}_V}\left( \lambda  \right)( {{\bf{\hat H}}_I^H{{\bf{U}}_I}{\bf{W}}_I^H} )} ( {{\bf{\hat H}}_I^H{{\bf{U}}_I}{\bf{W}}_I^H} )^H \right),
\end{align}
\begin{align}
{\rm{Tr}}\left( {{{\bf{V}}_E^{{\star}H}}{\bf{V}}_E^{\star}} \right)&={\rm{Tr}}\left( {{{\bf{\Theta }}_{VE}}\left( \lambda  \right)( {{\bf{\hat H}}_E^H{{\bf{U}}_E}{\bf{W}}_E^H} )}( {{\bf{\hat H}}_E^H{{\bf{U}}_E}{\bf{W}}_E^H} )^H{{\bf{\Theta }}^H_{VE}}\left( \lambda  \right) \right) \nonumber \\
&={\rm{Tr}}\left( {{{\bf{\Theta }}^H_{VE}}\left( \lambda  \right){{\bf{\Theta }}_{VE}}\left( \lambda  \right)( {{\bf{\hat H}}_E^H{{\bf{U}}_E}{\bf{W}}_E^H} )} ( {{\bf{\hat H}}_E^H{{\bf{U}}_E}{\bf{W}}_E^H} )^H \right).
\end{align}
Then $P(\lambda)$ becomes
\begin{align} \label{Plambdanew}
P(\lambda)={\rm{Tr}}\left( {{{\bf{\Theta }}^n_V}( {{\bf{\hat H}}_I^H{{\bf{U}}_I}{\bf{W}}_I^H} ) ( {{\bf{\hat H}}_I^H{{\bf{U}}_I}{\bf{W}}_I^H} )^H }\right) +{\rm{Tr}}\left( {{{\bf{\Theta }}^n_{VE}}( {{\bf{\hat H}}_E^H{{\bf{U}}_E}{\bf{W}}_E^H} ) ( {{\bf{\hat H}}_E^H{{\bf{U}}_E}{\bf{W}}_E^H} )^H }\right),
\end{align}
where
\begin{align} \label{ThetaVnew}
{{\bf{\Theta }}^n_V}&={{\bf{\Theta }}^H_V}\left( \lambda  \right){{\bf{\Theta }}_V}\left( \lambda  \right)=\left( {{{\bf{H}}_V} + \lambda {\bf{I}}} \right)^{ \dag H }\left( {{{\bf{H}}_V} + \lambda {\bf{I}}} \right)^{ \dag},
\end{align}
\begin{align} \label{ThetaVEnew}
{{\bf{\Theta }}^n_{VE}}&={{\bf{\Theta }}^H_{VE}}\left( \lambda  \right){{\bf{\Theta }}_{VE}}\left( \lambda  \right) =\left( {{{\bf{H}}_{VE}} + \lambda {\bf{I}}} \right)^{ \dag H}\left( {{{\bf{H}}_{VE}} + \lambda {\bf{I}}} \right)^{ \dag }.
\end{align}
To find the optimal ${\lambda^{\star}}\geq0$, we first check whether $\lambda=0$ is the optimal solution or not.  If
\begin{align} \label{lambdazero}
P(0)= {\rm{Tr}}\left( {{{\bf{V}}^{{\star}H}}(0){\bf{V}}^{\star}(0)} \right) + {\rm{Tr}}\left( {{\bf{V}}_E^{{\star}H}(0){{\bf{V}}_E}^{\star}(0)} \right)\le {P_{T}},
\end{align}
then the optimal solutions are given by ${{\bf{V}}^{\star}}={{\bf{V}}}(0)$ and ${{\bf{V}}_{E}^{\star}}={{\bf{V}}_{E}}(0)$. Otherwise, the optimal $\lambda^{\star}>0$ is the solution of the equation $P(\lambda)=0$.

It is ready to verify that ${{\bf{H}}_V}$ and ${{\bf{H}}_{VE}}$ is a positive semidefinite matrix. Let us define the rank of ${{\bf{H}}_V}$ and ${{\bf{H}}_{VE}}$ as $r_{V}={\rm{rank}}({\bf{H}}_{V})\le N_T$ and $r_{VE}={\rm{rank}}({\bf{H}}_{VE}) \le N_T$ respectively. By decomposing ${{\bf{H}}_V}$ and ${{\bf{H}}_{VE}}$ by using the singular value decomposition (SVD), we have
\begin{equation}\label{svdforHVHVE}
  {{\bf{H}}_V} = \left[ {{{\bf{P}}_{V,1}},{{\bf{P}}_{V,2}}} \right]{{\bf{\Sigma }}_V}{\left[ {{{\bf{P}}_{V,1}},{{\bf{P}}_{V,2}}} \right]^{\rm{H}}},{{\bf{H}}_{VE}} = \left[ {{{\bf{P}}_{{VE},1}},{{\bf{P}}_{{VE},2}}} \right]{{\bf{\Sigma }}_{VE}}{\left[ {{{\bf{P}}_{{VE},1}},{{\bf{P}}_{{VE},2}}} \right]^{\rm{H}}},
\end{equation}
where ${\bf{P}}_{V,1}$ comprises the first $r_V$ singular vectors associated with
the $r_V$ positive eigenvalues of ${{\bf{H}}_V}$, and ${\bf{P}}_{V,2}$ includes the last $N_T-r_V$ singular vectors associated with the $N_T-r_V$ zero-valued eigenvalues of ${{\bf{H}}_V}$, ${{\bm{\Sigma}} _V} = {\rm{diag}}\left\{ {{{\bm{\Sigma}} _{V,1}},{{\bf{0}}_{\left( {{N_T} - {r_V}} \right) \times \left( {{N_T} - {r_V}} \right)}}} \right\}$ with ${\bm{\Sigma}} _{V,1}$ representing the diagonal submatrix collecting the first $r_V$ positive eigenvalues. Similarly, the first $r_{VE}$ singular vectors corresponding to
the $r_{VE}$ positive eigenvalues of ${{\bf{H}}_{VE}}$ are contained in ${\bf{P}}_{VE,1}$, while the last $N_T-r_{VE}$ singular vectors corresponding to the $N_T-r_{VE}$ zero-valued eigenvalues of ${{\bf{H}}_{VE}}$ are held in ${\bf{P}}_{VE,2}$. ${{\bm{\Sigma}} _{VE}} = {\rm{diag}}\left\{ {{{\bm{\Sigma}} _{{VE},1}},{{\bf{0}}_{\left( {{N_T} - {r_{VE}}} \right) \times \left( {{N_T} - {r_{VE}}} \right)}}} \right\}$ is a diagonal matrix with ${\bm{\Sigma}} _{{VE},1}$ representing the diagonal submatrix gathering the first $r_{VE}$ positive eigenvalues. By defining ${{\bf{P}}_V} \buildrel \Delta \over =  \left[ {{{\bf{P}}_{V,1}},{{\bf{P}}_{V,2}}} \right]$ and ${{\bf{P}}_{VE}} \buildrel \Delta \over =  \left[ {{{\bf{P}}_{{VE},1}},{{\bf{P}}_{{VE},2}}} \right]$, and substituting \eqref{svdforHVHVE} into \eqref{ThetaVnew} and \eqref{ThetaVEnew}, $P(\lambda)$ becomes
\begin{align} \label{Plamda}
& P(\lambda) ={\rm {Tr}}\left({[{\left({{{\bf {P}}_{V}}{{\bf {\Sigma}}_{V}}{\bf {P}}_{V}^{H}+\lambda{{\bf {P}}_{V}}{\bf {P}}_{V}^{H}}\right)^{-1}}{\left({{{\bf {P}}_{V}}{{\bf {\Sigma}}_{V}}{\bf {P}}_{V}^{H}+\lambda{{\bf {P}}_{V}}{\bf {P}}_{V}^{H}}\right)^{-1}}]({{\bf {\hat{H}}}_{I}^{H}{{\bf {U}}_{I}}{\bf {W}}_{I}^{H}})({{\bf {\hat{H}}}_{I}^{H}{{\bf {U}}_{I}}{\bf {W}}_{I}^{H}})^{H}}\right)\nonumber \\
 & \!\!+\!\!{\rm {Tr}}\!\left(\!{[\!{\left(\!{{{\bf {P}}_{VE}}{{\bf {\Sigma}}_{VE}}{\bf {P}}_{VE}^{H}\!+\!\lambda{{\bf {P}}_{VE}}{\bf {P}}_{VE}^{H}}\!\right)^{-1}}\!\!{\left({{{\bf {P}}_{VE}}{{\bf {\Sigma}}_{VE}}{\bf {P}}_{VE}^{H}\!+\!\lambda{{\bf {P}}_{VE}}{\bf {P}}_{VE}^{H}}\right)^{-1}}\!]\!(\!{{\bf {\hat{H}}}_{E}^{H}{{\bf {U}}_{E}}{\bf {W}}_{E}^{H}}\!)\!({{\bf {\hat{H}}}_{E}^{H}{{\bf {U}}_{E}}{\bf {W}}_{E}^{H}})^{H}}\!\right)\nonumber \\
 & ={\rm {Tr}}\left({[{\left({{{\bf {\Sigma}}_{V}}+\lambda{\bf {I}}}\right)^{-2}}]{\bf {Z}}_{V}}\right)+{\rm {Tr}}\left({[{\left({{{\bf {\Sigma}}_{VE}}+\lambda{\bf {I}}}\right)^{-2}}]{\bf {Z}}_{VE}}\right)\nonumber \\
 & {=}\sum\limits _{i=1}^{r_{V}}\left[{\frac{{{\left[{{\bf {Z}}_{V}}\right]}_{i,i}}}{{{\left({{{\left[{{\Sigma}_{V}}\right]}_{i,i}}\!+\!\lambda}\right)}^{2}}}}\right]+\sum\limits _{i=1}^{r_{VE}}\left[{\frac{{{\left[{{\bf {Z}}_{VE}}\right]}_{i,i}}}{{{\left({{{\left[{{\Sigma}_{VE}}\right]}_{i,i}}\!+\!\lambda}\right)}^{2}}}}\right]+\sum\limits _{i={r_{V}}+1}^{{N_{T}}}{\left[{\frac{{{\left[{{\bf {Z}}_{V}}\right]}_{i,i}}}{{{\left({\lambda}\right)}^{2}}}}\right]}\!+\!\sum\limits _{i={r_{VE}}+1}^{{N_{T}}}{\left[{\frac{{{\left[{{\bf {Z}}_{VE}}\right]}_{i,i}}}{{{\left({\lambda}\right)}^{2}}}}\right]},
\end{align}
where ${{\bf{Z}}_{V}}={\bf{P}}_V^H( {{\bf{\hat H}}_I^H{{\bf{U}}_I}{\bf{W}}_I^H} ) ( {{\bf{\hat H}}_I^H{{\bf{U}}_I}{\bf{W}}_I^H} )^H {{\bf{P}}_V}$ and ${{\bf{Z}}_{VE}}={\bf{P}}_{VE}^H( {{\bf{\hat H}}_E^H{{\bf{U}}_E}{\bf{W}}_E^H} ) ( {{\bf{\hat H}}_E^H{{\bf{U}}_E}{\bf{W}}_E^H} )^H {{\bf{P}}_{VE}}$. ${\left[ {{{\bf{Z}}_V}} \right]}_{i,i}$, ${\left[ {{{\bf{Z}}_{VE}}} \right]}_{i,i}$, ${\left[ {{{\Sigma}}_{V}} \right]}_{i,i}$, and ${\left[ {{{\Sigma}}_{VE}} \right]}_{i,i}$ represent the $i$th diagonal element of matrices ${{{\bf{Z}}_V}}$, ${{\bf{Z}}_{VE}}$, ${{{\Sigma}}_{V}}$, and ${{{\Sigma}}_{VE}}$, respectively. The first line of (\ref{Plamda}) is obtained by substituting (\ref{svdforHVHVE}) into the expression of $P({\lambda})$ in (\ref{Plambdanew}). It can be verified from the last line of (\ref{Plamda}) that $P({\lambda})$ is a monotonically decreasing function.

Then, the optimal $\lambda^{\star}$ can be obtained by solving the following equation,
\begin{align}\label{lambdaequat}
\sum\limits_{i = 1}^{r_{V}}\!\left[{\frac{{{\left[ {{{\bf{Z}}_{V}}} \right]}_{i,i}}}{{{{\left( {{{\left[ {{{\Sigma}}_{V}} \right]}_{i,i}} + \lambda} \right)}^2}}}}\right]\!+\!\!\sum\limits_{i = 1}^{r_{VE}}\!\left[{\frac{{{\left[ {{{\bf{Z}}_{VE}}} \right]}_{i,i}}}{{{{\left( {{{\left[ {{{\Sigma}}_{VE}} \right]}_{i,i}} + \lambda} \right)}^2}}}}\right]\!+\!\! \sum\limits_{i = {r_{V}} + 1}^{{N_T}}\! {\left[{\frac{{{\left[ {{{\bf{Z}}_{V}}} \right]}_{i,i}}}{{{{\left( {\lambda} \right)}^2}}}}\right]}\!+ \!\!\sum\limits_{i = {r_{VE}} + 1}^{{N_T}}\! {\left[{\frac{{{\left[ {{{\bf{Z}}_{VE}}} \right]}_{i,i}}}{{{{\left( {\lambda} \right)}^2}}}}\right]}=P_{T}.
\end{align}
To solve it, the bisection search method is utilized. Since $P(\infty )=0$, the solution to Equation (\ref{lambdaequat}) must exist. The lower bound of $\lambda^{\star}$ is a positive value approaching zero, while the upper bound of $\lambda^{\star}$ is given by
\begin{equation}\label{xddfcerf}
  {\lambda^{\star}} < \sqrt {\frac{{\sum\limits_{i = 1}^{{N_T}} {{{\left[ {{{\bf{Z}}_V}} \right]}_{i,i}}} }+{\sum\limits_{i = 1}^{{N_T}} {{{\left[ {{{\bf{Z}}_{VE}}} \right]}_{i,i}}} }}{{{P_{T}}}}}  \buildrel \Delta \over = \lambda^{{\rm{ub}}}.
\end{equation}
which can be proved as
\begin{align}\label{asdftg}
 {P}(\lambda^{{\rm{ub}}})&=\sum\limits_{i = 1}^{r_{V}} {\frac{{{{\left[ {{{\bf{Z}}_V}} \right]}_{i,i}}}}{{{{\left({{{\left[{{{\Sigma}}_{V}} \right]}_{i,i}} + {\lambda^{{\rm{ub}}}}} \right)}^2}}}}+\sum\limits_{i = 1}^{r_{VE}} {\frac{{{{\left[ {{{\bf{Z}}_{VE}}}\right]}_{i,i}}}}{{{{\left( {{{\left[ {{{\Sigma}}_{VE}} \right]}_{i,i}} + {\lambda^{{\rm{ub}}}}} \right)}^2}}}}+\sum\limits_{i={r_{V}}+1}^{{N_{T}}}{\left[{\frac{{{\left[{{\bf{Z}}_{V}}\right]}_{i,i}}}{{{\left({\lambda^{{\rm{ub}}}}\right)}^{2}}}}\right]}\!+\!\sum\limits _{i={r_{VE}}+1}^{{N_{T}}}{\left[{\frac{{{\left[{{\bf{Z}}_{VE}}\right]}_{i,i}}}{{{\left({\lambda^{{\rm{ub}}}}\right)}^{2}}}}\right]}\nonumber \\
 & < \sum\limits_{i = 1}^{{N_T}} {\frac{{{{\left[ {{{\bf{Z}}_V}} \right]}_{i,i}}}}{{{{\left( {\lambda^{{\rm{ub}}}} \right)}^2}}}}+\sum\limits_{i = 1}^{{N_T}} {\frac{{{{\left[ {{{\bf{Z}}_{VE}}} \right]}_{i,i}}}}{{{{\left( {\lambda^{{\rm{ub}}}} \right)}^2}}}}  = {P_{T}}.
\end{align}

When the optimal $\lambda^{{\star}}$ is found, the optimal matrices ${{\bf{V}}^{{\star}}}$ and ${{\bf{V}}_{E}^{{\star}}}$ can be obtained by substituting $\lambda^\star$ into (\ref{optimalV}) and (\ref{optimalVE}).

\vspace{-0.4cm}\subsection{Optimizing the Phase Shifts $\mathbf{\Phi }$}\label{hwdi}
In this subsection, the phase shift matrix $\mathbf{\Phi }$ is optimized by fixing ${{\bf{V}}}$ and ${{\bf{V}}_E}$. The transmit power constraint in Problem \eqref{optorigVElowerbndSmp} is only related with ${{\bf{V}}}$ and ${{\bf{V}}_E}$, thus is removed. Then, the optimization problem for $\mathbf{\Phi }$ reduced from Problem \eqref{optorigVElowerbndSmp} is formulated as
\begin{subequations} \label{optproblemforfaimin}
\begin{align}
&\ \ \underset{{\bf{\Phi}}}{\mathop\text{min}} \ \ {g_{0}}(\mathbf{\Phi })\buildrel \Delta \over=-\text{Tr}({{\mathbf{W}}_{I}}{{\mathbf{V}}^{H}}{{{\mathbf{\hat{H}}}}_{I}}^{H}{{\mathbf{U}}_{I}})-\text{Tr}({{\mathbf{W}}_{I}}{{\mathbf{U}}_{I}}^{H}{{{\mathbf{\hat{H}}}}_{I}}\mathbf{V})+\text{Tr}({{\mathbf{V}}^{H}}{{\mathbf{H}}_{V}}\mathbf{V}) \nonumber \\
 &\quad \quad \quad \quad \quad \quad \ -\text{Tr}({{\mathbf{W}}_{E}}{{\mathbf{V}}_{E}}^{H}{{{\mathbf{\hat{H}}}}_{E}}^{H}{{\mathbf{U}}_{E}})-\text{Tr}({{\mathbf{W}}_{E}}{{\mathbf{U}}_{E}}^{H}{{{\mathbf{\hat{H}}}}_{E}}{{\mathbf{V}}_{E}})+\text{Tr}({{\mathbf{V}}_{E}}^{H}{{\mathbf{H}}_{VE}}{{\mathbf{V}}_{E}}) \label{optproblemforfaimina} \\
& \ \ \text{s.t.} \quad \left| {{{{\phi}} _m}} \right| = 1,m = 1, \cdots ,M. \label{optproblemforfaiminb}
\end{align}
\end{subequations}
By the aid of complex mathematical manipulations, which are given in details in Appendix \ref{newOFDeriv}, Problem (\ref{optproblemforfaimin}) can be transformed into a form that can facilitate the MM algorithm. Based on the derivations in Appendix \ref{newOFDeriv}, the OF ${g_{0}}(\mathbf{\Phi })$ can be equivalently transformed into
\begin{align} \label{eqg0forPhi}
{g_{0}}(\mathbf{\Phi })={\rm{Tr}}\left( {{{\bf{\Phi}}^H{\bf{D}}^H}} \right) + {\rm{Tr}}\left( {{\bf{\Phi D}}} \right) + {\rm{Tr}}\left[ {{{\bf{\Phi }}^H}{{\bf{B}}_{VE}}{\bf{\Phi }}{{\bf{C}}_{VE}}} \right] + {\rm{Tr}}\left( {{{\bf{\Phi }}^H}{{\bf{B}}_V}{\bf{\Phi }}{{\bf{C}}_V}} \right)+C_t,
\end{align}
where $C_t$, ${\bf{D}}$, ${{\bf{C}}_{VE}}$, ${{\bf{C}}_{V}}$, ${{\bf{B}}_{VE}}$ and ${{\bf{B}}_{V}}$ are constants for ${\bf{\Phi }}$, and are given in Appendix \ref{newOFDeriv}.

By exploiting the matrix properties in \cite[Eq. (1.10.6)]{zhang2017matrix}, the trace operators can be removed, and the third and fourth terms in (\ref{eqg0forPhi}) become as
\begin{subequations}\label{saddewde}
\begin{align}
{\rm{Tr}}\left( {{{\bm{\Phi}} ^{\rm{H}}}{\bf{B}}_{VE}{\bm{\Phi}} {\bf{C}}_{VE}} \right) = {{\bm{\phi}} ^{\rm{H}}}\left( {{\bf{B}}_{VE} \odot {{\bf{C}}_{VE}^{\rm{T}}}} \right){\bm{\phi}}, \\
{\rm{Tr}}\left( {{{\bm{\Phi}} ^{\rm{H}}}{\bf{B}}_{V}{\bm{\Phi}} {\bf{C}}_{V}} \right) = {{\bm{\phi}} ^{\rm{H}}}\left( {{\bf{B}}_{V} \odot {{\bf{C}}_{V}^{\rm{T}}}} \right){\bm{\phi}},
\end{align}
\end{subequations}
where ${\bm{\phi}} \buildrel \Delta \over = {\left[ {{e^{j{\theta _1}}}, \cdots ,{e^{j{\theta _m}}}, \cdots ,{e^{j{\theta _M}}}} \right]^{\rm{T}}}$ is a vector holding the diagonal elements of ${\bm{\Phi}}$.

Similarly, the trace operators can be removed for the first and second terms in (\ref{eqg0forPhi}) as
\begin{equation}\label{sdewf}
  {\rm{Tr}}\left( {{{\bm{\Phi}} ^{\rm{H}}}{{\bf{D}}^{\rm{H}}}} \right) = {{\bf{d}}^{\rm{H}}}({{\bm{\phi}}}^*), {\rm{Tr}}\left( {{\bm{\Phi}} {\bf{D}}} \right)={\bm{\phi}}^{\rm{T}}{\bf{d}},
\end{equation}
where ${\bf{d}} = {\left[ {{{\left[ {\bf{D}} \right]}_{1,1}}, \cdots ,{{\left[ {\bf{D}} \right]}_{M,M}}} \right]^{\rm{T}}}$ is a vector gathering the diagonal elements of matrix ${\bf{D}}$.

Hence, Problem (\ref{optproblemforfaimin}) can be rewritten as
\begin{subequations}\label{optproblemforlittlefaimin}
\begin{align}
&{\mathop {\min }\limits_{{\bm{\phi}}}  \quad {{\bm{\phi}} ^{\rm{H}}}{\bm{\Xi} }{\bm{\phi}} + {\bm{\phi}}^{\rm{T}}{\bf{d}} + {{\bf{d}}^{\rm{H}}}({{\bm{\phi}}}^*)}
\\
&\textrm{s.t.}\quad  \left| {{\phi _m}} \right| = 1 , m = 1, \cdots ,M,
\end{align}
\end{subequations}
where $\bm{\Xi}={\bf{B}}_{VE} \odot {{\bf{C}}_{VE}^{\rm{T}}}+{\bf{B}}_{V} \odot {{\bf{C}}_{V}^{\rm{T}}} $. $\bm{\Xi}$ is a semidefinite matrix, because it is a sum of two semidefinite matrices, both of which are Hadamard products of two semidefinite matrices. It is observed that ${\bf{B}}_{VE}$, ${{\bf{C}}_{VE}^{\rm{T}}}$, ${\bf{B}}_{V}$ and ${{\bf{C}}_{V}^{\rm{T}}}$ are semidefinite matrices. Then, the Hadamard products of ${{\bf{B}}_{VE} \odot {{\bf{C}}_{VE}^{\rm{T}}}}$ and ${{\bf{B}}_{V} \odot {{\bf{C}}_{V}^{\rm{T}}}}$ are semidefinite according to the Property (9) on Page 104 of \cite{zhang2017matrix}. Problem (\ref{optproblemforlittlefaimin}) can be further simplified as
 \begin{subequations}\label{appjig}
\begin{align}
&{\mathop {\min }\limits_{\bm{\phi}}  \quad f({\bm{\phi}})\buildrel \Delta \over = {{\bm{\phi}} ^{\rm{H}}}{\bm{\Xi}}{\bm{\phi}} +  2{\rm{Re}}\left\{ {{{\bm{\phi}} ^{\rm{H}}}({{\bf{d}}}^*)} \right\}}
\\
&\textrm{s.t.}\quad  \left| {{\phi _m}} \right| = 1 , m = 1, \cdots ,M. \label{dshxsdceur}
\end{align}
\end{subequations}
The Problem (\ref{appjig}) can be solved by the SDR technique \cite{wu2019intelligent} by transforming the unimodulus constraint into a rank-one constraint, however, the rank-one solution cannot always be obtained and the computation complexity is heavy for the SDR method. Thus, we propose to solve Problem (\ref{appjig}) efficiently by the MM algorithm as \cite{pan2019multicell}, where the closed-form solution can be obtained in each iteration. Details are omitted for simplicity.

\vspace{-0.4cm}\subsection{Overall Algorithm to Solve Problem (\ref{optorigVE})}
To sum up, the detailed execution of the overall BCD-MM algorithm proposed for solving Problem (\ref{optorigVE}) is provided in Algorithm \ref{bcd}. The MM algorithm is exploited for solving the optimal phase shifts ${\bf{\Phi}}^{(n+1)}$ of Problem (\ref{appjig}) in Step 5. The iteration process in MM algorithm ensures that the OF value of Problem (\ref{appjig}) decreases monotonically. Moreover, the BCD algorithm also guarantees that the OF value of Problem (\ref{optorigVElowerbndSmp}) monotonically decreases in each step and each iteration of Algorithm \ref{bcd}. Since the OF value in (\ref{optorigVElowerbndSmpa}) has a lower bound with the power limit, the convergence of Algorithm \ref{bcd} is guaranteed.

\begin{algorithm}
\caption{BCD-MM Algorithm}\label{bcd}
\begin{algorithmic}[1]
\STATE Parameter Setting. Set the maximum number of iterations $n_{\rm{max}}$ and the first iterative number $n=1$; Give the error tolerance $\varepsilon$.
\STATE Variables Initialization. Initialize the variables ${\bf{V}}^{(1)}$,
${\bf{V}}_{E}^{(1)}$ and ${\bf{\Phi}}^{(1)}$ in the feasible region; Compute the OF value of Problem (\ref{optorigVE}) as ${\rm{OF(}}{{\bf{V}}^{(1)}},{{\bf{V}}^{(1)}_{E}},{\bf{\Phi}}^{(1)}{\rm{)}}$;
\STATE Auxiliary Variables Calculation. Given ${\bf{V}}^{(n)} ,{{\bf{V}}^{(n)}_E}$, ${\bf{\Phi}}^{(n)}$, compute the optimal matrices ${{\bf{U}}^{(n)}_I},{{\bf{W}}^{(n)}_I},{{\bf{U}}^{(n)}_E},{{\bf{W}}^{(n)}_E},{{\bf{W}}^{(n)}_X}$ according to \eqref{optUI}, \eqref{optWI}, \eqref{optUE}, \eqref{optWE}, \eqref{optWX} respectively;
\STATE Matrices Optimization. Given ${{\bf{U}}^{(n)}_I},{{\bf{W}}^{(n)}_I},{{\bf{U}}^{(n)}_E},{{\bf{W}}^{(n)}_E},{{\bf{W}}^{(n)}_X}$, solve the optimal TPC matrix ${\bf{V}}^{(n+1)}$ and equivalent AN covariance matrix ${{\bf{V}}^{(n+1)}_E}$ of Problem (\ref{optVVE}) with the Lagrangian multiplier method;
 \STATE Phase Shifts Optimization. Given ${{\bf{U}}^{(n)}_I},{{\bf{W}}^{(n)}_I},{{\bf{U}}^{(n)}_E},{{\bf{W}}^{(n)}_E},{{\bf{W}}^{(n)}_X}$ and ${\bf{V}}^{(n+1)} ,{{\bf{V}}^{(n+1)}_E}$, solve the optimal phase shifts ${\bf{\Phi}}^{(n+1)}$ of Problem (\ref{appjig}) with the MM algorithm;
 \STATE Termination Check. If ${{\left| \!{{\rm{OF}}\!(\!{\bf{V}}^{(n+1)}\! ,\!{{\bf{V}}^{(n+1)}_E}\!,\!{\bf{\Phi}}^{(n+1)}\!)\! \!- \!\! {\rm{OF}}\!(\!{\bf{V}}^{(n)}\! ,\!{{\bf{V}}^{(n)}_E}\!,\!{\bf{\Phi}}^{(n)}\!)} \!\right|} / \!{{\rm{OF}}\!(\!{\bf{V}}^{(n+1)} \!,\!{{\bf{V}}^{(n+1)}_E}\!,\!{\bf{\Phi}}^{(n+1)}\!)}}\! < \varepsilon$ or $n\geq n_{\rm{max}}$, terminate.  Otherwise, update $n \leftarrow n + 1$  and jump to step 2.
\end{algorithmic}
\end{algorithm}

Based on the algorithm description, the complexity analysis of the proposed BCD-MM algorithm is performed. In Step 3, computing the decoding matrices ${{\bf{U}}^{(n)}_I}$ and ${{\bf{U}}^{(n)}_E}$ costs the complexity of ${\cal O}(N_I^3)+{\cal O}(N_E^3)$, while calculating the auxiliary matrices ${{\bf{W}}^{(n)}_I}$, ${{\bf{W}}^{(n)}_E}$, and ${{\bf{W}}^{(n)}_X}$ consumes the complexity of ${\cal O}(d^3)+{\cal O}(N_T^3)+{\cal O}(N_E^3)$. The complexity of calculating the TPC matrix ${\bf{V}}^{(n+1)} $ and AN covariance matrix ${{\bf{V}}^{(n+1)}_E}$ in Step 4 can be analyzed according to the specific process of Lagrangian multiplier method based on the fact that the complexity of computing product ${\bf{XY}}$ of complex matrices ${\bf{X}} \in {{\mathbb{C}}^{m \times n}}$ and ${\bf{Y}} \in {{\mathbb{C}}^{n \times p}}$ is ${\cal O}\left( {mnp} \right)$. By assuming that $N_T>N_I({\rm{or \ }} N_E)>d$, the complexity of computing the matrices $\{{{\mathbf{H}}_{V}},{{\mathbf{H}}_{VE}}\}$ in (\ref{HV}) and (\ref{HVE}) is ${\cal O}(N_T^3)+{\cal O}(2N_T^2d)+{\cal O}(2N_T^2N_E)$; while the complexity of calculating ${\bf{V}}^*$, ${\bf{V}}_E^*$ in (\ref{optimalV}) and (\ref{optimalVE}) is ${\cal O}(2N_T^3)$. The SVD decomposition of $\{{{\mathbf{H}}_{V}},{{\mathbf{H}}_{VE}}\}$ requires the computation complexity of ${\cal O}(2N_T^3)$, while calculating $\{{\bf{Z}}_V\}$ and $\{{\bf{Z}}_{VE}\}$ requires the complexity of ${\cal O}(N_T^2N_I)+{\cal O}(2N_T^3)$. The complexity of finding the Lagrangian multipliers $\{\lambda\}$ is negligible. Thus, the overall complexity for ${\bf{V}}^{(n+1)}$, ${\bf{V}}_E^{(n+1)}$ is about ${\cal O}({\rm{max}}\{2N_T^3,2N_T^2N_E\})$. In step 5, obtaining optimal ${\bf{\Phi}}^{(n+1)}$ by the MM algorithm need a complexity of $C_{MM}={\cal O}(M^3+T_{MM}M^2)$, where $T_{MM}$ is the iteration number for convergence. Based on the complexity required in Step 3, 4 and 5, the overall complexity $C_{\rm{BCD-MM}}$ of the BCD-MM algorithm can be evaluated by
 \begin{equation}\label{aefar}
   C_{\rm{BCD-MM}}={\cal O}({\rm{max}}\{2N_T^3,2N_T^2N_E,C_{MM}\}).
 \end{equation}

\section{Extension to the Multiple-IRs Scenario}
\subsection{Problem Formulation}
Consider a multicast extension where there are $L\geq2$ legitimate IRs, and they all intend to receive
the same message. The signal model for the MIMO multi-IRs wiretap channel scenario is
\begin{align}
{\bf {y}}_{I,l} ={\hat{{\bf {H}}}_{I,l}}({\bf {V}}{\bf {s}}+{\bf {n}})+{{\bf {n}}_{I,l}},l=1,\cdots,L,
\end{align}
 where ${{\hat{\bf{H}}}_{I,l}}\overset{\triangle}{=} {{\bf{H}}_{b,I,l}}+{{\bf{H}}_{R,I,l}}{\bf{\Phi}} {\bf{G}}$. The subscript $l$ indicates the $l$th IR, and the other notations are the same
as (\ref{eq4t}) and (\ref{eq6t}). Under these settings, the achievable
SR is given by \cite{liang2009information}
\begin{alignat}{1}
R_{s}{\rm {(}}{\bf {V}},{{\bf {V}}_{E}},{\bf {\Phi}}{\rm {)}} & =\underset{l=1,\cdots,L}{\min}\{{R_{I,l}}({\bf {V}},{\bf {\Phi}},{\bf {Z}})-{R_{E}}({\bf {V}},{\bf {\Phi}},{\bf {Z}})\},
\end{alignat}
where ${R_{I,l}}({\bf {V}},{\bf {\Phi}},{\bf {Z}})  ={\rm {log}}\left|{{\bf {I}}+{{\hat{{\bf {H}}}}_{I,l}}{\bf {V}}{{\bf {V}}^{H}}\hat{{\bf {H}}}_{I,l}^{H}{\bf {J}}_{I,l}^{-1}}\right|$ and ${{\bf {J}}_{I,l}}  \overset{\triangle}{=}{{\hat{{\bf {H}}}}_{I,l}}{\bf {Z}}{{\hat{{\bf {H}}}}_{I,l}}^{H}+\sigma_{I,l}^{2}{{\bf {I}}_{{{N}_{I}}}}$.

Then the multicast counterpart of the AN-aided SRM
problem (\ref{optorigVE}) is formulated as \begin{subequations} \label{multicastoptorigVE}
\begin{align}
 & \ \underset{{\bf {V}},{{\bf {V}}_{E}},{\bf {\Phi}}}{\mathop{\text{max}}}\ \ {R_{s}}{\rm {(}}{\bf {V}},{{\bf {V}}_{E}},{\bf {\Phi}}{\rm {)}}\label{multicastoptorigVE_a}\\
 & \ \ \text{s.t.}\quad\ {\rm {Tr(}}{\bf {V}}{{\bf {V}}^{H}}{\rm {+}}{{\bf {V}}_{E}}{{\bf {V}}_{E}^{H}}{\rm {)}}\le{P_{T}},\label{multicastoptorigVE_b}\\
 & \quad\quad\quad\!\left|{{\phi}_{m}}\right|=1,m=1,\cdots,M.\label{multicastoptorigVE_c}
\end{align}
\end{subequations}

The objective function of Problem (\ref{multicastoptorigVE}) can
be rewritten as \begin{subequations}\label{OFRs}
\begin{alignat}{1}
R_{s}{\rm {(}}{\bf {V}},{{\bf {V}}_{E}},{\bf {\Phi}}{\rm {)}} & =\underset{l=1,\cdots,L}{\min}\{\underbrace{{\rm {log}}\left|{{\bf {I}}_{{N_{I}}}+{{\hat{{\bf {H}}}}_{I,l}}{\bf {V}}{{\bf {V}}^{H}}\hat{{\bf {H}}}_{I,l}^{H}{{({{\hat{{\bf {H}}}}_{I,l}}{{\bf {V}}_{E}}{{\bf {V}}_{E}^{H}}{{\hat{{\bf {H}}}}_{I,l}}^{H}+\sigma_{I,l}^{2}{{\bf {I}}_{{N_{I}}}})}^{-1}}}\right|}_{{f_{1,l}}}\}\nonumber \\
 & \quad{\rm {+}}\underbrace{{\rm {log}}\left|{{{\bf {I}}_{{N_{E}}}}+{{\hat{{\bf {H}}}}_{E}}{{\bf {V}}_{E}}{{\bf {V}}_{E}^{H}}{{\hat{{\bf {H}}}}_{E}}^{H}(\sigma_{E}^{2}{{\bf {I}}_{{N_{E}}}})^{-1}}\right|}_{{f_{2}}}\nonumber \\
 & \quad\underbrace{-{\rm {log}}\left|{{{\bf {I}}_{{N_{E}}}}+\sigma_{E}^{-2}{{\hat{{\bf {H}}}}_{E}}({\bf {V}}{{\bf {V}}^{H}}+{{\bf {V}}_{E}}{{\bf {V}}_{E}^{H}})\hat{{\bf {H}}}_{E}^{H}}\right|}_{{f_{3}}},\\
 & =\underset{l=1,\cdots,L}{\min}\{\mathop{\text{max}}\limits _{{{\bf {U}}_{I,l}},{{\bf {W}}_{I,l}}\succeq0}h_{1,l}({{\bf {U}}_{I,l}},{\bf {V}},{{\bf {V}}_{E}},{{\bf {W}}_{I,l}})\}+\mathop{\text{max}}\limits _{{{\bf {U}}_{E}},{{\bf {W}}_{E}}\succeq0}h_{2}({{\bf {U}}_{E}},{{\bf {V}}_{E}},{{\bf {W}}_{E}})\nonumber \\
 & \quad+\mathop{\text{max}}\limits _{{{\bf {W}}_{X}}\succeq0}h_{3}({\bf {V}},{{\bf {V}}_{E}},{{\bf {W}}_{X}}).\label{eq:OFRsminmax}
\end{alignat}
\end{subequations}

The lower bound to the first term of (\ref{eq:OFRsminmax}) can be found as \begin{subequations}\label{RIminmax}
\begin{alignat}{1}
 & \underset{l=1,\cdots,L}{\min}\{\mathop{\text{max}}\limits _{{{\bf {U}}_{I,l}},{{\bf {W}}_{I,l}}\succeq0}h_{1,l}({{\bf {U}}_{I,l}},{\bf {V}},{{\bf {V}}_{E}},{{\bf {W}}_{I,l}})\}\label{Rspart1} \\
 & \geq \mathop{\text{max}}\limits _{\{{{\bf {U}}_{I,l}},{{\bf {W}}_{I,l}}\succeq0\}_{l=1}^{L}}\{\underset{l=1,\cdots,L}{\min}h_{1,l}({{\bf {U}}_{I,l}},{\bf {V}},{{\bf {V}}_{E}},{{\bf {W}}_{I,l}})\},\label{eq:RIminmaxfinal}
\end{alignat}
\end{subequations}
where \eqref{eq:RIminmaxfinal} holds due to the fact that $\underset{x}{\min}\ \underset{y}{\max}f(x,y)\geq\underset{y}{\max}\ \underset{x}{\min}f(x,y)$
for any function $f(x,y)$. Here by exchanging the positions of $\mathop{\text{max}}\limits _{\{{{\bf {U}}_{I,l}},{{\bf {W}}_{I,l}}\succeq0\}_{l=1}^{L}}$
and $\underset{l=1,\cdots,L}{\min}$ in \eqref{Rspart1},
we can find a lower bound to $R_{s}{\rm {(}}{\bf {V}},{{\bf {V}}_{E}},{\bf {\Phi}}{\rm {)}}$ as
\begin{alignat}{1}
& f_{ms}({\bf {V}},{{\bf {V}}_{E}},\{{{\bf {U}}_{I,l}},{{\bf {W}}_{I,l}}\}_{l=1}^{L},{{\bf {U}}_{E}},{{\bf {W}}_{E}},{{\bf {W}}_{X}}) \nonumber \\
& \triangleq\mathop{\text{max}}\limits _{{\bf {V}},{{\bf {V}}_{E}},\{{{\bf {U}}_{I,l}},{{\bf {W}}_{I,l}}\succeq0\}_{l=1}^{L},{{\bf {U}}_{E}},{{\bf {W}}_{E}}\succeq0,{{\bf {W}}_{X}}\succeq0}\{\underset{l=1,\cdots,L}{\min}h_{1,l}({{\bf {U}}_{I,l}},{\bf {V}},{{\bf {V}}_{E}},{{\bf {W}}_{I,l}})\nonumber \\
 & \quad+h_{2}({{\bf {U}}_{E}},{{\bf {V}}_{E}},{{\bf {W}}_{E}})+h_{3}({\bf {V}},{{\bf {V}}_{E}},{{\bf {W}}_{X}})\}. \label{fms}
\end{alignat}
We simplify Problem (\ref{multicastoptorigVE}) by maximizing a lower bound to its original objective as follows,
\begin{subequations} \label{multicastoptorigVEfms}
\begin{align}
 & \ \underset{{\bf {V}},{{\bf {V}}_{E}},{\bf {\Phi}},\{{{\bf {U}}_{I,l}},{{\bf {W}}_{I,l}}\succeq0\}_{l=1}^{L},{{\bf {U}}_{E}},{{\bf {W}}_{E}}\succeq0,{{\bf {W}}_{X}}\succeq0}{\mathop{\text{max}}}f_{ms}({\bf {V}},{{\bf {V}}_{E}},\{{{\bf {U}}_{I,l}},{{\bf {W}}_{I,l}}\}_{l=1}^{L},{{\bf {U}}_{E}},{{\bf {W}}_{E}},{{\bf {W}}_{X}})\label{multicastoptorigVEfms_a}\\
 & \ \ \text{s.t.}\quad\ {\rm {Tr(}}{\bf {V}}{{\bf {V}}^{H}}{\rm {+}}{{\bf {V}}_{E}}{{\bf {V}}_{E}^{H}}{\rm {)}}\le{P_{T}},\label{multicastoptorigVEfms_b}\\
 & \quad\quad\quad\!\left|{{\phi}_{m}}\right|=1,m=1,\cdots,M.\label{multicastoptorigVEfms_c}
\end{align}
\end{subequations}
To solve the multicast AN-aided SRM problem in (\ref{multicastoptorigVEfms}), a BCD-QCQP-CCP algorithm is proposed.
\subsection{BCD Iterations for Problem (\ref{multicastoptorigVEfms})}
The equivalent SRM problem in (\ref{multicastoptorigVEfms}) provides a desirable formulation for BCD algorithm. In particular, one can
show that problem (\ref{multicastoptorigVEfms}) is convex w.r.t. either ${\bf {V}},{{\bf {V}}_{E}}$ or ${\bf {\Phi}}$ or $\{{{\bf {U}}_{I,l}}$, ${{\bf {W}}_{I,l}}\}_{l=1}^{L}$, ${{\bf {U}}_{E}}$, ${{\bf {W}}_{E}}$, ${{\bf {W}}_{X}}$. By fixing ${\bf {\Phi}}$, the iteration process for problem (\ref{multicastoptorigVEfms}) is as follows. Let ${\bf {V}}^{n}$, ${{\bf {V}}_{E}^{n}}$, $\{{{\bf {U}}_{I,l}^{n}},{{\bf {W}}_{I,l}^{n}}\}_{l=1}^{L},{{\bf {U}}_{E}^{n}},{{\bf {W}}_{E}^{n}},{{\bf {W}}_{X}^{n}}$ denote the BCD iterate at the $n$th iteration. The BCD iterates are generated via \begin{subequations}\label{BCD-iteration}
\begin{alignat}{1}
 & {{\bf {U}}_{I,l}^{n}}=\text{arg}\mathop{\text{max}}\limits _{{{\bf {U}}_{I,l}}}h_{1,l}({{\bf {U}}_{I,l}},{\bf {V}}^{n},{{\bf {V}}_{E}^{n}},{{\bf {W}}_{I,l}^{n}})\nonumber \\
 & \qquad\:\:{\rm {=}}({\hat{{\bf {H}}}_{I,l}}{{\bf {V}}_{E}^{n}}{{\bf {V}}_{E}^{nH}}{\hat{{\bf {H}}}_{I,l}}^{H}{\rm {+}}\sigma_{I,l}^{2}{{\bf {I}}_{{N_{I}}}}{\rm {+}}{\hat{{\bf {H}}}_{I,l}}{\bf {V}}^{n}{{\bf {V}}^{nH}}{\hat{{\bf {H}}}_{I,l}}^{H})^{-1}{\hat{{\bf {H}}}_{I,l}}{\bf {V}}^{n}, \label{BCD-iteration_a} \\
 & {{\bf {W}}_{I,l}^{n}}=\text{arg}\mathop{\text{max}}\limits _{{{\bf {W}}_{I,l}}\succeq0}h_{1,l}({{\bf {U}}_{I,l}^{n}},{\bf {V}}^{n},{{\bf {V}}_{E}^{n}},{{\bf {W}}_{I,l}}){\rm {=[}}{{\bf {E}}_{I,l}}({{\bf {U}}_{I,l}^{n}},{\bf {V}}^{n},{{\bf {V}}_{E}^{n}}){]^{-1}}\nonumber \\
 & {\rm {=}}\text{[}({{\bf {U}}_{I,l}^{{n}H}}{\hat{{\bf {H}}}_{I,l}}{\bf {V}}^{n}-{\bf {I}}_{d}){({{\bf {U}}_{I,l}^{{n}H}}{\hat{{\bf {H}}}_{I,l}}{\bf {V}}^{n}-{\bf {I}}_{d})^{H}}+{{\bf {U}}_{I,l}^{{n}H}}({\hat{{\bf {H}}}_{I,l}}{{\bf {V}}_{E}^{n}}{{\bf {V}}_{E}^{nH}}{\hat{{\bf {H}}}_{I,l}}^{H}{\rm {+}}\sigma_{I,l}^{2}{{\bf {I}}_{{N_{I}}}}){{\bf {U}}_{I,l}^{{n}}}]^{-1}, \label{BCD-iteration_b} \\
 & {{\bf {U}}_{E}^{n}}{\rm {=}}\text{arg}\mathop{\text{max}}\limits _{{{\bf {U}}_{E}}}h_{2}({{\bf {U}}_{E}},{{\bf {V}}_{E}^{n}},{{\bf {W}}_{E}^{n}})\nonumber \\
 & \qquad\:\:{\rm {=}}(\sigma_{E}^{2}{{\bf {I}}_{{N_{E}}}}{\rm {+}}{\hat{{\bf {H}}}_{E}}{{\bf {V}}_{E}^{n}}{{\bf {V}}_{E}^{nH}}{\hat{{\bf {H}}}_{E}}^{H})^{-1}{\hat{{\bf {H}}}_{E}}{{\bf {V}}_{E}^{n}}, \label{BCD-iteration_c} \\
 & {{\bf {W}}_{E}^{n}}{=}\text{arg}\mathop{\text{max}}\limits _{{{\bf {W}}_{E}}\succeq0}h_{2}({{\bf {U}}_{E}^{n}},{{\bf {V}}_{E}^{n}},{{\bf {W}}_{E}}){\rm {=[}}{{\bf {E}}_{E}}({{\bf {U}}_{E}^{n}},{{\bf {V}}_{E}^{n}}){]^{-1}}\nonumber \\
 & \qquad\:\:{\rm {=}}[({{\bf {U}}_{E}}^{{n}H}{{\hat{{\bf {H}}}}_{E}}{\bf {V}}_{E}^{n}-{\bf {I}}_{N_{T}}){({{\bf {U}}_{E}^{{n}H}}{{\hat{{\bf {H}}}}_{E}}{\bf {V}}_{E}^{n}-{\bf {I}}_{N_{T}})^{H}}+{{\bf {U}}_{E}^{{n}H}}(\sigma_{E}^{2}{{\bf {I}}_{{N_{E}}}}){{\bf {U}}_{E}^{n}}]^{-1}, \label{BCD-iteration_d} \\
 & {{\bf {W}}_{X}^{n}}{\rm {=}}\text{arg}\mathop{\text{max}}\limits _{{{\bf {W}}_{X}}\succeq0}h_{3}({\bf {V}}^{n},{{\bf {V}}_{E}^{n}},{{\bf {W}}_{X}}){\rm {=[}}{{\bf {E}}_{X}}({\bf {V}}^{n},{{\bf {V}}_{E}^{n}}){]^{-1}}\nonumber \\
 & \qquad\:\:{\rm {=}}[{{{\bf {I}}_{{N_{E}}}}+\sigma_{E}^{-2}{{\hat{{\bf {H}}}}_{E}}({\bf {V}}^{n}{{\bf {V}}^{nH}}+{{\bf {V}}_{E}^{n}}{{\bf {V}}_{E}^{nH}})\hat{{\bf {H}}}_{E}^{H}}]^{-1}. \label{BCD-iteration_e}
\end{alignat}
\end{subequations}

The parameters ${\bf {V}}^{n}$, ${{\bf {V}}_{E}^{n}}$, ${\bf {\Phi}}^{n}$ are obtained
by solving the following problem \begin{subequations} \label{multicastoptorigVEfai}
\begin{align}
 & \ \underset{{\bf {V}},{{\bf {V}}_{E}},{\bf {\Phi}}}{\mathop{\text{max}}}\ \ \underset{l=1,\cdots,L}{\min}\{h_{1,l}({{\bf {U}}_{I,l}},{\bf {V}},{{\bf {V}}_{E}},{{\bf {W}}_{I,l}})\}\nonumber \\
 & \quad\quad\quad\quad\quad\quad+h_{2}({{\bf {U}}_{E}},{{\bf {V}}_{E}},{{\bf {W}}_{E}})+h_{3}({\bf {V}},{{\bf {V}}_{E}},{{\bf {W}}_{X}})\label{multicastoptorigVEfai_a}\\
 & \ \ \text{s.t.}\quad\ {\rm {Tr(}}{\bf {V}}{{\bf {V}}^{H}}{\rm {+}}{{\bf {V}}_{E}}{{\bf {V}}_{E}^{H}}{\rm {)}}\le{P_{T}},\label{multicastoptorigVEfai_b}\\
 & \quad\quad\quad\!\left|{{\phi}_{m}}\right|=1,m=1,\cdots,M.\label{multicastoptorigVEfai_c}
\end{align}
\end{subequations}
\subsection{Optimizing the Matrices ${\bf {V}}$ and ${{\bf {V}}_{E}}$}
By fixing ${\bf {\Phi}}$, Problem (\ref{multicastoptorigVEfai})
can be written more compactly as \begin{subequations} \label{multicastoptorigVVEfaiconvx}
\begin{align}
 & \ \underset{{\bf {V}},{{\bf {V}}_{E}}}{\mathop{\text{min}}}\ \ \underset{l=1,\cdots,L}{\max}\{-\text{Tr}({{\bf {W}}_{I,l}}{{\bf {V}}^{H}}{{\hat{{\bf {H}}}}_{I,l}}^{H}{{\bf {U}}_{I,l}})-\text{Tr}({{\bf {W}}_{I,l}}{{\bf {U}}_{I,l}}^{H}{{\hat{{\bf {H}}}}_{I,l}}{\bf {V}})\nonumber \\
 & \quad\quad\quad\quad\quad\quad\quad+\text{Tr}({{\mathbf{V}}^{H}}{{\mathbf{H}}_{V,l}^{i}}\mathbf{V})+\text{Tr}({{\mathbf{V}}_{E}^{H}}{{\mathbf{H}}_{VE,l}^{i}}{{\mathbf{V}}_{E}})-{C}_{l}\}\nonumber \\
 & \quad\quad\quad\quad\quad\quad\quad-\text{Tr}({{\mathbf{W}}_{E}}{{\mathbf{V}}_{E}^{H}}{{\mathbf{\hat{H}}}_{E}}^{H}{{\mathbf{U}}_{E}})-\text{Tr}({{\mathbf{W}}_{E}}{{\mathbf{U}}_{E}^{H}}{{\mathbf{\hat{H}}}_{E}}{{\mathbf{V}}_{E}})\nonumber \\
 & \quad\quad\quad\quad\quad\quad\quad+\text{Tr}({{\mathbf{V}}^{H}}{{\mathbf{H}}_{V}^{e}}\mathbf{V})+\text{Tr}({{\mathbf{V}}_{E}^{H}}{{\mathbf{H}}_{VE}^{e}}{{\mathbf{V}}_{E}})\label{multicastoptorigVVEfaiconvx_a}\\
 & \ \ \ \ \ \text{s.t.}\quad\ {\rm {Tr(}}{\bf {V}}{{\bf {V}}^{H}}{\rm {+}}{{\bf {V}}_{E}}{{\bf {V}}_{E}^{H}}{\rm {)}}\le{P_{T}},\label{multicastoptorigVVEfaiconvx_b}
\end{align}
\end{subequations}
where \begin{subequations}
\begin{alignat}{1}
 & {{\mathbf{H}}_{V}^{e}}\text{(}\mathbf{\Phi}\text{)}=\sigma_{E}^{-2}\mathbf{\hat{H}}_{E}^{H}{{\mathbf{W}}_{X}}{{\mathbf{\hat{H}}}_{E}},\\
 & {{\mathbf{H}}_{VE}^{e}}\text{(}\mathbf{\Phi}\text{)}={{\mathbf{\hat{H}}}_{E}}^{H}{{\mathbf{U}}_{E}}{{\mathbf{W}}_{E}}{{\mathbf{U}}_{E}^{H}}{{\mathbf{\hat{H}}}_{E}}+\sigma_{E}^{-2}{{\mathbf{\hat{H}}}_{E}}^{H}{{\mathbf{W}}_{X}}{{\mathbf{\hat{H}}}_{E}},\\
 & {{\mathbf{H}}_{V,l}^{i}}\text{(}\mathbf{\Phi}\text{)}={{\mathbf{\hat{H}}}_{I,l}}^{H}{{\mathbf{U}}_{I,l}}{{\mathbf{W}}_{I,l}}{{\mathbf{U}}_{I,l}^{H}}{{\mathbf{\hat{H}}}_{I,l}},\\
 & {{\mathbf{H}}_{VE,l}^{i}}\text{(}\mathbf{\Phi}\text{)}={{\mathbf{\hat{H}}}_{I,l}}^{H}{{\mathbf{U}}_{I,l}}{{\mathbf{W}}_{I,l}}{{\mathbf{U}}_{I,l}^{H}}{{\mathbf{\hat{H}}}_{I,l}},\\
 & {C}_{l}=\log\left|{{\bf {W}}_{I,l}}\right|+d-\text{Tr}({{\bf {W}}_{I,l}}+\sigma_{I,l}^{2}{{\bf {W}}_{I,l}}{{\bf {U}}_{I,l}}^{H}{{\bf {U}}_{I,l}}).
\end{alignat}
\end{subequations}

The Problem (\ref{multicastoptorigVVEfaiconvx}) is a convex QCQP problem, we can
obtain its optimal solution using a general-purpose convex
optimization solver.
\subsection{Optimizing the Phase Shifts $\mathbf{\Phi}$}
By fixing ${\bf {V}},{{\bf {V}}_{E}}$, the optimization problem for
the phase shift matrix $\mathbf{\Phi}$ reduced from Problem (\ref{multicastoptorigVVEfaiconvx})
is formulated as
\begin{subequations} \label{multicastoptorigfaiconvx}
\begin{align}
 & \ \underset{\mathbf{\Phi}}{\mathop{\text{min}}}\ \ {g_{0}}(\mathbf{\Phi})\triangleq\underset{l=1,\cdots,L}{\max}\{-\text{Tr}({{\bf {W}}_{I,l}}{{\bf {V}}^{H}}{{\hat{{\bf {H}}}}_{I,l}}^{H}{{\bf {U}}_{I,l}})-\text{Tr}({{\bf {W}}_{I,l}}{{\bf {U}}_{I,l}}^{H}{{\hat{{\bf {H}}}}_{I,l}}{\bf {V}})\nonumber \\
 & \quad\quad\quad\quad\quad\quad\quad+\text{Tr}({{\mathbf{V}}^{H}}{{\mathbf{H}}_{V,l}^{i}}\mathbf{V})+\text{Tr}({{\mathbf{V}}_{E}^{H}}{{\mathbf{H}}_{VE,l}^{i}}{{\mathbf{V}}_{E}})-{C}_{l}\}\nonumber \\
 & \quad\quad\quad\quad\quad\quad\quad-\text{Tr}({{\mathbf{W}}_{E}}{{\mathbf{V}}_{E}^{H}}{{\mathbf{\hat{H}}}_{E}}^{H}{{\mathbf{U}}_{E}})-\text{Tr}({{\mathbf{W}}_{E}}{{\mathbf{U}}_{E}^{H}}{{\mathbf{\hat{H}}}_{E}}{{\mathbf{V}}_{E}})\nonumber \\
 & \quad\quad\quad\quad\quad\quad\quad+\text{Tr}({{\mathbf{V}}^{H}}{{\mathbf{H}}_{V}^{e}}\mathbf{V})+\text{Tr}({{\mathbf{V}}_{E}^{H}}{{\mathbf{H}}_{VE}^{e}}{{\mathbf{V}}_{E}})\label{multicastoptorigfaiconvx_a}\\
 & \ \ \ \ \ \text{s.t.}\quad\ \!\left|{{\phi}_{m}}\right|=1,m=1,\cdots,M.\label{multicastoptorigfaiconvx_b}
\end{align}
\end{subequations}

By complex mathematical manipulations, the OF ${g_{0}}(\mathbf{\Phi})$
can be equivalently transformed into
\begin{alignat}{1}
{g_{0}}(\mathbf{\Phi}) & \triangleq\underset{l=1,\cdots,L}{\max}\{{g_{0,l}^{i}}(\mathbf{\Phi})\}+{g_{0}^{e}}(\mathbf{\Phi}),
\end{alignat}
where
\begin{subequations}
\begin{alignat}{1}
{g_{0,l}^{i}}(\mathbf{\Phi}) & ={\rm {Tr}}\left({{\bf {\Phi}}^{H}{\bf {D}}_{l}^{iH}}\right)+{\rm {Tr}}\left({\bf {\Phi}}{\bf {D}}_{l}^{i}\right)+{\rm {Tr}}\left[{{\bf {\Phi}}{{\bf {C}}_{VE}}{{\bf {\Phi}}^{H}}{{\bf {B}}_{VE,l}^{i}}}\right]+{\rm {Tr}}\left({{\bf {\Phi}}{{\bf {C}}_{V}}{{\bf {\Phi}}^{H}}{{\bf {B}}_{V,l}^{i}}}\right)+C_{l}^{i},\\
{g_{0}^{e}}(\mathbf{\Phi}) & ={\rm {Tr}}\left({{\bf {\Phi}}^{H}{\bf {D}}^{eH}}\right)+{\rm {Tr}}\left({\bf {\Phi}}{\bf {D}}^{e}\right)+{\rm {Tr}}\left[{{\bf {\Phi}}{{\bf {C}}_{VE}}{{\bf {\Phi}}^{H}}{{\bf {B}}_{VE}^{e}}}\right]+{\rm {Tr}}\left({{\bf {\Phi}}{{\bf {C}}_{V}}{{\bf {\Phi}}^{H}}{{\bf {B}}_{V}^{e}}}\right)+C^{e},
\end{alignat}
\end{subequations}
and
\begin{subequations}
\begin{alignat}{1}
{\bf {D}}_{l}^{i} & ={\bf {G}}{{\bf {V}}_{X}}{\bf {H}}_{b,I,l}^{H}{{\bf {M}}_{I,l}}{{\bf {H}}_{R,I,l}}-{\bf {GV}}{{\bf {W}}_{I,l}}{\bf {U}}_{I,l}^{H}{{\bf {H}}_{R,I,l}},\\
{{\bf {C}}_{VE}} & ={\bf {G}}{{\bf {V}}_{E}}{\bf {V}}_{E}^{H}{{\bf {G}}^{H}},\\
{{\bf {C}}_{V}} & ={\bf {G}}{\bf {V}}{\bf {V}}^{H}{{\bf {G}}^{H}},\\
{{\bf {B}}_{VE,l}^{i}} & =\left({{\bf {H}}_{R,I,l}^{H}{{\bf {U}}_{I,l}}{{\bf {W}}_{I,l}}{\bf {U}}_{I,l}^{H}{{\bf {H}}_{R,I,l}}}\right),\\
{{\bf {B}}_{V,l}^{i}} & =\left({{\bf {H}}_{R,I,l}^{H}{{\bf {U}}_{I,l}}{{\bf {W}}_{I,l}}{\bf {U}}_{I,l}^{H}{{\bf {H}}_{R,I,l}}}\right),\\
C_{l}^{i} & ={\rm {Tr}}\left[{{\bf {H}}_{b,I,l}}{{\bf {V}}_{X}}{\bf {H}}_{b,I,l}^{H}{{\bf {M}}_{I,l}}\right]\!+\!{\rm {Tr}}\left[{{{\bf {U}}_{I,l}}{\bf {W}}_{I,l}^{H}{{\bf {V}}^{H}}{\bf {H}}_{b,I,l}^{H}}\right]\!+\!{\rm {Tr}}\left[{{{\bf {H}}_{b,I,l}}{\bf {V}}{{\bf {W}}_{I,l}}{\bf {U}}_{I,l}^{H}}\right]-{C}_{l},\\
{\bf {M}}_{I,l} & ={{\bf {U}}_{I,l}}{{\bf {W}}_{I,l}}{\bf {U}}_{I,l}^{H},\\
{\bf {D}}^{e} & =\sigma_{E}^{-2}{\bf {G}}{{\bf {V}}_{X}}{\bf {H}}_{b,E}^{H}{{\bf {W}}_{X}}{{\bf {H}}_{R,E}}+{\bf {G}}{{\bf {V}}_{E}}{\bf {V}}_{E}^{H}{\bf {H}}_{b,E}^{H}{{\bf {M}}_{E}}{{\bf {H}}_{R,E}}-{\bf {G}}{{\bf {V}}_{E}}{{\bf {W}}_{E}}{\bf {U}}_{E}^{H}{{\bf {H}}_{R,E}},\\
{{\bf {B}}_{VE}^{e}} & =\left({\sigma_{E}^{-2}{\bf {H}}_{R,E}^{H}{{\bf {W}}_{X}}{{\bf {H}}_{R,E}}+{\bf {H}}_{R,E}^{H}{{\bf {U}}_{E}}{{\bf {W}}_{E}}{\bf {U}}_{E}^{H}{{\bf {H}}_{R,E}}}\right),\\
{{\bf {B}}_{V}^{e}} & =\left({\sigma_{E}^{-2}{\bf {H}}_{R,E}^{H}{{\bf {W}}_{X}}{{\bf {H}}_{R,E}}}\right),\\
C^{e} & =\sigma_{E}^{-2}{\rm {Tr}}\left[{{\bf {H}}_{b,E}}{{\bf {V}}_{X}}{\bf {H}}_{b,E}^{H}{{\bf {W}}_{X}}\right]+{\rm {Tr}}\left[{{\bf {H}}_{b,E}}{{\bf {V}}_{E}}{\bf {V}}_{E}^{H}{\bf {H}}_{b,E}^{H}{{\bf {M}}_{E}}\right] \nonumber \\
 & +{\rm {Tr}}\left[{{{\bf {U}}_{E}}{\bf {W}}_{E}^{H}{\bf {V}}_{E}^{H}{\bf {H}}_{b,E}^{H}}\right]+{\rm {Tr}}\left[{{{\bf {H}}_{b,E}}{{\bf {V}}_{E}}{{\bf {W}}_{E}}{\bf {U}}_{E}^{H}}\right].
\end{alignat}
\end{subequations}

Similarly, Problem (\ref{multicastoptorigfaiconvx}) can be further simplified as
\begin{subequations}\label{findfaiprepareforMM}
\begin{alignat}{1}
\underset{{\bm{\phi}}}{\mathop{\text{min}}} & \underset{l=1,\cdots,L}{\max}\{{{\bm{\phi}}^{{\rm {H}}}}{\bm{\Xi}_{l}^{i}}{\bm{\phi}}+2\textrm{Re}[{\bm{\phi}}^{{\rm {H}}}{\bf {d}}_{l}^{i*}]+C_{l}^{i}\}+{{\bm{\phi}}^{{\rm {H}}}}{\bm{\Xi}^{e}}{\bm{\phi}}+2\textrm{Re}[{\bm{\phi}}^{{\rm {H}}}{\bf {d}}^{e*}]+C^{e}\\
\text{s.t.} & \quad\quad\ \!\left|{{\phi}_{m}}\right|=1,m=1,\cdots,M,
\end{alignat}
\end{subequations}
where
\begin{subequations}
\begin{alignat}{1}
{\bm{\Xi}_{l}^{i}} & ={\bf {B}}_{VE,l}^{i}\odot{{\bf {C}}_{VE}^{{\rm {T}}}}+{\bf {B}}_{V,l}^{i}\odot{{\bf {C}}_{V}^{{\rm {T}}}},\\
{\bm{\Xi}^{e}} & ={\bf {B}}_{VE}^{e}\odot{{\bf {C}}_{VE}^{{\rm {T}}}}+{\bf {B}}_{V}^{e}\odot{{\bf {C}}_{V}^{{\rm {T}}}},\\
{\bf {d}}_{l}^{i} & ={\left[{{{\left[{\bf {D}}_{l}^{i}\right]}_{1,1}},\cdots,{{\left[{\bf {D}}_{l}^{i}\right]}_{M,M}}}\right]^{{\rm {T}}}},\\
{\bf {d}}^{e} & ={\left[{{{\left[{\bf {D}}^{e}\right]}_{1,1}},\cdots,{{\left[{\bf {D}}^{e}\right]}_{M,M}}}\right]^{{\rm {T}}}}.
\end{alignat}
\end{subequations}
By using the Lemma 1 in \cite{pan2019multicell}, Problem (\ref{findfaiprepareforMM}) will be recast as \begin{subequations}\label{findfaiMMformorg}
\begin{alignat}{1}
\underset{{\bm{\phi}}}{\mathop{\max}} & \underset{l=1,\cdots,L}{\min}\{2\textrm{Re}[{{\bm{\phi}}^{{\rm {H}}}}{\bf {q}}_{l}^{i,t}]-C_{q,l}^{i,t}\}+2\textrm{Re}[{{\bm{\phi}}^{{\rm {H}}}}{\bf {q}}^{e,t}]-C_{q}^{e,t}\\
\text{s.t.} & \quad\quad\ \!\left|{{\phi}_{m}}\right|=1,m=1,\cdots,M,
\end{alignat}
\end{subequations}
where
\begin{subequations}
\begin{alignat}{1}
{\bf {q}}_{l}^{i,t} & =\left({{\lambda_{l,{\rm {\max}}}^{i}}{{\bf {I}}_{M}}-{\bm{\Xi}_{l}^{i}}}\right){{\bm{\phi}}^{t}}-{\bf {d}}_{l}^{i*},\\
C_{q,l}^{i,t} & =2M{\lambda_{l,{\rm {\max}}}^{i}}-{\left({{\bm{\phi}}^{t}}\right)^{{\rm {H}}}}\left({\bm{\Xi}_{l}^{i}}\right){{\bm{\phi}}^{t}}+C_{l}^{i},\\
{\bf {q}}^{e,t} & =\left({{\lambda_{{\rm {\max}}}^{e}}{{\bf {I}}_{M}}-{\bm{\Xi}^{e}}}\right){{\bm{\phi}}^{t}}-{\bf {d}}^{e*},\\
C_{q}^{e,t} & =2M{\lambda_{{\rm {\max}}}^{e}}-{\left({{\bm{\phi}}^{t}}\right)^{{\rm {H}}}}\left({\bm{\Xi}^{e}}\right){{\bm{\phi}}^{t}}+C^{e},
\end{alignat}
\end{subequations}
and ${\lambda_{l,{\rm {\max}}}^{i}}$ is the the maximum eigenvalue
of ${\bm{\Xi}_{l}^{i}}$, and ${\lambda_{{\rm {\max}}}^{e}}$ is the
maximum eigenvalue of ${\bm{\Xi}^{e}}$. By defining ${\bf {q}}_{l}^{ie,t} \triangleq {\bf {q}}_{l}^{i,t}+{\bf {q}}^{e,t}$ and $C_{q,l}^{ie,t}\triangleq C_{q,l}^{i,t}+C_{q}^{e,t}$, Problem (\ref{findfaiMMformorg})
can be rewritten as \begin{subequations}\label{findfaiMMform2}
\begin{alignat}{1}
\underset{{\bm{\phi}}}{\mathop{\max}} & \underset{l=1,\cdots,L}{\min}\{2\textrm{Re}[{{\bm{\phi}}^{{\rm {H}}}}{\bf {q}}_{l}^{ie,t}]-C_{q,l}^{ie,t}\}\\
\text{s.t.} & \quad\quad\ \!\left|{{\phi}_{m}}\right|=1,m=1,\cdots,M,
\end{alignat}
\end{subequations}
which is equivalent to the following problem
\begin{subequations}
\begin{alignat}{1}
\underset{{\bm{\phi}},z}{\mathop{\max}} & \quad z\\
\text{s.t.}\ \ & 2\textrm{Re}[{{\bm{\phi}}^{{\rm {H}}}}{\bf {q}}_{l}^{ie,t}]-C_{q,l}^{ie,t}\geq z, l=1,\cdots,L,\\
 & \left|{{\phi}_{m}}\right|=1,m=1,\cdots,M. \label{unitmodulusCCP}
\end{alignat}
\end{subequations}

We note that the above problem is still non-convex due to the unit-modulus
constraints. To deal with the non-convex constraints, the penalty
CCP method is applied. Following the penalty CCP framework, the constraint
(\ref{unitmodulusCCP}) are firstly equivalently rewritten as $-1\leq\left|{{\phi}_{m}}\right|^{2}\leq1,m=1,\cdots,M$.
The non-convex parts of the resulting constraints are then linearized
by $\left|{{\phi}_{m}^{\text{(n)}}}\right|^{2}-2\textrm{Re}({{\phi}_{m}^{\text{(*)}}}{{\phi}_{m}^{\text{(n)}}})\leq-1,m=1,\cdots,M$.
We finally have the following convex subproblem of ${\bm{\phi}}$
as
\begin{subequations}\label{ccpfai}
\begin{alignat}{1}
\underset{{\bm{\phi}},z}{\mathop{\max}} & \quad z-\lambda^{(t)}\left\Vert \mathbf{b}\right\Vert _{1}\\
\text{s.t.} & \ 2\textrm{Re}[{{\bm{\phi}}^{{\rm {H}}}}{\bf {q}}_{l}^{ie,t}]-C_{q,l}^{ie,t}\geq z,\\
 & \left|{{\phi}_{m}^{\text{(n)}}}\right|^{2}-2\textrm{Re}({{\phi}_{m}^{\text{(*)}}}{{\phi}_{m}^{\text{(n)}}})\leq b_{m}-1,m=1,\cdots,M,\\
 & \left|{{\phi}_{m}}\right|^{2}\leq1+b_{M+m},m=1,\cdots,M.
\end{alignat}
\end{subequations}
where $\mathbf{b}=[b_{1},\cdots,b_{2M}]^{T}$ are slack variables
imposed over the equivalent linear constraints of the unit-modulus
constraints, and $\left\Vert \mathbf{b}\right\Vert _{1}$ is the penalty
term in the OF. $\left\Vert \mathbf{b}\right\Vert _{1}$
is scaled by the regularization factor $\lambda^{(t)}$ to control
the feasibility of the constraints. The specific steps of the penalty CCP method can be referred in \cite{zhou2020framework}.
\section{Simulation Results}\label{simlresult}
\begin{figure}
\centering
\includegraphics[width=3.4in]{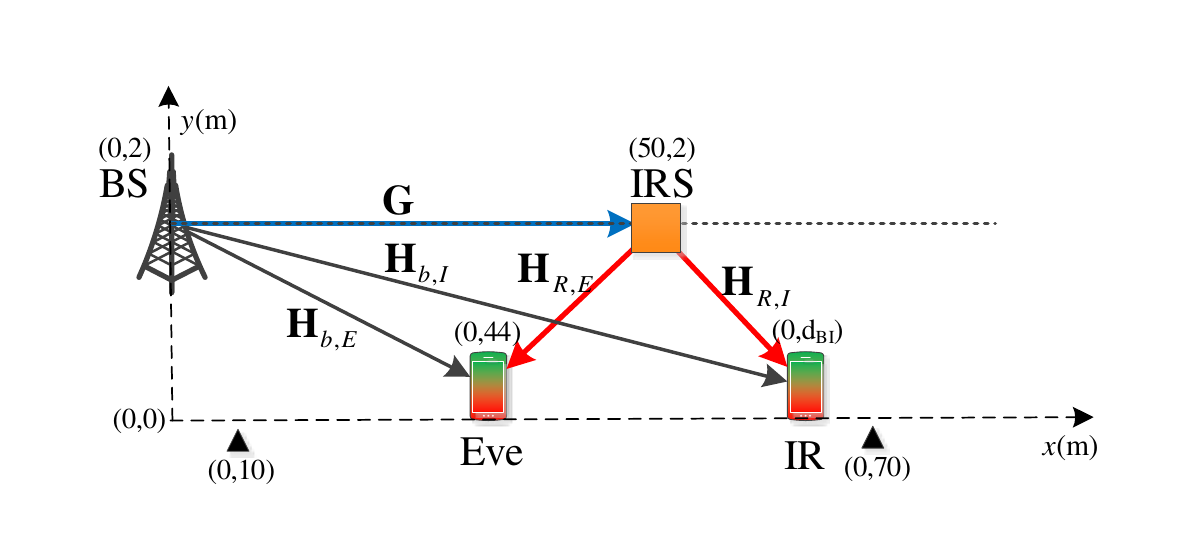}
\caption{The three-terminal MIMO communication scenario in simulation.}\vspace{-0.8cm}
\label{fig2add}
\end{figure}

In this section, numerical simulations are carried out to evaluate the assistance of the IRS on the AN-aided MIMO secure communication system. We focus on the scenario of the standard three-terminal MIMO Guassian wiretap channel shown in Fig.~\ref{fig2add}, where there are one BS, one legitimate IR and one Eve, all with multiple antennas. The distance from the BS to the IRS is $d_{BR}=50$ m. We assume that the line connecting the IR and Eve is parallel to the line connecting the BS and the IRS, and that the vertical distance between them is $d_{v}=2$ m.
The large-scale path loss is modeled as ${\rm{PL = }}{{\rm{P}}{{\rm{L}}_0} - 10\alpha {{\log }_{10}}\left( {\frac{d}{{{d_0}}}} \right)}$, where ${\rm{P}}{{\rm{L}}_0}$ is the path loss at the reference distance $d_0=1$ m, $\alpha$ is the path loss exponent, $d$ is the link distance. In our simulations, we set ${\rm{P}}{{\rm{L}}_0}=-30$ dB. The path loss exponents of the links from BS to Eve, from BS to IR, from IRS to Eve and from IRS to IR are ${\alpha _{{\rm{BE}}}}=3.5$, ${\alpha _{{\rm{BI}}}}=3.5$, ${\alpha _{{\rm{RE}}}}=2.5$ and ${\alpha _{{\rm{RI}}}}=2.5$ respectively. The path-loss exponents of the link from BS to IRS is set to be ${\alpha _{{\rm{BR}}}}= 2.2$, which means that the IRS is well-located, and the path loss is negligible in this link.

For the direct channels from the BSs to the Eve and IR, the small-scale fading is assumed to be Rayleigh fading due to extensive scatters. However, for the IRS-related channels, the small-scale fading is assumed to be Rician fading. Specifically, the small-scale channel can be modeled as
\begin{alignat}{1}
\tilde{\mathbf{H}} & =\left(\sqrt{\frac{\beta}{1+\beta}}\tilde{\mathbf{H}}^{LOS}+\sqrt{\frac{1}{1+\beta}}\tilde{\mathbf{H}}^{NLOS}\right),
\end{alignat}
where $\beta$ is the Rican factor, $\tilde{\mathbf{H}}^{LOS}$ denotes
the deterministic line of sight (LoS) component of the IRS-related
channel, and $\tilde{\mathbf{H}}^{NLOS}$ denotes the non-LoS (NLoS)
component of the IRS-related channel, which is modeled as Rayleigh
fading. By assuming the antennas at the BS, IRS, Eve and IR are arranged
in a uniform linear array (ULA), the $\tilde{\mathbf{H}}^{LOS}$ can
be modeled as $\tilde{\mathbf{H}}^{LOS}=\mathbf{a}_{r}\mathbf{a}_{t}^{H}$,
where $\mathbf{a}_{t}$ and $\mathbf{a}_{r}$ are the steering vectors
of the transmitting and receiving arrays respectively. The $\mathbf{a}_{t}$
and $\mathbf{a}_{r}$ are defined as,
\begin{subequations}\label{steeringvectr}
\begin{alignat}{1}
\mathbf{a}_{t} & =\left[\begin{array}{cccc}
1, & \exp(j2\pi\frac{d_{t}}{\lambda}\sin\varphi_{t}), & \cdots, & \exp(j2\pi\frac{d_{t}}{\lambda}(N_{t}-1)\sin\varphi_{t})\end{array}\right]^{T},\\
\mathbf{a}_{r} & =\left[\begin{array}{cccc}
1, & \exp(j2\pi\frac{d_{r}}{\lambda}\sin\varphi_{r}), & \cdots, & \exp(j2\pi\frac{d_{r}}{\lambda}(N_{r}-1)\sin\varphi_{r})\end{array}\right]^{T}.
\end{alignat}
\end{subequations}

In \eqref{steeringvectr}, $\lambda$ is the wavelength; $d_{t}$
and $d_{r}$ are the element intervals of the transmitting and receiving
array; $\varphi_{t}$ and $\varphi_{r}$ are the angle of departure
and the angle of arrival; $N_{t}$ and $N_{r}$ are the number of
antennas/elements at the transmitter and receiver, respectively. We
set $\frac{d_{t}}{\lambda}=\frac{d_{r}}{\lambda}=0.5$, and $\varphi_{t}=\tan^{-1}(\frac{y_{r}-y_{t}}{x_{r}-x_{t}})$,
$\varphi_{r}=\pi-\varphi_{t}$, where $(x_{t},y_{t})$ is the location
of the transmitter, and $(x_{r},y_{r})$ is the location of the receiver.

If not specified, the simulation parameters are set as follows. The IR's noise power and the Eve's noise power are $\sigma _{I}^{2}=-75$ dBm and $\sigma _{E}^{2}=-75$ dBm. The numbers of BS antennas, IR antennas, and Eve antennas are $N_T=4$, $N_I=2$, and $N_E=2$ respectively. There are $d=2$ data streams and $M=50$ IRS reflection elements. The transmit power limit is $P_{T}=15$ dBm, and the error tolerance is $\varepsilon=10^{-6}$. The horizontal distance between the BS and the Eve is $d_{BE}=44$ m. The horizontal distance between the BS and the IR is selected from $d_{BI}=[10 \ \text{m},70 \ \text{m}]$. The channels are realized 200 times independently to average the simulation results.

\subsection{Convergence Analysis}
\begin{figure}
\begin{minipage}[t]{0.495\linewidth}
\centering
\includegraphics[width=2.6in]{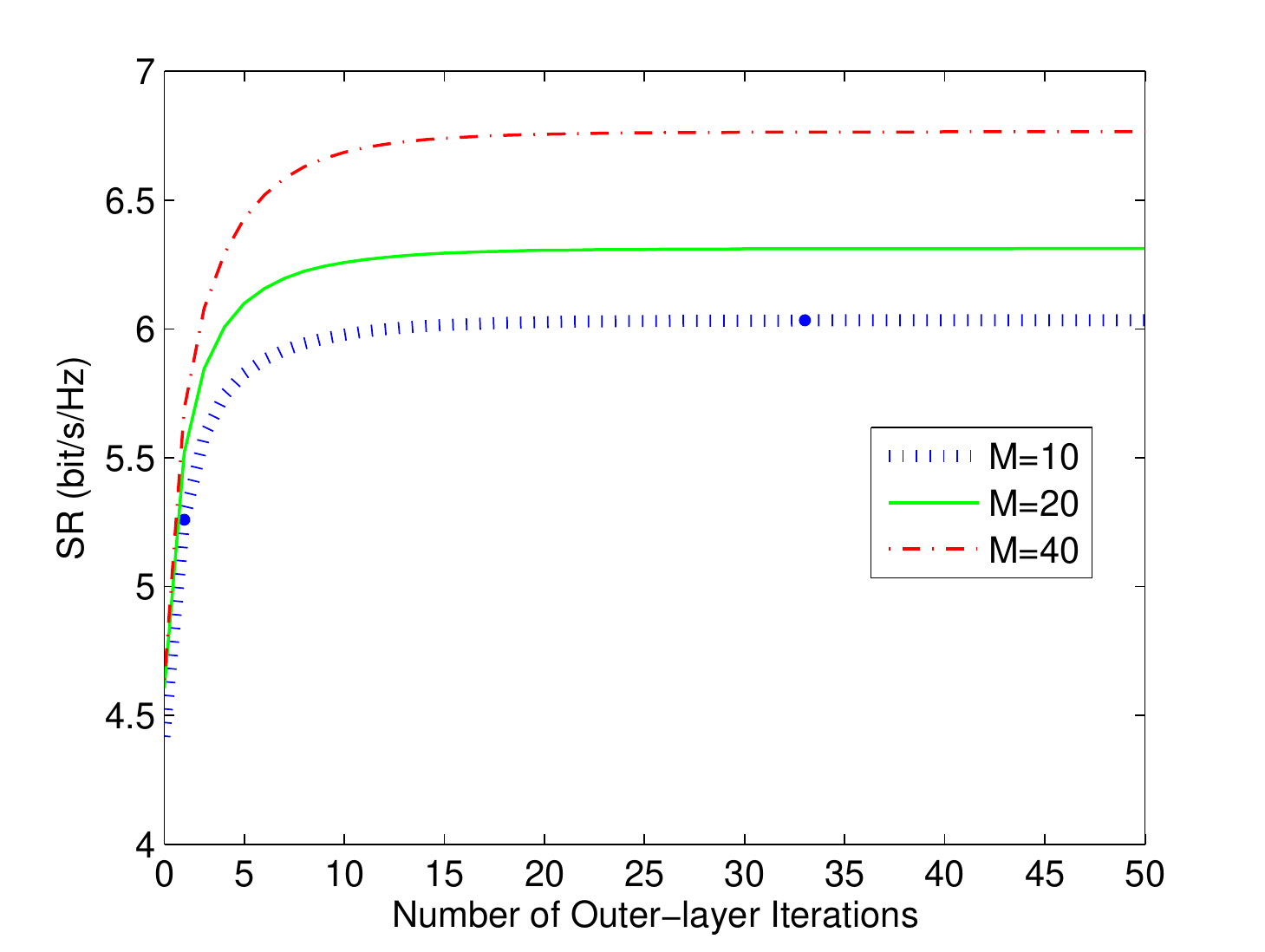}\vspace{-0.6cm}
\caption{Convergence behaviour of the BCD algorithm.}
\label{fig1simu}
\end{minipage}%
\hfill
\begin{minipage}[t]{0.495\linewidth}
\centering
\includegraphics[width=2.6in]{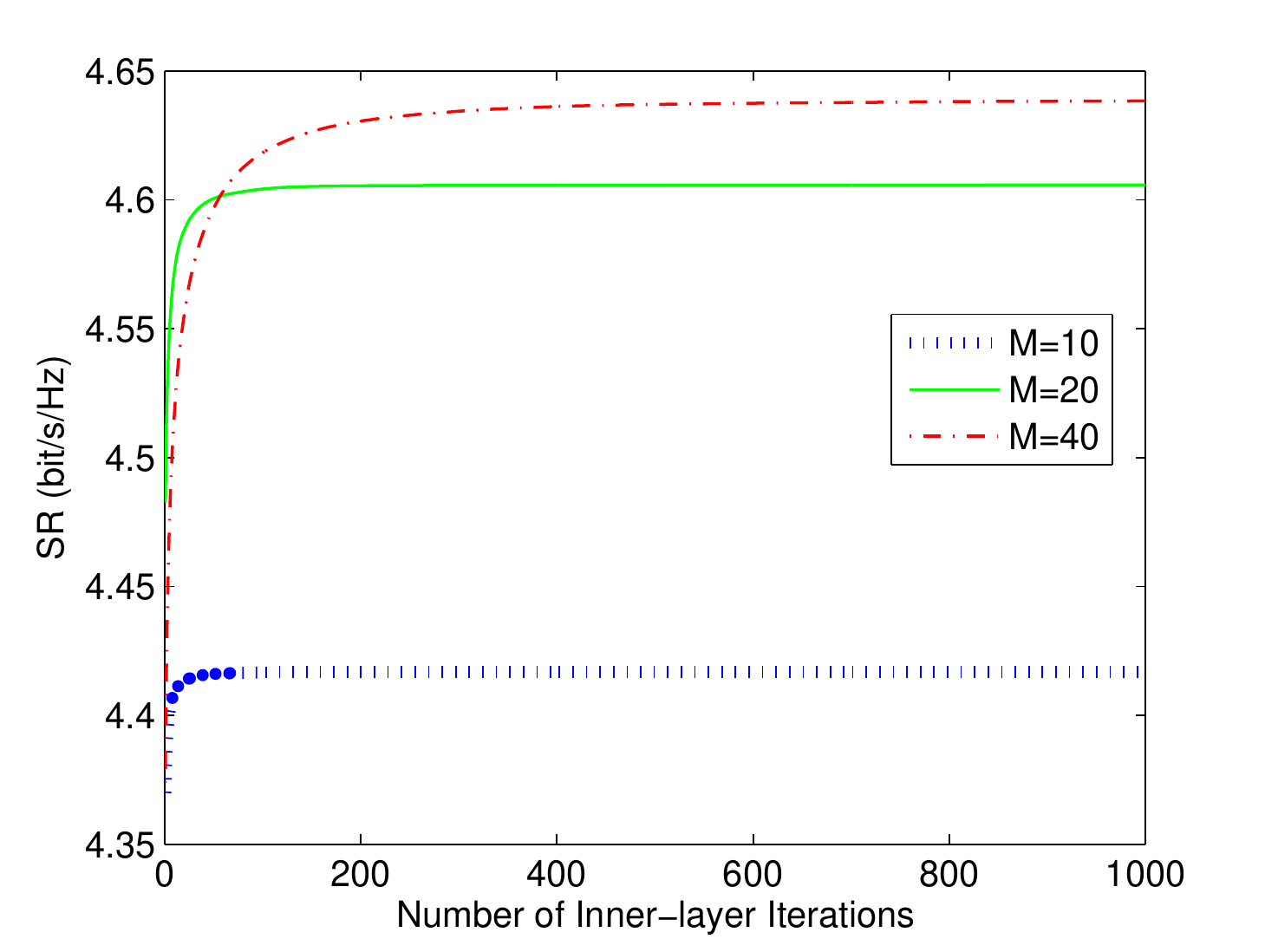}\vspace{-0.6cm}
\caption{Convergence behaviour of the MM algorithm.}
\label{fig2simu}
\end{minipage}\vspace{-0.7cm}
\end{figure}

The convergence performance of the proposed BCD-MM algorithm is investigated. The iterations of the BCD algorithm are termed as outer-layer iterations, while the iterations of the MM algorithm are termed as the inner-layer iterations. Fig.~\ref{fig1simu} shows three examples of convergence behaviour for $M=$10, 20 and 40 phase shifts of IRS. In Fig.~\ref{fig1simu}, the SR increases versus the iteration number, and finally reaches a stable value. It is shown that the algorithm converges quickly, almost with 20 iterations, which demonstrates the efficiency of the proposed algorithm. Moreover, a larger converged SR value is reached with a larger $M$, which means that better security can be obtained by using more IRS elements. However, more IRS elements bring a heavier computation, which is demonstrated in Fig.~\ref{fig1simu} in the form of a slower convergence speed with more phase shifters.


Specifically, we evaluate the convergence performance of the MM algorithm used for solving the optimal IRS phase shifts. The inner-layer iterative process of the MM algorithm in the first iteration of the BCD algorithm is shown in Fig.~\ref{fig2simu}. The SR value increases as the iteration number increases, and finally converges to a stable value. According with the convergency performance in the out-layer iteration, similar conclusions can be drawn for the inner-layer iteration, which is that a higher converged SR value can be obtained with more phase shifts but at the cost of lower convergence speed. The reason for the lower convergence speed with larger $M$ value is that more optimization variables are introduced, which require more computation complexity.

\subsection{Performance Evaluation}
In this subsection, our proposed algorithm is evaluated by comparing the simulation results to three schemes of
\begin{enumerate}
  \item  \textbf{RandPhase}: The phase shifts of the IRS are randomly selected from $[0,2\pi]$. In this scheme, the MM algorithm is skipped, and only the TPC matrix and AN covariance matrix are optimized.
  \item \textbf{No-IRS}: Without the IRS, the channel matrices of IRS related links become zero matrices, which is ${\bf{H}}_{R,I}={\bf{0}}$, ${\bf{H}}_{R,E}={\bf{0}}$ and ${{\bf{G}}}={\bf{0}}$. This scheme results a conventional AN-aided communication system, and only the TPC matrix and AN covariance matrix need to be optimized.
      \item \textbf{BCD-QCQP-SDR}: The BCD algorithm is utilized. However, the TPC matrix and the AN covariance matrix is optimized by tackling Problem (\ref{optorigVElowerbndSmpNOfai}) as a QCQP problem, which is solved by the general CVX solvers, e.g. Sedumi or Mosek. The phase shifts of IRS are optimized by solving Problem (\ref{appjig}) with the SDR technique.
\end{enumerate}
\subsubsection{Impact of Transmit Power}
\begin{figure}
\begin{minipage}[t]{0.495\linewidth}
\centering
\includegraphics[width=2.6in]{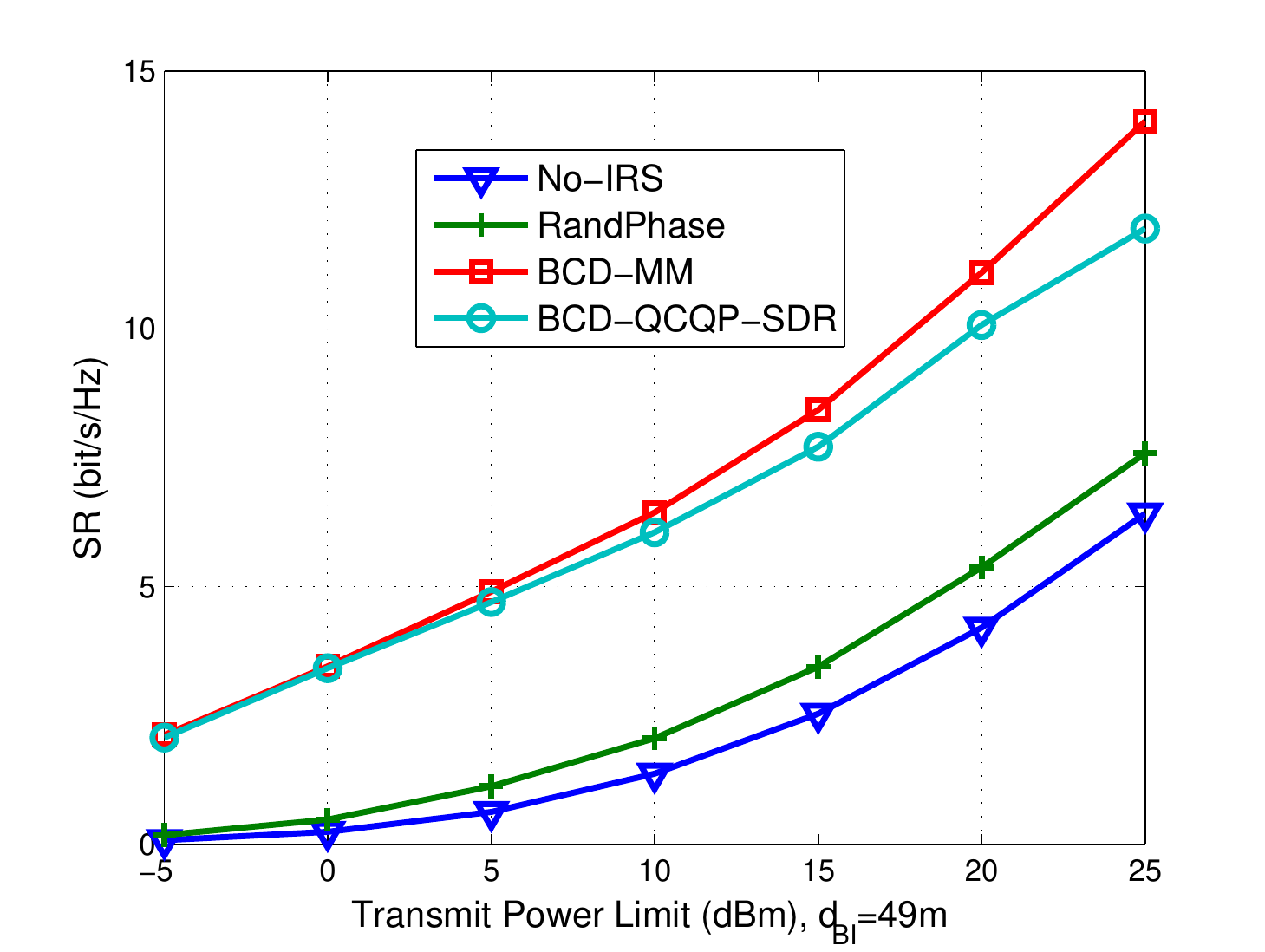}\vspace{-0.6cm}
\caption{Achievable SR versus the transmit power limit.}
\label{fig3simu}
\end{minipage}%
\hfill
\begin{minipage}[t]{0.495\linewidth}
\centering
\includegraphics[width=2.6in]{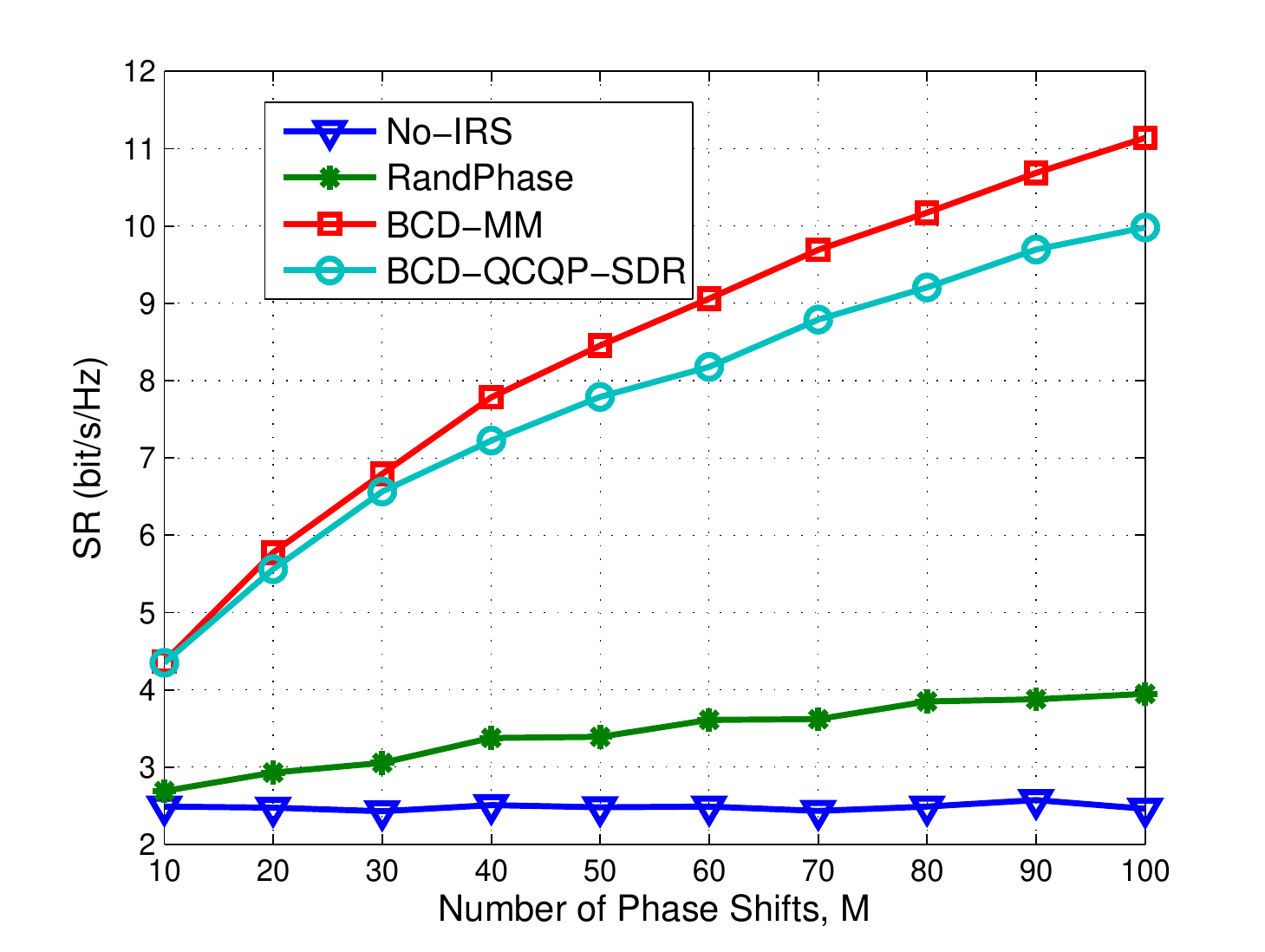}\vspace{-0.6cm}
\caption{Achievable SR versus the number of phase shifts $M$.}
\label{fig4simu}
\end{minipage}\vspace{-0.7cm}
\end{figure}

To evaluate the impact of the transmit power limit $P_T$, the average SRs versus the transmit power limit for various schemes are given in Fig.~\ref{fig3simu}, which demonstrates that the achieved SRs of three schemes increase as the power limit $P_T$ increases. It is observed that the BCD-MM algorithm significantly outperforms the other three benchmark schemes over the entire range of transmit power limits. By comparing the RandPhase scheme to the No-IRS scheme, we find that the RandPhase scheme is better than the No-IRS scheme for obtaining higher SR, and that the SR gap increases with the power limit $P_T$. The reason is that, for the RandPhase scheme, the IR is closer to the IRS than the Eve is, and more signal power from the IRS can be acquired by the IR than that by the Eve, while for the No-IRS scheme, the IR is further from the BS than the Eve is, and less signal power from the BS can be acquired by the IR than that by the Eve. This comparison result signifies that even the phase shifts of IRS are random, the IRS can enhance the system security. In comparison to the no-IRS scheme, the SR gain achieved by the proposed algorithm is very obvious, and increases greatly with the power limit $P_T$, which confirms the effectiveness and benefits of employing the IRS. By comparing the proposed scheme and the RandPhase scheme, we find that the security gain obtained for the proposed scheme is much greater than that for the RandPhase scheme. That's because the phase shifts of IRS are properly designed to enhance the signal received at the IR more constructively, and weaken the signal received at the Eve more destructively. This comparison result emphasizes that optimizing the phase shifts of IRS is important and necessary. By comparing the proposed BCD-MM algorithm and the BCD-QCQP-SDR algorithm, we observe that in terms of the SR performance, the proposed BCD-MM algorithm is better BCD-QCQP-SDR, and the performance gain increases with the $P_T$. Moreover, the proposed BCD-MM algorithm is much more efficient than the BCD-QCQP-SDR algorithm. The superiority of the proposed algorithm is further validated.
\subsubsection{Impact of the Phase Shifts Number}
The averaged SR performance of four schemes with various phase shifts number $M$ is shown in Fig.~\ref{fig4simu}, which demonstrates that the proposed BCD-MM algorithm is significantly superior to the other three schemes. We observe that the SR achieved by the BCD-MM scheme obviously increases with $M$, while the RandPhase scheme only shows a slight improvement as $M$ increases, and the No-IRS scheme has very low SRs irrelative with $M$. Larger the element number $M$ of IRS is, more significant the performance gain obtained by the proposed algorithm is. For example, when $M$ is small as $M=10$, the SR gain of the BCD-MM relative to the No-IRS is only 1.3 bit/s/Hz, while this SR gain becomes 9.5 bit/s/Hz when $M$ increases to $M=100$. The performance gain for the proposed algorithm originates from two perspectives. On the one hand, a higher array gain can be obtained by increasing $M$, since more signal power can be received at the IRS with larger $M$. On the other hand, a higher reflecting beamforming gain can be obtained by increasing $M$, which means that the sum of coherently adding the reflected signals at the IRS elements increases with $M$ by appropriately designing the phase shifts. However, only the array gain can be exploited by the RandPhase scheme, thus the SRs for it increase very slowly, and remain at much lower values than those for the proposed algorithm. These results further confirm that more security improvements can be archived by using a large IRS with more reflect elements and optimizing the phase shifts properly, however there may bring the computation complexity problem. In comparison to the BCD-QCQP-SDR algorithm, the proposed BCD-MM algorithm can achieve the higher SR, and the SR performance gap increases with $M$.

\subsubsection{Impact of the relative location of IRS}
\begin{figure}
\begin{minipage}[t]{0.495\linewidth}
\centering
\includegraphics[width=2.6in]{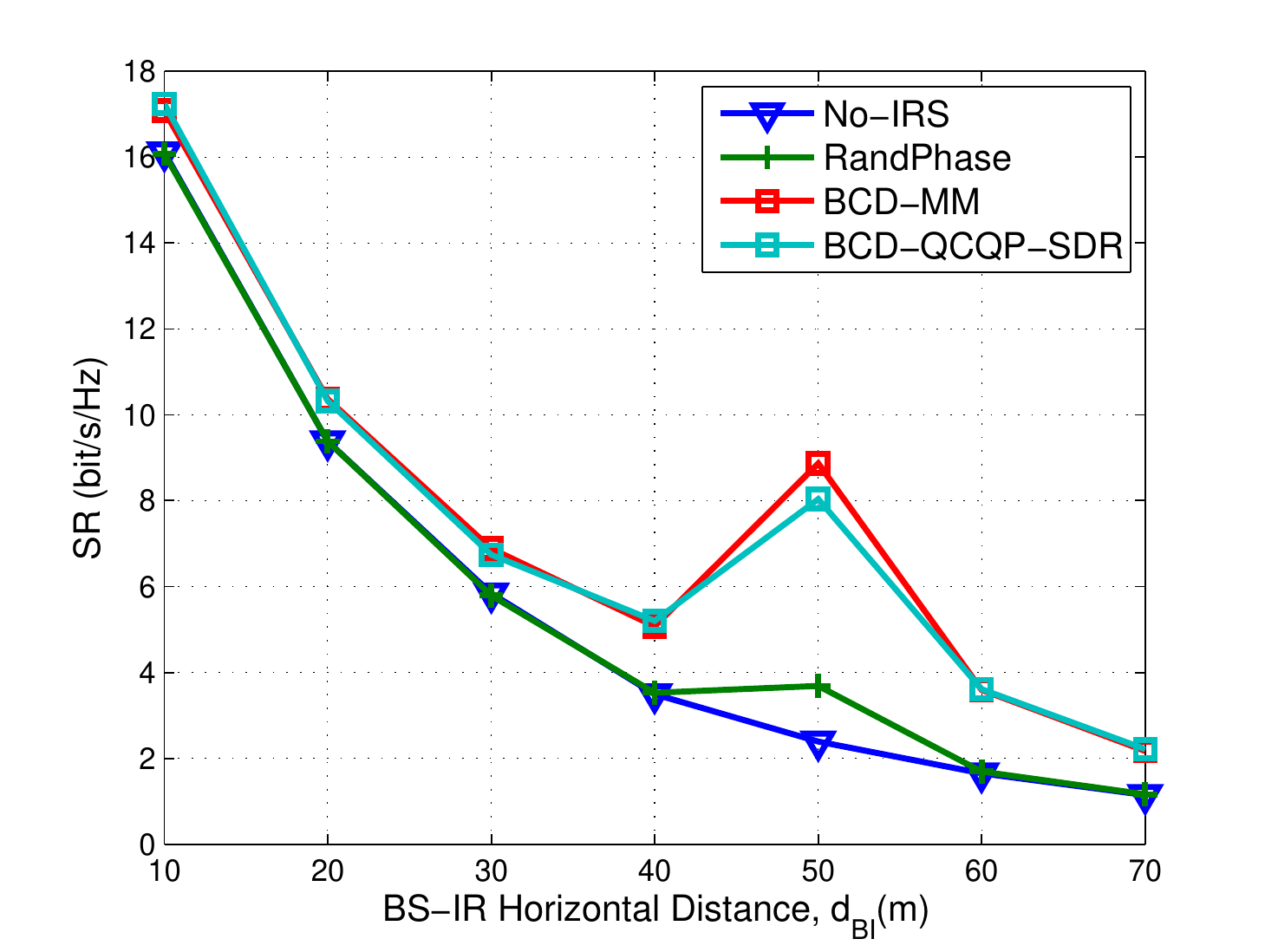}\vspace{-0.6cm}
\caption{Achievable SR versus the location of the IR $d_{\rm{BI}}$.}
\label{fig5simu}
\end{minipage}%
\hfill
\begin{minipage}[t]{0.495\linewidth}
\centering
\includegraphics[width=2.6in]{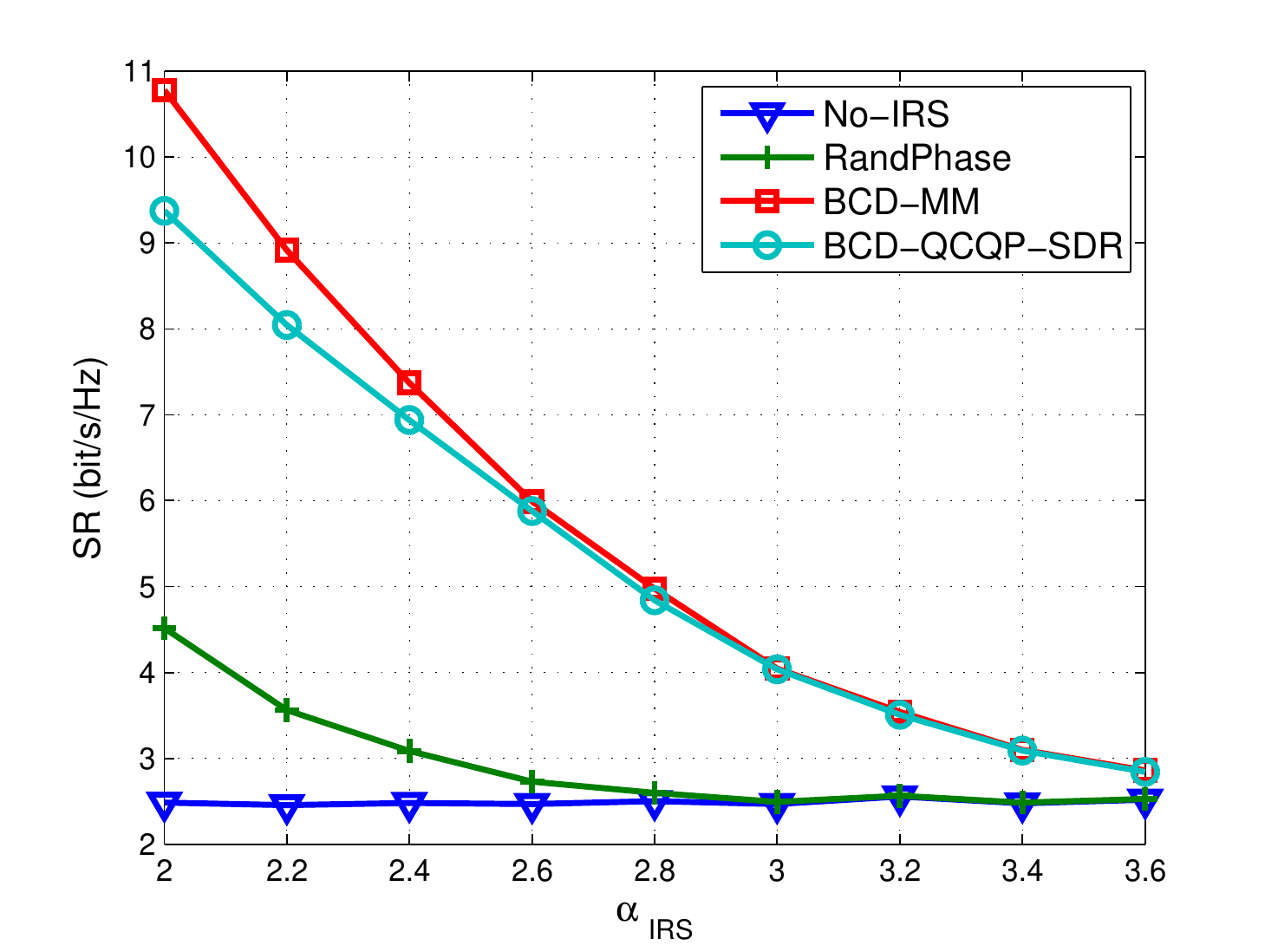}\vspace{-0.6cm}
\caption{Achievable SR versus the path loss exponent of IRS-related links.}
\label{fig6simu}
\end{minipage}\vspace{-0.7cm}
\end{figure}
Fig.~\ref{fig5simu} illustrates the achieved SRs for four schemes with various BS-IR horizontal distance $d_{BI}$, where the BS-Eve distance is fixed to be $d_{BE} = 44$ m. It is observed that the proposed BCD-MM algorithm is the best among the four schemes for obtaining the highest SR value. When the IR moves far away from the BS, the SRs decrease for the four schemes, however, the SRs achieved for the RandPhase, the proposed BCD-MM algorithm and the BCD-QCQP-SDR algorithm increase greatly when the IR approaches the IRS. The achieved SRs at different BS-IR distances of the RandPhase scheme and the no-IRS scheme are almost the same, except for $d_{BI} \in [40 \text{m}, 50 \text{m}]$, in which case the IRS brings a prominent
security enhancement when IR becomes close to it even with random IRS phase shifts. Similarly, the proposed BCD-MM algorithm and the BCD-QCQP-SDR algorithm can achieve almost the same SRs, except for $d_{BI} \in [40 \text{m}, 50 \text{m}]$, in which case the IR is close to the IRS, and the proposed BCD-MM algorithm is superior to the BCD-QCQP-SDR algorithm. For other BS-IR distances where the IR is far from the IRS, the SRs of RandPhase scheme are similar with those of the No-IRS scheme due to the not fully explored potential of IRS. By optimizing the phase shifts of IRS, the SRs are enhanced at different BS-IRS distances. And the SR gain of the proposed BCD-MM algorithm over the RandPhase scheme increases when the IR moves close to the IRS ($d_{BI} \in [40 \text{m}, 50 \text{m}]$). This signifies that as long as the IRS is deployed close to the IR, significant security enhancement can be achieved by the IRS in an AN-aided MIMO communication system. Moreover, it is highly recommended that the IRS phase shifts should be optimized to prevent the system security degrading into the level of No-IRS scheme.
\subsubsection{Impact of the Path Loss Exponent of IRS-related Links}
In the above simulations, the path loss exponents of the IRS-related links (including the BS-IRS link, IRS-IR link and IRS-Eve link) are set to be low by assuming that the IRS is properly located to obtain clean channels without heavy blockage. Practically, such kind of settings may not always be sensible due to the real-field environment. Thus, it is necessary to investigate the security gain brought by the IRS and our proposed algorithm with higher values of IRS-related path loss exponents. For the sake of analysis, we assume the path-loss exponents of the links from BS to IRS, from IRS to IR and from IRS to Eve are the same as ${\alpha _{{\rm{BR}}}}={\alpha _{{\rm{RI}}}}={\alpha _{{\rm{RE}}}} \buildrel \Delta \over = {\alpha _{{\rm{IRS}}}}$. Then, the achieved SRs versus the path-loss exponent ${\alpha _{{\rm{IRS}}}}$ of IRS-related links are shown in Fig.~\ref{fig6simu}, which demonstrates that the SR obtained by the BCD-MM algorithm decreases as ${\alpha _{{\rm{IRS}}}}$ increases, and finally drops to the same SR value which is achieved by the RandPhase, BCD-QCQP-SDR and No-IRS schemes. The reason is that larger ${\alpha _{{\rm{IRS}}}}$ means more severe signal attenuation in the IRS-related links, and more weakened signal received and reflected at the IRS. In comparison to the BCD-QCQP-SDR algorithm, the proposed BCD-MM algorithm can achieve the higher SR when the channel state of the IRS related channels is good, i.e., ${\alpha _{{\rm{IRS}}}}$ is low, and achieve almost the same SR when ${\alpha _{{\rm{IRS}}}}$ is large. Similarly, the performance gains brought by our proposed algorithm over the RandPhase and No-IRS schemes is significant with a small ${\alpha _{{\rm{IRS}}}}$. Specifically, for ${\alpha _{{\rm{IRS}}}}=2$ (almost ideal channels), the security gain is up to 9.6 bit/s/Hz over the No-IRS scheme, and 6.8 bit/s/Hz over the RandPhase scheme. Therefore, the security gain of IRS-assisted systems depends on the channel conditions of the IRS-related links. This suggests that it is much preferred to deploy the IRS with fewer obstacles, in which case, the performance gain brought by the IRS can be explored thoroughly. Fig.~\ref{fig6simu} also shows that when ${\alpha _{{\rm{IRS}}}}$ is small, the RandPhase scheme can obtain security gain over the No-IRS scheme, but this security gain decreases to zero when ${\alpha _{{\rm{IRS}}}}$ becomes large. However, the SR gain of the RandPhase scheme over the No-IRS scheme is almost negligible in comparison to the SR gain of the proposed scheme over the No-IRS scheme, which demonstrates that the necessity of jointly optimizing the TPC matrix, AN covariance matrix and the phase shifts at the IRS.

\subsubsection{Impact of the Number of Data Streams}
\begin{figure}
\begin{minipage}[t]{0.495\linewidth}
\centering
\includegraphics[width=2.6in]{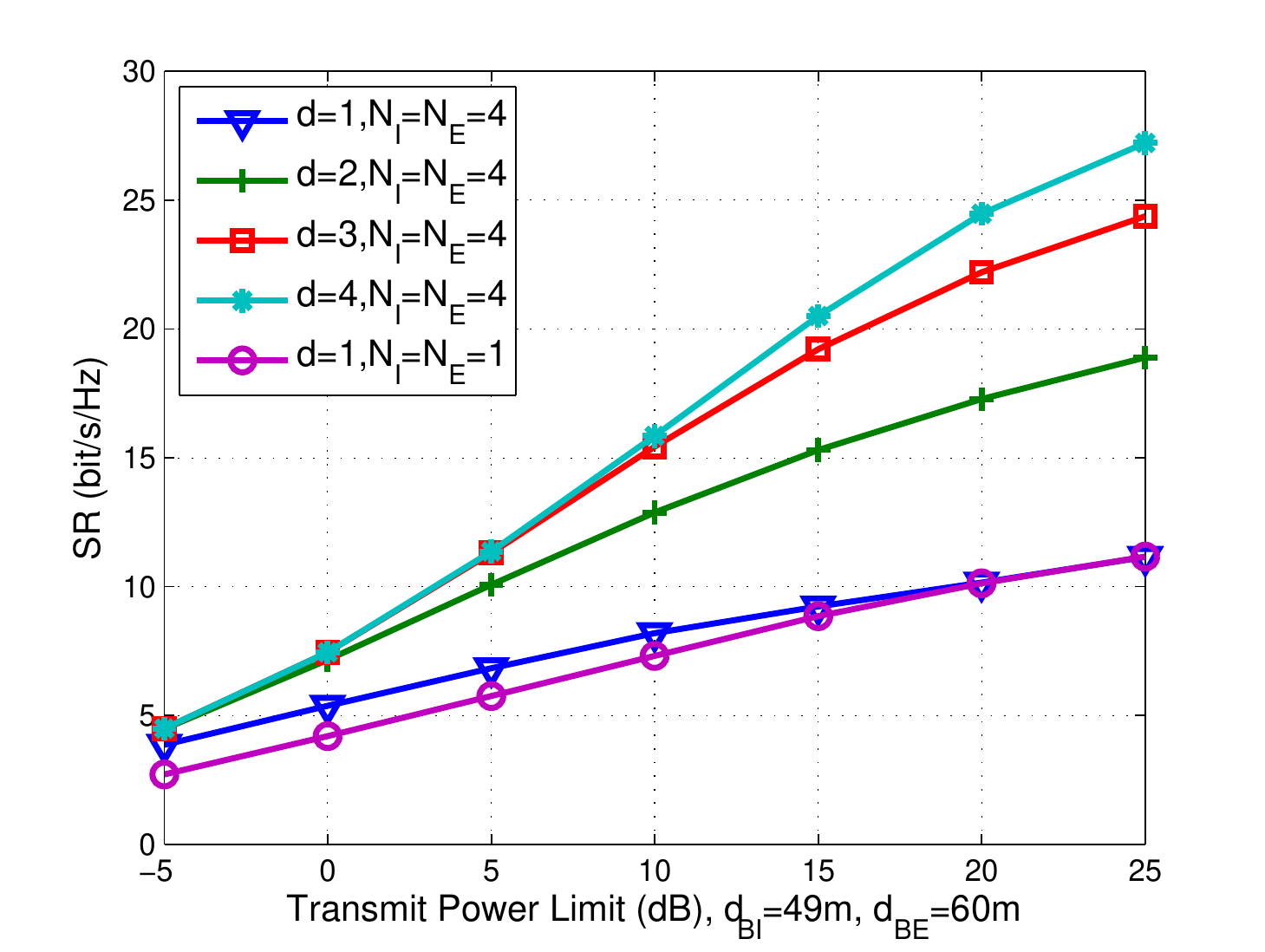}\vspace{-0.6cm}
\caption{Achievable SR versus the transmit power limit for various numbers of data streams.}
\label{figdatastreams}
\end{minipage}%
\hfill
\begin{minipage}[t]{0.495\linewidth}
\centering
\includegraphics[width=2.6in]{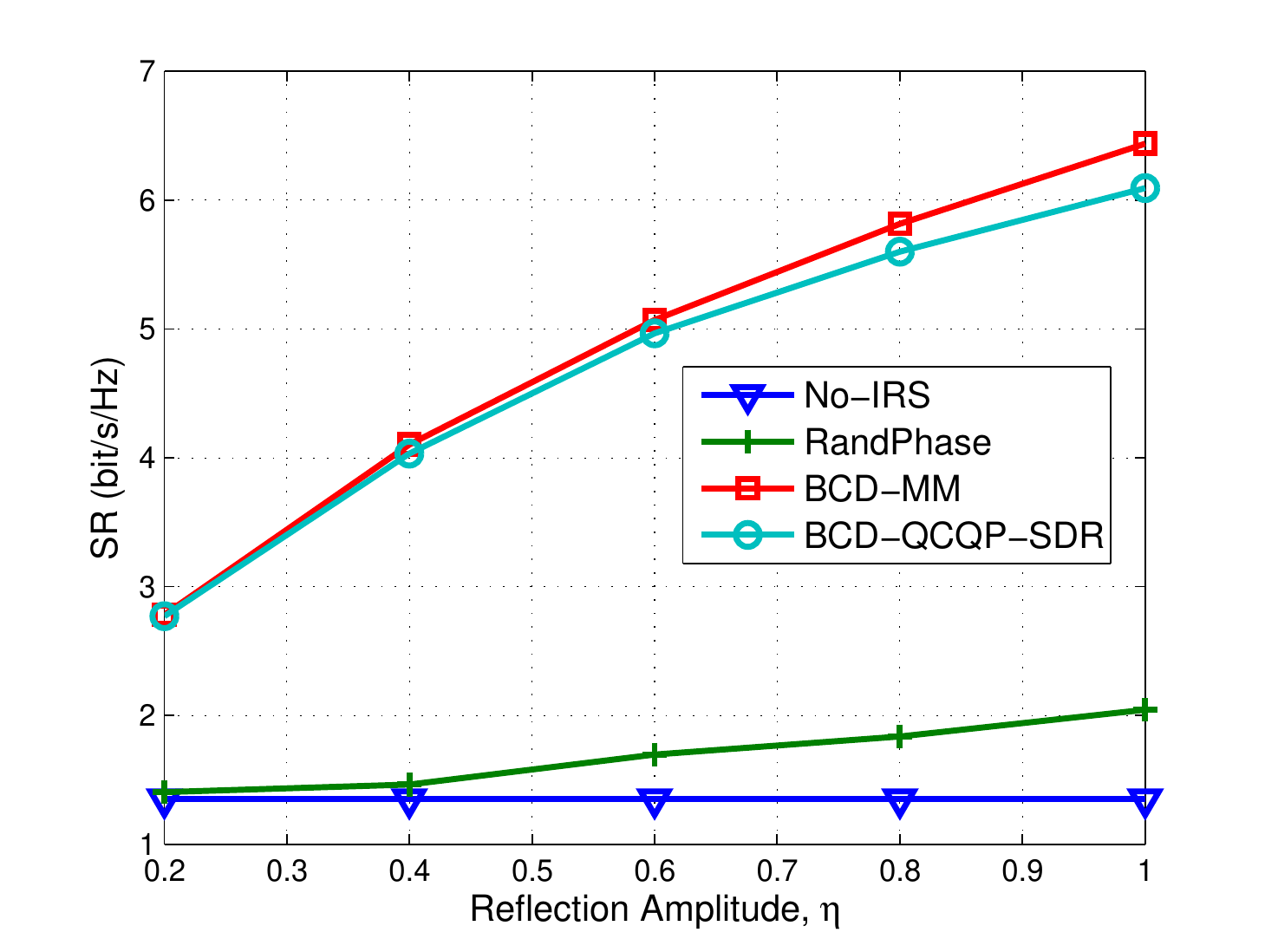}\vspace{-0.6cm}
\caption{Achievable SR versus the reflection amplitude $\eta$.}
\label{figSRvsAmplitude}
\end{minipage}\vspace{-0.7cm}
\end{figure}

Compared with the MISO scenario, a significant advantage of the MIMO scenario is that multiple data streams can be transmitted to the users. To evaluate the impact of the number of data streams on the SR, the average SRs versus the transmit power limit for various numbers of data streams are given in Fig.~\ref{figdatastreams}.

The number of transmit antennas is $N_T=4$. The Rician fading channels are utilized. The path loss exponents are ${\alpha _{{\rm{BR}}}}= 2.2$, ${\alpha _{{\rm{BE}}}}=3.5$, ${\alpha _{{\rm{BI}}}}=2.5$, ${\alpha _{{\rm{RE}}}}=3.5$ and ${\alpha _{{\rm{RI}}}}=2.5$ respectively. The Rician factor is $\beta=3$. The number of phase shifts is $M=50$. As shown in Fig.~\ref{figdatastreams}, the SR increases with the transmit power limit and larger number of data streams result the higher SR. When the transmit power limit is low, marginal performance gains
are achieved by increasing the number $d$ of data streams. When the transmit power limit is high, significant performance gains can be achieved by increasing the number $d$ of data streams. This means that a greater number of data streams ensure the higher SR, and the performance gain expands with the transmit power limit. For the case of $d=1$, the SR performance of $N_I=N_E=4$ and $N_I=N_E=1$ is further compared. It is revealed that the SR obtained by four receiving antennas is higher than the SR obtained by one single receiving antenna when the transmit power limit is relatively low. With the increase of transmit power limit, the SR performance gain brought by multiple receiving antennas decreases. When the transmit power limit is high enough, the SR performance is saturated, and the SR performance of the multiple receiving antennas and single receiving antenna becomes the same.
\subsubsection{Impact of the Reflection Amplitude}
Due to the manufactural and hardware reasons, the signals reflected by the IRS may be attenuated. Then, in Fig.~\ref{figSRvsAmplitude}, we study the impact of the reflection amplitude on the security performance. The transmit power limit is 10dBm. We assume that the reflection amplitudes of all the IRS elements are same as $\eta$, and that the phase shift matrix of the IRS is rewritten as ${\bf{\Phi}} =\eta\text{diag }\!\!\{\!\!\text{ }{{\phi }_{1}},\cdots ,{{\phi }_{m}},\cdots ,{{\phi }_{M}}\text{ }\!\!\}\!\!\text{ }$. As expected, the SR achieved by the IRS-aided scheme increases with $\eta$ due to less power loss. As $\eta$ increases, the superiority of the proposed BCD-MM algorithm over the other algorithms becomes more obvious. The reflection amplitude has a great impact on the security performance. Specifically, when $\eta$ increases from 0.2 to 1, the SR increases over 3.6 bit/s/Hz for the proposed BCD-MM algorithm.
\subsubsection{Impact of the Discrete Phase Shifts}
\begin{figure}
\begin{minipage}[t]{0.495\linewidth}
\centering
\includegraphics[width=2.6in]{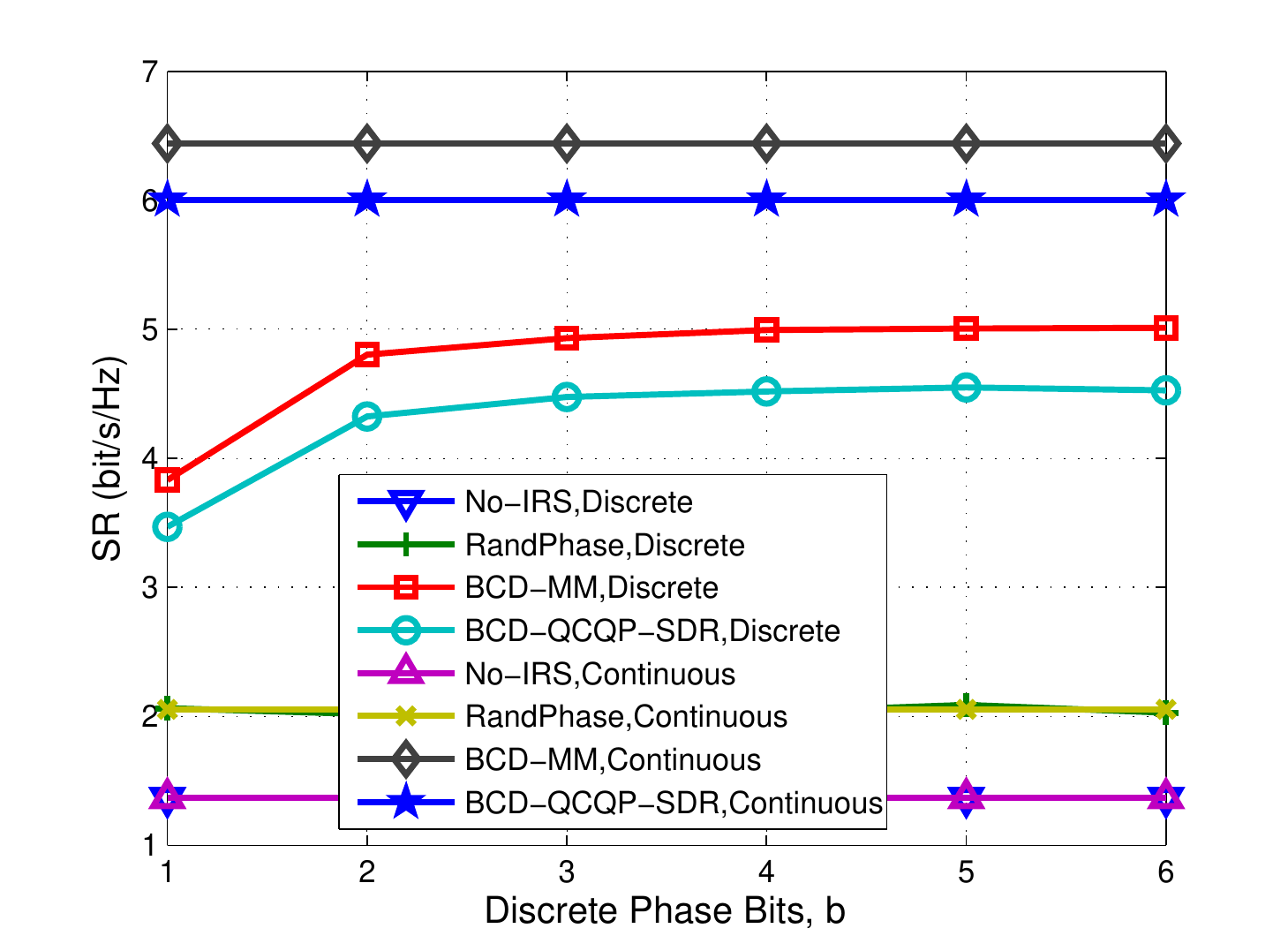}\vspace{-0.6cm}
\caption{Achievable SR versus the discrete phase bits $b$.}
\label{figSRvsDiscretePhasebits}
\end{minipage}%
\hfill
\begin{minipage}[t]{0.495\linewidth}
\centering
\includegraphics[width=2.6in]{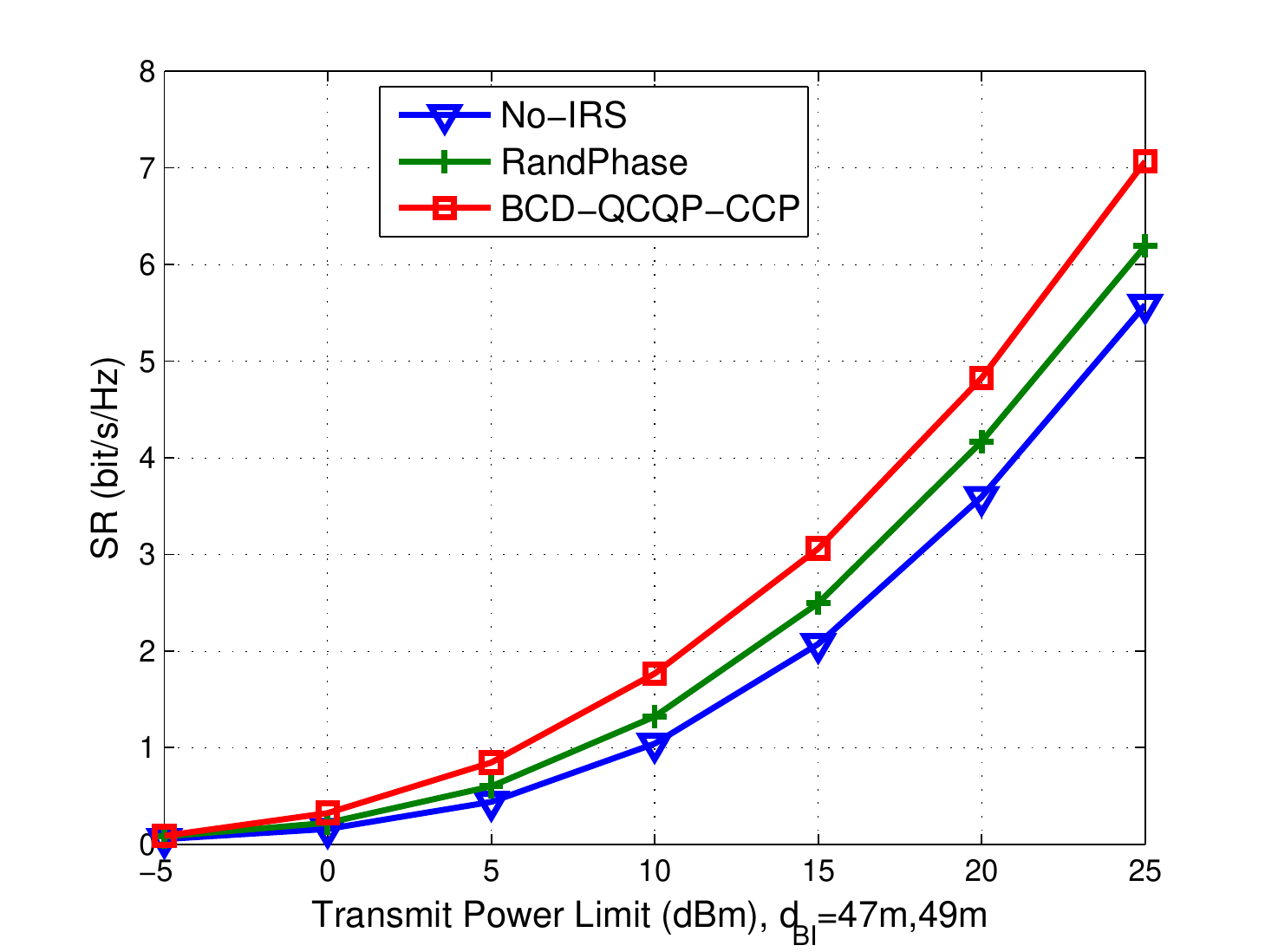}\vspace{-0.6cm}
\caption{Achievable SR versus the transmit power limit for multiple IRs.}
\label{figSRvspowMultIRs}
\end{minipage}\vspace{-0.7cm}
\end{figure}

In practice, it is difficult to realize continuous phase shifts at the reflecting elements of the IRS due to the high manufacturing cost. It is more cost-effective to implement only discrete phase shifts with a small number of control bits for each element, e.g., 1-bit for two-level (0 or $\pi$) phase shifts. Thus, the impact of the controlling bits $b$ of the discrete phase shifts on the security performance is investigated in Fig.~\ref{figSRvsDiscretePhasebits}. The transmit power limit is 10dBm. It is shown that the SR with continuous phase shifts of the IRS is higher than those with discrete phase shifts. The limited discrete phase shifts inevitably cause SR performance degradation. The SR of the IRS with discrete phase shifts increases with the number of controlling bits $b$, and becomes saturated when $b\ge4$, which means that the SR loss is inevitable even when the number of controlling bits $b$ is high. For the proposed BCD-MM algorithm, the maximum SR gap between the continuous phase shifts and the discrete phase shifts is 1.4 bit/s/Hz.

\subsubsection{Multiple IRs Scenario}
Finally, we consider the multiple IRs scenario to investigate the security enhancement brought by the IRS on the AN-aided MIMO communication systems. The horizontal distances between the BS and the two IRs are selected as $d_{BI,1}=47$m and $d_{BI,2}=49$m. Considering the heavy computational load, the element number of the IRS is assumed to be $M=20$. The proposed BCD-QCQP-CCP algorithm is utilized to perform the joint optimization of the TPC matrix, AN covariance matrix and the phase shifts of the IRS. The achieved SRs for the proposed algorithm, the random IRS scheme and the No-IRS scheme are shown in Fig.~\ref{figSRvspowMultIRs}. By comparing with the Random IRS scheme and the No-IRS scheme, the proposed BCD-QCQP-CCP algorithm can optimize the phase shifts of the IRS, thus achieve the higher SR. The SR gain increases with the power limit $P_T$. However, the performance gain is not as much as that in Fig.~\ref{fig3simu}. On the one hand, the element number of the IRS is set to be lower than that in Fig.~\ref{fig3simu} due to the heavy computational load. On the other hand, it is more difficult for the IRS to adjust the phase shifts to guarantee higher SRs for more legitimate IRs.
\section{Conclusions}\label{conclu}

In this paper, we propose to enhance the security of AN-aided MIMO secure communication systems by exploiting an IRS. To exploit the IRS efficiently, we formulate a SRM problem by jointly optimizing the TPC matrix at the BS, the covariance matrix of AN and phase shifts at the IRS with the constraints of transmit power limit and unit-modulus of phase shifts. To solve this non-convex problem, we propose to use the BCD algorithm to decouple the optimization variables, and optimize them iteratively. The optimal TPC matrix and AN covariance matrix were obtained in semi-closed form by the Lagrange multiplier method, and the phase shifts at the IRS were obtained in closed form by an efficient MM algorithm. Various simulations validated that significant security gains can be achieved by the proposed algorithm with IRS. Furthermore, useful suggestions for choosing and deploying the IRS are provided.

\begin{appendices}
\vspace{-0.5cm}\section{Derivation of the Problem \eqref{optorigVElowerbndSmp}}\label{firstconstrntDeriv}
By substituting $h_1({{\bf{U}}_I},{\bf{V}},{{\bf{V}}_E},{{\bf{W}}_I})$ of (\ref{lowboundh1f1}), $h_2({{\bf{U}}_E},{{\bf{V}}_E},{{\bf{W}}_E})$ of (\ref{lowboundh2f2}) and $h_3({{\bf{V}}},{{\bf{V}}_E},{{\bf{W}}_X})$ of (\ref{lowboundh3f3}) into (\ref{lowboundCANlow}), we have
\begin{align} \label{lowCAN}
&{{\rm{C}}^{l}_{AN}}({{\bf{U}}_I},{{\bf{W}}_I},{{\bf{U}}_E},{{\bf{W}}_E},{{\bf{W}}_X},{\bf{V}},{{\bf{V}}_E},{\bf{\Phi}})=\log \left| {{{\bf{W}}_I}} \right| - {\rm{Tr}}({{\bf{W}}_I}{{\bf{E}}_I}({{\bf{U}}_I},{\bf{V}},{{\bf{V}}_E})) + \log \left| {{{\bf{W}}_E}} \right| \nonumber \\
& \qquad  \qquad  \qquad   - {\rm{Tr}}({{\bf{W}}_E}{{\bf{E}}_{E}}({{\bf{U}}_E},{{\bf{V}}_E})) + \log \left| {{{\bf{W}}_X}} \right| - {\rm{Tr}}({{\bf{W}}_X}{{\bf{E}}_{X}}({{\bf{V}}},{{\bf{V}}_E}))+d+N_t+N_E \nonumber \\
&\qquad=C_{g_{0}}-{\underbrace{ {\rm{Tr}}({{\bf{W}}_I}{{\bf{E}}_I}({{\bf{U}}_I},{\bf{V}},{{\bf{V}}_E}))}_{g_{1}}}-{\underbrace{  {\rm{Tr}}({{\bf{W}}_E}{{\bf{E}}_{E}}({{\bf{U}}_E},{{\bf{V}}_E}))}_{g_{2}}}- {\underbrace{{\rm{Tr}}({{\bf{W}}_X}{{\bf{E}}_{X}}({{\bf{V}}},{{\bf{V}}_E}))}_{g_{3}}},
\end{align}
where $C_{g_{0}}\buildrel \Delta \over =\log \left| {{{\bf{W}}_I}} \right|+ \log \left| {{{\bf{W}}_E}} \right|+ \log \left| {{{\bf{W}}_X}} \right|+d+N_t+N_E$.

$C_{g_{0}}$ contains the constant terms irrelated with ${\bf{V}},{{\bf{V}}_E},{\bf{\Phi}}$. By putting matrix functions ${{\bf{E}}_I}$, ${{\bf{E}}_E}$ and ${{\bf{E}}_X}$ into \eqref{lowCAN}, we expand ${g_{1}}$, ${g_{2}}$, and ${g_{3}}$ respectively as follows.

(1) ${{g}_{1}}$ can be reformulated as
\begin{align}
  {{g}_{1}}&=\text{Tr}({{\bf{W}}_{I}}[({\bf{I}}-{{\bf{U}}_{I}}^{H}{{{\hat{\bf{H}}}}_{I}}{\bf{V}}){{({\bf{I}}-{{\bf{U}}_{I}}^{H}{{{\hat{\bf{H}}}}_{I}}{\bf{V}})}^{H}}+{{\bf{U}}_{I}}^{H}({{{\hat{\bf{H}}}}_{I}}{{\bf{V}}_{E}}{{\bf{V}}_{E}}^{H}{{{\hat{\bf{H}}}}_{I}}^{H}+\sigma _{I}^{2}{{\bf{I}}_{{{N}_{I}}}}){{\bf{U}}_{I}}])\nonumber \\
 &=\text{Tr}({{\bf{W}}_{I}}[({\bf{I}}-{{\bf{V}}^{H}}{{{\hat{\bf{H}}}}_{I}}^{H}{{\bf{U}}_{I}}-{{\bf{U}}_{I}}^{H}{{{\hat{\bf{H}}}}_{I}}{\bf{V}}+{{\bf{U}}_{I}}^{H}{{{\hat{\bf{H}}}}_{I}}{\bf{V}}{{\bf{V}}^{H}}{{{\hat{\bf{H}}}}_{I}}^{H}{{\bf{U}}_{I}}) \nonumber \\
 &\quad +({{\bf{U}}_{I}}^{H}{{{\hat{\bf{H}}}}_{I}}{{\bf{V}}_{E}}{{\bf{V}}_{E}}^{H}{{{\hat{\bf{H}}}}_{I}}^{H}{{\bf{U}}_{I}}\text{+}{{\bf{U}}_{I}}^{H}\sigma _{I}^{2}{{\bf{I}}_{{{N}_{I}}}}{{\bf{U}}_{I}})]). \label{eq29t}
\end{align}
By gathering the constant terms related with ${{\bf{W}}_{I}},{{\bf{U}}_{I}}$ in ${{C}}_{g_{1}}$, ${{g}_{1}}$ can be simplified as
\begin{align}
{{g}_{1}}&=-\text{Tr}({{\bf{W}}_{I}}{{\bf{V}}^{H}}{{{\hat{\bf{H}}}}_{I}}^{H}{{\bf{U}}_{I}})-\text{Tr}({{\bf{W}}_{I}}{{\bf{U}}_{I}}^{H}{{{\hat{\bf{H}}}}_{I}}{\bf{V}})+\text{Tr}({{\bf{V}}^{H}}{{{\hat{\bf{H}}}}_{I}}^{H}{{\bf{U}}_{I}}{{\bf{W}}_{I}}{{\bf{U}}_{I}}^{H}{{{\hat{\bf{H}}}}_{I}}{\bf{V}})\nonumber \\ &  \quad +\text{Tr}({{\bf{V}}_{E}}^{H}{{{\hat{\bf{H}}}}_{I}}^{H}{{\bf{U}}_{I}}{{\bf{W}}_{I}}{{\bf{U}}_{I}}^{H}{{{\hat{\bf{H}}}}_{I}}{{\bf{V}}_{E}})+{{C}}_{g_{1}}, \label{eq30t}
\end{align}
where ${{C}}_{g_{1}}\buildrel \Delta \over =\text{Tr}({{\bf{W}}_{I}}+\sigma _{I}^{2}{{\bf{W}}_{I}}{{\bf{U}}_{I}}^{H}{{\bf{U}}_{I}})$.

(2) ${{g}_{2}}$ can be reformulated as
\begin{align}
  {{g}_{2}}&=\text{Tr}({{\mathbf{W}}_{E}}[(\mathbf{I}-{{\mathbf{U}}_{E}}^{H}{{{\mathbf{\hat{H}}}}_{E}}{{\mathbf{V}}_{E}}){{(\mathbf{I}-{{\mathbf{U}}_{E}}^{H}{{{\mathbf{\hat{H}}}}_{E}}{{\mathbf{V}}_{E}})}^{H}}+\sigma _{E}^{2}{{\mathbf{U}}_{E}}^{H}{{\mathbf{U}}_{E}}]) \nonumber \\
 &=\text{Tr}({{\mathbf{W}}_{E}}[(\mathbf{I}-{{\mathbf{V}}_{E}}^{H}{{{\mathbf{\hat{H}}}}_{E}}^{H}{{\mathbf{U}}_{E}}-{{\mathbf{U}}_{E}}^{H}{{{\mathbf{\hat{H}}}}_{E}}{{\mathbf{V}}_{E}}+{{\mathbf{U}}_{E}}^{H}{{{\mathbf{\hat{H}}}}_{E}}{{\mathbf{V}}_{E}}{{\mathbf{V}}_{E}}^{H}{{{\mathbf{\hat{H}}}}_{E}}^{H}{{\mathbf{U}}_{E}}\nonumber \\
 &\quad +\sigma _{E}^{2}{{\mathbf{U}}_{E}}^{H}{{\mathbf{U}}_{E}}]). \label{eq31t}
\end{align}

By gathering the constant terms related with ${{\mathbf{W}}_{E}},{{\mathbf{U}}_{E}}$ in ${{C}}_{g_{2}}$, ${{g}_{2}}$ can be simplified as
\begin{align}
 {{g}_{2}}=-\text{Tr}({{\mathbf{W}}_{E}}{{\mathbf{V}}_{E}}^{H}{{{\mathbf{\hat{H}}}}_{E}}^{H}{{\mathbf{U}}_{E}})-\text{Tr}({{\mathbf{W}}_{E}}{{\mathbf{U}}_{E}}^{H}{{{\mathbf{\hat{H}}}}_{E}}{{\mathbf{V}}_{E}})+\text{Tr}({{\mathbf{V}}_{E}}^{H}{{{\mathbf{\hat{H}}}}_{E}}^{H}{{\mathbf{U}}_{E}}{{\mathbf{W}}_{E}}{{\mathbf{U}}_{E}}^{H}{{{\mathbf{\hat{H}}}}_{E}}{{\mathbf{V}}_{E}})+{{C}}_{g_{2}}, \label{eq32t}
\end{align}
where ${{C}}_{g_{2}}\buildrel \Delta \over =\text{Tr}({{\bf{W}}_{E}}+\sigma _{E}^{2}{{\bf{W}}_{E}}{{\bf{U}}_{E}}^{H}{{\bf{U}}_{E}})$.

(3) ${{g}_{3}}$ can be reformulated as
\begin{align}
{{g}_{3}}=\text{Tr}({{\mathbf{W}}_{X}}({{{\bf{I}}_{{N_E}}} + \sigma _E^{-2}{{\hat {\bf{H}}}_E}({\bf{V}}{{\bf{V}}^H}+{{\bf{V}}_E}{{\bf{V}}_E}^H)\hat {\bf{H}}_E^H})).
\label{eq33t}
\end{align}

By gathering the constant terms related with ${{\mathbf{W}}_{X}}$ in ${{C}}_{g_{3}}$, ${{g}_{3}}$ can be simplified as
\begin{align}
  {{g}_{3}}=\sigma _E^{-2}\text{Tr}({{\mathbf{V}}^{H}}\mathbf{\hat{H}}_{E}^{H}{{\mathbf{W}}_{X}}{{{\mathbf{\hat{H}}}}_{E}}\mathbf{V})+\sigma _E^{-2}\text{Tr}({{\mathbf{V}}_{E}}^{H}{{{\mathbf{\hat{H}}}}_{E}}^{H}{{\mathbf{W}}_{X}}{{{\mathbf{\hat{H}}}}_{E}}{{\mathbf{V}}_{E}})+{{C}}_{g_{3}},
\label{eq34t}
\end{align}
where ${{C}}_{g_{3}}\buildrel \Delta \over =\text{Tr}({{\bf{W}}_{X}})$.

By substituting \eqref{eq30t}, \eqref{eq32t} and \eqref{eq34t} into \eqref{lowCAN}, we have
\begin{align} \label{lowCANwithconstants}
&{{\rm{C}}^{l}_{AN}}({{\bf{U}}_I},{{\bf{W}}_I},{{\bf{U}}_E},{{\bf{W}}_E},{{\bf{W}}_X},{\bf{V}},{{\bf{V}}_E},{\bf{\Phi}})=\text{Tr}({{\mathbf{W}}_{I}}{{\mathbf{V}}^{H}}{{{\mathbf{\hat{H}}}}_{I}}^{H}{{\mathbf{U}}_{I}})
+\text{Tr}({{\mathbf{W}}_{I}}{{\mathbf{U}}_{I}}^{H}{{{\mathbf{\hat{H}}}}_{I}}\mathbf{V})\nonumber \\
 & -\text{Tr}({{\mathbf{V}}^{H}}{{{\mathbf{\hat{H}}}}_{I}}^{H}{{\mathbf{U}}_{I}}{{\mathbf{W}}_{I}}{{\mathbf{U}}_{I}}^{H}{{{\mathbf{\hat{H}}}}_{I}}\mathbf{V}) -\text{Tr}({{\mathbf{V}}_{E}}^{H}{{{\mathbf{\hat{H}}}}_{I}}^{H}{{\mathbf{U}}_{I}}{{\mathbf{W}}_{I}}{{\mathbf{U}}_{I}}^{H}{{{\mathbf{\hat{H}}}}_{I}}{{\mathbf{V}}_{E}}) +\text{Tr}({{\mathbf{W}}_{E}}{{\mathbf{V}}_{E}}^{H}{{{\mathbf{\hat{H}}}}_{E}}^{H}{{\mathbf{U}}_{E}})\nonumber \\
 & +\text{Tr}({{\mathbf{W}}_{E}}{{\mathbf{U}}_{E}}^{H}{{{\mathbf{\hat{H}}}}_{E}}{{\mathbf{V}}_{E}}) -\text{Tr}({{\mathbf{V}}_{E}}^{H}{{{\mathbf{\hat{H}}}}_{E}}^{H}{{\mathbf{U}}_{E}}{{\mathbf{W}}_{E}}{{\mathbf{U}}_{E}}^{H}{{{\mathbf{\hat{H}}}}_{E}}{{\mathbf{V}}_{E}})
 -\sigma _E^{-2}\text{Tr}({{\mathbf{V}}^{H}}{{{\mathbf{\hat{H}}}}}_{E}^{H}{{\mathbf{W}}_{X}}{{{\mathbf{\hat{H}}}}_{E}}\mathbf{V})\nonumber \\
 & -\sigma _E^{-2}\text{Tr}({{\mathbf{V}}_{E}}^{H}{{{\mathbf{\hat{H}}}}_{E}}^{H}{{\mathbf{W}}_{X}}{{{\mathbf{\hat{H}}}}_{E}}{{\mathbf{V}}_{E}})+C_{g},
\end{align}
where $C_{g}\buildrel \Delta \over =C_{g_{0}}-C_{g_{1}}-C_{g_{2}}-C_{g_{3}}$.

Equation \eqref{lowCANwithconstants} can be rewritten more compactly as
\begin{align}
  & {{\rm{C}}^{l}_{AN}}({{\bf{U}}_I},{{\bf{W}}_I},{{\bf{U}}_E},{{\bf{W}}_E},{{\bf{W}}_X},{\bf{V}},{{\bf{V}}_E},{\bf{\Phi}})=C_{g}+\text{Tr}({{\mathbf{W}}_{I}}{{\mathbf{V}}^{H}}{{{\mathbf{\hat{H}}}}_{I}}^{H}{{\mathbf{U}}_{I}})+\text{Tr}({{\mathbf{W}}_{I}}{{\mathbf{U}}_{I}}^{H}{{{\mathbf{\hat{H}}}}_{I}}\mathbf{V})\nonumber \\
 &\quad-\text{Tr}({{\mathbf{V}}^{H}}{{\mathbf{H}}_{V}}\mathbf{V}) +\text{Tr}({{\mathbf{W}}_{E}}{{\mathbf{V}}_{E}}^{H}{{{\mathbf{\hat{H}}}}_{E}}^{H}{{\mathbf{U}}_{E}})+\text{Tr}({{\mathbf{W}}_{E}}{{\mathbf{U}}_{E}}^{H}{{{\mathbf{\hat{H}}}}_{E}}{{\mathbf{V}}_{E}})-\text{Tr}({{\mathbf{V}}_{E}}^{H}{{\mathbf{H}}_{VE}}{{\mathbf{V}}_{E}}).
\label{eq37t}
\end{align}
where
\begin{align}
{{\mathbf{H}}_{V}}={{\mathbf{\hat{H}}}_{I}}^{H}{{\mathbf{U}}_{I}}{{\mathbf{W}}_{I}}{{\mathbf{U}}_{I}}^{H}{{\mathbf{\hat{H}}}_{I}}+\sigma _E^{-2}\mathbf{\hat{H}}_{E}^{H}{{\mathbf{W}}_{X}}{{\mathbf{\hat{H}}}_{E}}.
\label{eq38t}
\end{align}
\begin{align}
{{\mathbf{H}}_{VE}}={{\mathbf{\hat{H}}}_{I}}^{H}{{\mathbf{U}}_{I}}{{\mathbf{W}}_{I}}{{\mathbf{U}}_{I}}^{H}{{\mathbf{\hat{H}}}_{I}}+{{\mathbf{\hat{H}}}_{E}}^{H}{{\mathbf{U}}_{E}}{{\mathbf{W}}_{E}}{{\mathbf{U}}_{E}}^{H}{{\mathbf{\hat{H}}}_{E}}+\sigma _E^{-2}{{\mathbf{\hat{H}}}_{E}}^{H}{{\mathbf{W}}_{X}}{{\mathbf{\hat{H}}}_{E}}.
\label{eq39t}
\end{align}
By substituting \eqref{eq37t} into Problem \eqref{optorigVElowerbnd}, and removing the constant term $C_{g}$, we arrive at the Problem \eqref{optorigVElowerbndSmp}.

\vspace{-0.5cm}\section{Derivation of the new OF form in \eqref{eqg0forPhi}}\label{newOFDeriv}

The objective function of Problem \eqref{optproblemforfaimin} is
\begin{align} \label{firstconstraint}
{g_{0}}(\mathbf{V},{{\mathbf{V}}_{E}},\mathbf{\Phi })=&-\text{Tr}({{\mathbf{W}}_{I}}{{\mathbf{V}}^{H}}{{{\mathbf{\hat{H}}}}_{I}}^{H}{{\mathbf{U}}_{I}})-\text{Tr}({{\mathbf{W}}_{I}}{{\mathbf{U}}_{I}}^{H}{{{\mathbf{\hat{H}}}}_{I}}\mathbf{V})+\text{Tr}({{\mathbf{V}}^{H}}{{\mathbf{H}}_{V}}\mathbf{V}) \nonumber \\
 &-\text{Tr}({{\mathbf{W}}_{E}}{{\mathbf{V}}_{E}}^{H}{{{\mathbf{\hat{H}}}}_{E}}^{H}{{\mathbf{U}}_{E}})-\text{Tr}({{\mathbf{W}}_{E}}{{\mathbf{U}}_{E}}^{H}{{{\mathbf{\hat{H}}}}_{E}}{{\mathbf{V}}_{E}})+\text{Tr}({{\mathbf{V}}_{E}}^{H}{{\mathbf{H}}_{VE}}{{\mathbf{V}}_{E}}).
\end{align}

The third term of \eqref{firstconstraint} is
\begin{align} \label{thirdoffirstconst}
{\rm{Tr}}\left( {{{\bf{V}}^H}{{\bf{H}}_V}{\bf{V}}} \right)& = {\rm{Tr}}\left[ {{{\bf{V}}^H}\left( {{\bf{\hat H}}_I^H{{\bf{U}}_I}{{\bf{W}}_I}{\bf{U}}_I^H{{{\bf{\hat H}}}_I} + \sigma _E^{-2}{\bf{\hat H}}_E^H{{\bf{W}}_X}{{{\bf{\hat H}}}_E}} \right){\bf{V}}} \right] \nonumber \\
&= {\rm{Tr}}\left[ {{{{\bf{\hat H}}}_I}{\bf{V}}{{\bf{V}}^H}{\bf{\hat H}}_I^H{{\bf{U}}_I}{{\bf{W}}_I}{\bf{U}}_I^H} \right] +\sigma _E^{-2} {\rm{Tr}}\left[ {{{{\bf{\hat H}}}_E}{\bf{V}}{{\bf{V}}^H}{\bf{\hat H}}_E^H{{\bf{W}}_X}} \right].
\end{align}

The sixth term of \eqref{firstconstraint} is
\begin{align} \label{sixoffirstconst}
{\rm{Tr}}\left( {{\bf{V}}_E^H{{\bf{H}}_{VE}}{{\bf{V}}_E}} \right) =& {\rm{Tr}}\left[ {{\bf{V}}_E^H\left( {{\bf{\hat H}}_I^H{{\bf{U}}_I}{{\bf{W}}_I}{\bf{U}}_I^H{{{\bf{\hat H}}}_I} + {\bf{\hat H}}_E^H{{\bf{U}}_E}{{\bf{W}}_E}{\bf{U}}_E^H{{{\bf{\hat H}}}_E} +\sigma _E^{-2} {\bf{\hat H}}_E^H{{\bf{W}}_X}{{{\bf{\hat H}}}_E}} \right){{\bf{V}}_E}} \right] \nonumber \\
=& {\rm{Tr}}\left[ {{{{\bf{\hat H}}}_I}{{\bf{V}}_E}{\bf{V}}_E^H{\bf{\hat H}}_I^H{{\bf{U}}_I}{{\bf{W}}_I}{\bf{U}}_I^H} \right] + {\rm{Tr}}\left[ {{{{\bf{\hat H}}}_E}{{\bf{V}}_E}{\bf{V}}_E^H{\bf{\hat H}}_E^H{{\bf{U}}_E}{{\bf{W}}_E}{\bf{U}}_E^H} \right] \nonumber \\
&+ \sigma _E^{-2} {\rm{Tr}}\left[ {{{{\bf{\hat H}}}_E}{{\bf{V}}_E}{\bf{V}}_E^H{\bf{\hat H}}_E^H{{\bf{W}}_X}} \right].
\end{align}
The summation of Equation \eqref{thirdoffirstconst} and Equation \eqref{sixoffirstconst} is
\begin{align} \label{third_sixoffirstconst}
{\rm{Tr}}\left( {{{\bf{V}}^H}{{\bf{H}}_V}{\bf{V}}} \right) + {\rm{Tr}}\left( {{\bf{V}}_E^H{{\bf{H}}_{VE}}{{\bf{V}}_E}} \right) =& {\rm{Tr}}\left[ {{{{\bf{\hat H}}}_I}\left( {{\bf{V}}{{\bf{V}}^H} + {{\bf{V}}_E}{\bf{V}}_E^H} \right){\bf{\hat H}}_I^H{{\bf{U}}_I}{{\bf{W}}_I}{\bf{U}}_I^H} \right] \nonumber \\
& + \sigma _E^{-2}{\rm{Tr}}\left[ {{{{\bf{\hat H}}}_E}\left( {{\bf{V}}{{\bf{V}}^H} + {{\bf{V}}_E}{\bf{V}}_E^H} \right){\bf{\hat H}}_E^H{{\bf{W}}_X}} \right] \nonumber \\
&+ {\rm{Tr}}\left[ {{{{\bf{\hat H}}}_E}{{\bf{V}}_E}{\bf{V}}_E^H{\bf{\hat H}}_E^H{{\bf{U}}_E}{{\bf{W}}_E}{\bf{U}}_E^H} \right].
\end{align}
By defining ${{\bf{V}}_X}=\left( {{\bf{V}}{{\bf{V}}^H} + {{\bf{V}}_E}{\bf{V}}_E^H} \right)$ and ${{\bf{M}}_I}={{\bf{U}}_I}{{\bf{W}}_I}{\bf{U}}_I^H$, {the first part} of \eqref{third_sixoffirstconst} can be derived as
\begin{align} \label{Part1_third_sixoffirstconst}
&{\rm{Tr}}\left[ {{{{\bf{\hat H}}}_I}\left( {{\bf{V}}{{\bf{V}}^H} + {{\bf{V}}_E}{\bf{V}}_E^H} \right){\bf{\hat H}}_I^H{{\bf{U}}_I}{{\bf{W}}_I}{\bf{U}}_I^H} \right]\nonumber \\
&={\rm{Tr}}\left[ {{{{\bf{\hat H}}}_I}{{\bf{V}}_X}{\bf{\hat H}}_I^H{{\bf{M}}_I}} \right] \nonumber \\
&= {\rm{Tr}}\left[ {\left( {{{\bf{H}}_{b,I}} + {{\bf{H}}_{R,I}}{\bf{\Phi G}}} \right){{\bf{V}}_X}\left( {{\bf{H}}_{b,I}^H + {{\bf{G}}^H}{{\bf{\Phi }}^H}{\bf{H}}_{R,I}^H} \right){{\bf{M}}_I}} \right]\nonumber \\
&= {\rm{Tr}}\left[ {\left( {{{\bf{H}}_{b,I}}{{\bf{V}}_X}{\bf{H}}_{b,I}^H + {{\bf{H}}_{b,I}}{{\bf{V}}_X}{{\bf{G}}^H}{{\bf{\Phi }}^H}{\bf{H}}_{R,I}^H + {{\bf{H}}_{R,I}}{\bf{\Phi G}}{{\bf{V}}_X}{\bf{H}}_{b,I}^H + {{\bf{H}}_{R,I}}{\bf{\Phi G}}{{\bf{V}}_X}{{\bf{G}}^H}{{\bf{\Phi }}^H}{\bf{H}}_{R,I}^H} \right){{\bf{M}}_I}} \right]\nonumber \\
&= {\rm{Tr}}[ {{\bf{H}}_{b,I}}{{\bf{V}}_X}{\bf{H}}_{b,I}^H{{\bf{M}}_I} + {{\bf{H}}_{b,I}}{{\bf{V}}_X}{{\bf{G}}^H}{{\bf{\Phi }}^H}{\bf{H}}_{R,I}^H{{\bf{M}}_I} + {{\bf{H}}_{R,I}}{\bf{\Phi G}}{{\bf{V}}_X}{\bf{H}}_{b,I}^H{{\bf{M}}_I} \nonumber \\
&\quad+ {{\bf{H}}_{R,I}}{\bf{\Phi G}}{{\bf{V}}_X}{{\bf{G}}^H}{{\bf{\Phi }}^H}{\bf{H}}_{R,I}^H{{\bf{M}}_I} ].
\end{align}
The derivation in \eqref{Part1_third_sixoffirstconst} can be used for the second and third parts of \eqref{third_sixoffirstconst}.

Based on the derivation in \eqref{Part1_third_sixoffirstconst}, it is obvious that {the second part} of \eqref{third_sixoffirstconst} can be derived as
\begin{align} \label{Part2_third_sixoffirstconst}
&\sigma _E^{-2}{\rm{Tr}}\left[ {{{{\bf{\hat H}}}_E}\left( {{\bf{V}}{{\bf{V}}^H} + {{\bf{V}}_E}{\bf{V}}_E^H} \right){\bf{\hat H}}_E^H{{\bf{W}}_X}} \right] \nonumber \\
&=\sigma _E^{-2}{\rm{Tr}}[ {{\bf{H}}_{b,E}}{{\bf{V}}_X}{\bf{H}}_{b,E}^H{{\bf{W}}_X}+{{\bf{H}}_{b,E}}{{\bf{V}}_X}{{\bf{G}}^H}{{\bf{\Phi }}^H}{\bf{H}}_{R,E}^H{{\bf{W}}_X}+ {{\bf{H}}_{R,E}}{\bf{\Phi G}}{{\bf{V}}_X}{\bf{H}}_{b,E}^H{{\bf{W}}_X}\nonumber \\
&\quad+ {{\bf{H}}_{R,E}}{\bf{\Phi G}}{{\bf{V}}_X}{{\bf{G}}^H}{{\bf{\Phi }}^H}{\bf{H}}_{R,E}^H{{\bf{W}}_X}].
\end{align}

Based on the derivation in \eqref{Part1_third_sixoffirstconst} and by defining ${{\bf{M}}_E}={{\bf{U}}_E}{{\bf{W}}_E}{\bf{U}}_E^H$, it is obvious that {the third part} of \eqref{third_sixoffirstconst} can be derived as
\begin{align}\label{Part3_third_sixoffirstconst}
&{\rm{Tr}}\left[ {{{{\bf{\hat H}}}_E}{{\bf{V}}_E}{\bf{V}}_E^H{\bf{\hat H}}_E^H{{\bf{U}}_E}{{\bf{W}}_E}{\bf{U}}_E^H} \right]\nonumber \\
&={\rm{Tr}}\left[ {{{{\bf{\hat H}}}_E}\left( {{{\bf{V}}_E}{\bf{V}}_E^H} \right){\bf{\hat H}}_E^H{{\bf{M}}_E}} \right] \nonumber \\
&= {\rm{Tr}}[ {{\bf{H}}_{b,E}}{{\bf{V}}_E}{\bf{V}}_E^H{\bf{H}}_{b,E}^H{{\bf{M}}_E}+ {{\bf{H}}_{b,E}}{{\bf{V}}_E}{\bf{V}}_E^H{{\bf{G}}^H}{{\bf{\Phi }}^H}{\bf{H}}_{R,E}^H{{\bf{M}}_E}+ {{\bf{H}}_{R,E}}{\bf{\Phi G}}{{\bf{V}}_E}{\bf{V}}_E^H{\bf{H}}_{b,E}^H{{\bf{M}}_E} \nonumber \\
&\quad + {{\bf{H}}_{R,E}}{\bf{\Phi G}}{{\bf{V}}_E}{\bf{V}}_E^H{{\bf{G}}^H}{{\bf{\Phi }}^H}{\bf{H}}_{R,E}^H{{\bf{M}}_E}].
\end{align}

By adding \eqref{Part1_third_sixoffirstconst}, \eqref{Part2_third_sixoffirstconst} and \eqref{Part3_third_sixoffirstconst}, and gathering constant terms irreverent with ${\bf{\Phi}}$, Equation \eqref{third_sixoffirstconst} becomes
\begin{align}\label{third_sixoffirstconst_form2}
&{\rm{Tr}}\left( {{{\bf{V}}^H}{{\bf{H}}_V}{\bf{V}}} \right) + {\rm{Tr}}\left( {{\bf{V}}_E^H{{\bf{H}}_{VE}}{{\bf{V}}_E}} \right) \nonumber \\
&= {\rm{Tr}}\left[ {{{\bf{\Phi }}^H}\left( {{\bf{H}}_{R,I}^H{{\bf{M}}_I}{{\bf{H}}_{b,I}}{{\bf{V}}_X}{{\bf{G}}^H} + \sigma _E^{-2} {\bf{H}}_{R,E}^H{{\bf{W}}_X}{{\bf{H}}_{b,E}}{{\bf{V}}_X}{{\bf{G}}^H} + {\bf{H}}_{R,E}^H{{\bf{M}}_E}{{\bf{H}}_{b,E}}{{\bf{V}}_E}{\bf{V}}_E^H{{\bf{G}}^H}} \right)} \right]  \nonumber \\
&\quad+{\rm{Tr}}\left[ {{\bf{\Phi }}\left( {{\bf{G}}{{\bf{V}}_X}{\bf{H}}_{b,I}^H{{\bf{M}}_I}{{\bf{H}}_{R,I}} + \sigma _E^{-2} {\bf{G}}{{\bf{V}}_X}{\bf{H}}_{b,E}^H{{\bf{W}}_X}{{\bf{H}}_{R,E}} + {\bf{G}}{{\bf{V}}_E}{\bf{V}}_E^H{\bf{H}}_{b,E}^H{{\bf{M}}_E}{{\bf{H}}_{R,E}}} \right)} \right] \nonumber  \\
& \quad + {\rm{Tr}}\left[ {{\bf{\Phi G}}{{\bf{V}}_X}{{\bf{G}}^H}{{\bf{\Phi }}^H}\left( {{\bf{H}}_{R,I}^H{{\bf{M}}_I}{{\bf{H}}_{R,I}} + \sigma _E^{-2} {\bf{H}}_{R,E}^H{{\bf{W}}_X}{{\bf{H}}_{R,E}}} \right)} \right] \nonumber  \\
&\quad+{\rm{Tr}}\left[ {{\bf{\Phi G}}{{\bf{V}}_E}{\bf{V}}_E^H{{\bf{G}}^H}{{\bf{\Phi }}^H}{\bf{H}}_{R,E}^H{{\bf{M}}_E}{{\bf{H}}_{R,E}}} \right]+C_{{t}_1},
\end{align}
where
\begin{align} \label{Ct1}
C_{{t}_1}={\rm{Tr}}\left[ {{\bf{H}}_{b,I}}{{\bf{V}}_X}{\bf{H}}_{b,I}^H{{\bf{M}}_I}\right]+\sigma _E^{-2}{\rm{Tr}}\left[{{\bf{H}}_{b,E}}{{\bf{V}}_X}{\bf{H}}_{b,E}^H{{\bf{W}}_X}\right]+{\rm{Tr}}\left[ {{\bf{H}}_{b,E}}{{\bf{V}}_E}{\bf{V}}_E^H{\bf{H}}_{b,E}^H{{\bf{M}}_E}\right].
\end{align}
The first term of ${g_{0}}(\mathbf{V},{{\mathbf{V}}_{E}},\mathbf{\Phi })$ is derived as
\begin{align} \label{firstoffirstconst}
{\rm{Tr}}\left( {{{\bf{W}}_I}{{\bf{V}}^H}{\bf{\hat H}}_I^H{{\bf{U}}_I}} \right) \!=\! {\rm{Tr}}\left( {{{\bf{U}}_I}{\bf{W}}_I^H{{\bf{V}}^H}{\bf{\hat H}}_I^H} \right)\! = \!\underbrace { {\rm{Tr}}\left[ {{{\bf{U}}_I}{\bf{W}}_I^H{{\bf{V}}^H}{\bf{H}}_{b,I}^H} \right]}_{C_{{t}_2}(\text{constant for }\mathbf{\Phi})}\! +\! {\rm{Tr}}\left[ {{\bf{H}}_{R,I}^H{{\bf{U}}_I}{\bf{W}}_I^H{{\bf{V}}^H}{{\bf{G}}^H}{{\bf{\Phi }}^H}} \right].
\end{align}
The second term of ${g_{0}}(\mathbf{V},{{\mathbf{V}}_{E}},\mathbf{\Phi })$ is derived as
\begin{align} \label{secondoffirstconst}
&{\rm{Tr}}\left( {{{\bf{W}}_I}{\bf{U}}_I^H{{{\bf{\hat H}}}_I}{\bf{V}}} \right) = {\rm{Tr}}\left( {{{{\bf{\hat H}}}_I}{\bf{V}}{{\bf{W}}_I}{\bf{U}}_I^H} \right) = {\rm{Tr}}\left[ {\left( {{{\bf{H}}_{b,I}} + {{\bf{H}}_{R,I}}{\bf{\Phi G}}} \right){\bf{V}}{{\bf{W}}_I}{\bf{U}}_I^H} \right] \nonumber \\
&=\underbrace { {\rm{Tr}}\left[ {{{\bf{H}}_{b,I}}{\bf{V}}{{\bf{W}}_I}{\bf{U}}_I^H} \right]}_{C_{{t}_3}(\text{constant for }\mathbf{\Phi})} + {\rm{Tr}}\left[ {{\bf{\Phi GV}}{{\bf{W}}_I}{\bf{U}}_I^H{{\bf{H}}_{R,I}}} \right].
\end{align}
The fourth term of ${g_{0}}(\mathbf{V},{{\mathbf{V}}_{E}},\mathbf{\Phi })$ is derived as
\begin{align} \label{fourthoffirstconst}
&{\rm{Tr}}\left( {{{\bf{W}}_E}{\bf{V}}_E^H{\bf{\hat H}}_E^H{{\bf{U}}_E}} \right)={\rm{Tr}}\left( {{{\bf{U}}_E}{{\bf{W}}_E^H}{\bf{V}}_E^H{\bf{\hat H}}_E^H} \right)  \nonumber \\
& =\underbrace { {\rm{Tr}}\left[ {{{\bf{U}}_E}{\bf{W}}_E^H{\bf{V}}_E^H{\bf{H}}_{b,E}^H} \right]}_{C_{{t}_4}(\text{constant for }\mathbf{\Phi})} + {\rm{Tr}}\left[ {{\bf{H}}_{R,E}^H{{\bf{U}}_E}{\bf{W}}_E^H{\bf{V}}_E^H{{\bf{G}}^H}{{\bf{\Phi }}^H}} \right].
\end{align}
The fifth term of ${g_{0}}(\mathbf{V},{{\mathbf{V}}_{E}},\mathbf{\Phi })$ is derived as
\begin{align} \label{fifthoffirstconst}
&{\rm{Tr}}\left( {{{\bf{W}}_E}{\bf{U}}_E^H{{{\bf{\hat H}}}_E}{{\bf{V}}_E}} \right) = {\rm{Tr}}\left( {{{{\bf{\hat H}}}_E}{{\bf{V}}_E}{{\bf{W}}_E}{\bf{U}}_E^H} \right) \nonumber \\
&= \underbrace { {\rm{Tr}}\left[ {{{\bf{H}}_{b,E}}{{\bf{V}}_E}{{\bf{W}}_E}{\bf{U}}_E^H} \right]}_{C_{{t}_5}(\text{constant for }\mathbf{\Phi})} + {\rm{Tr}}\left[ {{\bf{\Phi G}}{{\bf{V}}_E}{{\bf{W}}_E}{\bf{U}}_E^H{{\bf{H}}_{R,E}}} \right].
\end{align}
By including the first term in \eqref{firstoffirstconst}, the second term in \eqref{secondoffirstconst}, the fourth term in \eqref{fourthoffirstconst}, the fifth term in \eqref{fifthoffirstconst}, and the sum of the third and six terms in \eqref{third_sixoffirstconst_form2} of ${g_{0}}(\mathbf{V},{{\mathbf{V}}_{E}},\mathbf{\Phi })$ and gathering constant terms irreverent with ${\bf{\Phi}}$, we have
\begin{align}
&{g_{0}}(\mathbf{\Phi }) = -\rm{Equation \ }\eqref{firstoffirstconst}-\rm{Equation \ }\eqref{secondoffirstconst}-\rm{Equation \ }\eqref{fourthoffirstconst}-\rm{Equation \ }\eqref{fifthoffirstconst}+\rm{Equation \ }\eqref{third_sixoffirstconst_form2} \nonumber \\
&={\rm{Tr}}\left[ {{{\bf{\Phi }}^H}\left( \begin{array}{l}
{\bf{H}}_{R,I}^H{{\bf{M}}_I}{{\bf{H}}_{b,I}}{{\bf{V}}_X}{{\bf{G}}^H} +\sigma _E^{-2} {\bf{H}}_{R,E}^H{{\bf{W}}_X}{{\bf{H}}_{b,E}}{{\bf{V}}_X}{{\bf{G}}^H} + {\bf{H}}_{R,E}^H{{\bf{M}}_E}{{\bf{H}}_{b,E}}{{\bf{V}}_E}{\bf{V}}_E^H{{\bf{G}}^H} \nonumber \\
 -{\bf{H}}_{R,I}^H{{\bf{U}}_I}{\bf{W}}_I^H{{\bf{V}}^H}{{\bf{G}}^H} - {\bf{H}}_{R,E}^H{{\bf{U}}_E}{\bf{W}}_E^H{\bf{V}}_E^H{{\bf{G}}^H}
\end{array} \right)} \right]\nonumber \\
 &\quad + {\rm{Tr}}\left[ {{\bf{\Phi }}\left( \begin{array}{l}
{\bf{G}}{{\bf{V}}_X}{\bf{H}}_{b,I}^H{{\bf{M}}_I}{{\bf{H}}_{R,I}} +\sigma _E^{-2} {\bf{G}}{{\bf{V}}_X}{\bf{H}}_{b,E}^H{{\bf{W}}_X}{{\bf{H}}_{R,E}} + {\bf{G}}{{\bf{V}}_E}{\bf{V}}_E^H{\bf{H}}_{b,E}^H{{\bf{M}}_E}{{\bf{H}}_{R,E}}\nonumber \\
 - {\bf{GV}}{{\bf{W}}_I}{\bf{U}}_I^H{{\bf{H}}_{R,I}} - {\bf{G}}{{\bf{V}}_E}{{\bf{W}}_E}{\bf{U}}_E^H{{\bf{H}}_{R,E}}
\end{array} \right)} \right]\nonumber \\
&\quad + {\rm{Tr}}\left[ {{\bf{\Phi G}}{{\bf{V}}_E}{\bf{V}}_E^H{{\bf{G}}^H}{{\bf{\Phi }}^H}\left( {{\bf{H}}_{R,I}^H{{\bf{M}}_I}{{\bf{H}}_{R,I}} + \sigma _E^{-2} {\bf{H}}_{R,E}^H{{\bf{W}}_X}{{\bf{H}}_{R,E}} + {\bf{H}}_{R,E}^H{{\bf{M}}_E}{{\bf{H}}_{R,E}}} \right)} \right] \nonumber \\
&\quad + {\rm{Tr}}\left[ {{\bf{\Phi GV}}{{\bf{V}}^H}{{\bf{G}}^H}{{\bf{\Phi }}^H}\left( {{\bf{H}}_{R,I}^H{{\bf{M}}_I}{{\bf{H}}_{R,I}} + \sigma _E^{-2} {\bf{H}}_{R,E}^H{{\bf{W}}_X}{{\bf{H}}_{R,E}}} \right)} \right]+C_t\nonumber \\
& ={\rm{Tr}}\left[ {{{\bf{\Phi }}^H}\left( \begin{array}{l}
{\bf{H}}_{R,I}^H{{\bf{M}}_I}{{\bf{H}}_{b,I}}{{\bf{V}}_X}{{\bf{G}}^H} +\sigma _E^{-2} {\bf{H}}_{R,E}^H{{\bf{W}}_X}{{\bf{H}}_{b,E}}{{\bf{V}}_X}{{\bf{G}}^H} + {\bf{H}}_{R,E}^H{{\bf{M}}_E}{{\bf{H}}_{b,E}}{{\bf{V}}_E}{\bf{V}}_E^H{{\bf{G}}^H} \nonumber \\
 -{\bf{H}}_{R,I}^H{{\bf{U}}_I}{\bf{W}}_I^H{{\bf{V}}^H}{{\bf{G}}^H} - {\bf{H}}_{R,E}^H{{\bf{U}}_E}{\bf{W}}_E^H{\bf{V}}_E^H{{\bf{G}}^H}
\end{array} \right)} \right]\nonumber \\
& \quad+ {\rm{Tr}}\left[ {{\bf{\Phi }}\left( \begin{array}{l}
{\bf{G}}{{\bf{V}}_X}{\bf{H}}_{b,I}^H{{\bf{M}}_I}{{\bf{H}}_{R,I}} +\sigma _E^{-2} {\bf{G}}{{\bf{V}}_X}{\bf{H}}_{b,E}^H{{\bf{W}}_X}{{\bf{H}}_{R,E}} + {\bf{G}}{{\bf{V}}_E}{\bf{V}}_E^H{\bf{H}}_{b,E}^H{{\bf{M}}_E}{{\bf{H}}_{R,E}}\nonumber \\
 - {\bf{GV}}{{\bf{W}}_I}{\bf{U}}_I^H{{\bf{H}}_{R,I}} - {\bf{G}}{{\bf{V}}_E}{{\bf{W}}_E}{\bf{U}}_E^H{{\bf{H}}_{R,E}}
\end{array} \right)} \right]\nonumber \\
& \quad+ {\rm{Tr}}\left[ {{\bf{\Phi G}}{{\bf{V}}_E}{\bf{V}}_E^H{{\bf{G}}^H}{{\bf{\Phi }}^H}\left( {{\bf{H}}_{R,I}^H{{\bf{U}}_I}{{\bf{W}}_I}{\bf{U}}_I^H{{\bf{H}}_{R,I}} +\sigma _E^{-2} {\bf{H}}_{R,E}^H{{\bf{W}}_X}{{\bf{H}}_{R,E}} + {\bf{H}}_{R,E}^H{{\bf{U}}_E}{{\bf{W}}_E}{\bf{U}}_E^H{{\bf{H}}_{R,E}}} \right)} \right]\nonumber \\
&\quad+ {\rm{Tr}}\left[ {{\bf{\Phi GV}}{{\bf{V}}^H}{{\bf{G}}^H}{{\bf{\Phi }}^H}\left( {{\bf{H}}_{R,I}^H{{\bf{U}}_I}{{\bf{W}}_I}{\bf{U}}_I^H{{\bf{H}}_{R,I}} +\sigma _E^{-2} {\bf{H}}_{R,E}^H{{\bf{W}}_X}{{\bf{H}}_{R,E}}} \right)} \right]+C_t,
\end{align}
where
\begin{align}
C_t=C_{{t}_1}+C_{{t}_2}+C_{{t}_3}+C_{{t}_4}+C_{{t}_5}.
\end{align}
Then ${g_{0}}(\mathbf{\Phi })$ becomes
\begin{align}
{g_{0}}(\mathbf{\Phi })={\rm{Tr}}\left( {{{\bf{\Phi}}^H{\bf{D}}^H}} \right) + {\rm{Tr}}\left( {{\bf{\Phi D}}} \right) + {\rm{Tr}}\left[ {{\bf{\Phi }}{{\bf{C}}_{VE}}{{\bf{\Phi }}^H}{{\bf{B}}_{VE}}} \right] + {\rm{Tr}}\left( {{\bf{\Phi }}{{\bf{C}}_V}{{\bf{\Phi }}^H}{{\bf{B}}_V}} \right)+C_t \nonumber \\
={\rm{Tr}}\left( {{{\bf{\Phi}}^H{\bf{D}}^H}} \right) + {\rm{Tr}}\left( {{\bf{\Phi D}}} \right) + {\rm{Tr}}\left[ {{{\bf{\Phi }}^H}{{\bf{B}}_{VE}}{\bf{\Phi }}{{\bf{C}}_{VE}}} \right] + {\rm{Tr}}\left( {{{\bf{\Phi }}^H}{{\bf{B}}_V}{\bf{\Phi }}{{\bf{C}}_V}} \right)+C_t,\label{eq82t}
\end{align}
where
\begin{subequations}
\begin{align}
{\bf{D}}&={\bf{G}}{{\bf{V}}_X}{\bf{H}}_{b,I}^H{{\bf{M}}_I}{{\bf{H}}_{R,I}} + \sigma _E^{-2} {\bf{G}}{{\bf{V}}_X}{\bf{H}}_{b,E}^H{{\bf{W}}_X}{{\bf{H}}_{R,E}} + {\bf{G}}{{\bf{V}}_E}{\bf{V}}_E^H{\bf{H}}_{b,E}^H{{\bf{M}}_E}{{\bf{H}}_{R,E}}\nonumber \\
&\quad  - {\bf{GV}}{{\bf{W}}_I}{\bf{U}}_I^H{{\bf{H}}_{R,I}} - {\bf{G}}{{\bf{V}}_E}{{\bf{W}}_E}{\bf{U}}_E^H{{\bf{H}}_{R,E}}, \label{eq83ta} \\
{{\bf{C}}_{VE}}&={\bf{G}}{{\bf{V}}_E}{\bf{V}}_E^H{{\bf{G}}^H}, \label{eq83tb} \\
{{\bf{C}}_{V}}&={\bf{G}}{{\bf{V}}}{\bf{V}}^H{{\bf{G}}^H}, \label{eq83tc} \\
{{\bf{B}}_{VE}}&=\left( {{\bf{H}}_{R,I}^H{{\bf{U}}_I}{{\bf{W}}_I}{\bf{U}}_I^H{{\bf{H}}_{R,I}} + \sigma _E^{-2} {\bf{H}}_{R,E}^H{{\bf{W}}_X}{{\bf{H}}_{R,E}} + {\bf{H}}_{R,E}^H{{\bf{U}}_E}{{\bf{W}}_E}{\bf{U}}_E^H{{\bf{H}}_{R,E}}} \right), \label{eq83td} \\
{{\bf{B}}_{V}}&=\left( {{\bf{H}}_{R,I}^H{{\bf{U}}_I}{{\bf{W}}_I}{\bf{U}}_I^H{{\bf{H}}_{R,I}} + \sigma _E^{-2} {\bf{H}}_{R,E}^H{{\bf{W}}_X}{{\bf{H}}_{R,E}}} \right). \label{eq83te}
\end{align} \label{eq83t}
\end{subequations}
\end{appendices}

\
\





\vspace{-0.5cm}
\bibliographystyle{IEEEtran}
\bibliography{SRre}

\begin{thebibliography}{10}
\providecommand{\url}[1]{#1}
\csname url@samestyle\endcsname
\providecommand{\newblock}{\relax}
\providecommand{\bibinfo}[2]{#2}
\providecommand{\BIBentrySTDinterwordspacing}{\spaceskip=0pt\relax}
\providecommand{\BIBentryALTinterwordstretchfactor}{4}
\providecommand{\BIBentryALTinterwordspacing}{\spaceskip=\fontdimen2\font plus
\BIBentryALTinterwordstretchfactor\fontdimen3\font minus
  \fontdimen4\font\relax}
\providecommand{\BIBforeignlanguage}[2]{{%
\expandafter\ifx\csname l@#1\endcsname\relax
\typeout{** WARNING: IEEEtran.bst: No hyphenation pattern has been}%
\typeout{** loaded for the language `#1'. Using the pattern for}%
\typeout{** the default language instead.}%
\else
\language=\csname l@#1\endcsname
\fi
#2}}
\providecommand{\BIBdecl}{\relax}
\BIBdecl

\bibitem{saad2019vision}
\BIBentryALTinterwordspacing
W.~Saad, M.~Bennis, and M.~Chen, ``A vision of 6{G} wireless systems:
  Applications, trends, technologies, and open research problems.'' [Online].
  Available: \url{https://arxiv.org/abs/1902.10265}
\BIBentrySTDinterwordspacing

\bibitem{wang2019energy}
\BIBentryALTinterwordspacing
Q.~Wang, F.~Zhou, R.~Q. Hu, and Y.~Qian, ``Energy-efficient beamforming and
  cooperative jamming in {IRS}-assisted {MISO} networks.'' [Online]. Available:
  \url{https://arxiv.org/abs/1911.05133}
\BIBentrySTDinterwordspacing

\bibitem{liao2010qos}
W.~C. Liao, T.~H. Chang, W.~K. Ma, and C.~Y. Chi, ``Qo{S}-based transmit
  beamforming in the presence of eavesdroppers: An optimized
  artificial-noise-aided approach,'' \emph{IEEE Trans. Signal Process.},
  vol.~59, no.~3, pp. 1202--1216, 2010.

\bibitem{wu2018survey}
Y.~Wu, A.~Khisti, C.~Xiao, G.~Caire, K.~K. Wong, and X.~Gao, ``A survey of
  physical layer security techniques for 5{G} wireless networks and challenges
  ahead,'' \emph{IEEE J. Sel. Areas Commun.}, vol.~36, no.~4, pp. 679--695,
  2018.

\bibitem{wyner1975wire}
A.~D. Wyner, ``The wire-tap channel,'' \emph{Bell Syst. Tech. J.}, vol.~54,
  no.~8, pp. 1355--1387, 1975.

\bibitem{csiszar1978broadcast}
I.~Csisz{\'a}r and J.~Korner, ``Broadcast channels with confidential
  messages,'' \emph{IEEE Trans. Inf. Theory}, vol.~24, no.~3, pp. 339--348,
  1978.

\bibitem{khisti2010secure}
\BIBentryALTinterwordspacing
A.~Khisti and G.~Wornell, ``Secure transmission with multiple antennas {II}:
  The {MIMOME} wiretap channel.'' [Online]. Available:
  \url{https://arxiv.org/abs/1006.5879}
\BIBentrySTDinterwordspacing

\bibitem{oggier2011secrecy}
F.~Oggier and B.~Hassibi, ``The secrecy capacity of the {MIMO} wiretap
  channel,'' \emph{IEEE Trans. Inf. Theory}, vol.~57, no.~8, pp. 4961--4972,
  2011.

\bibitem{mukherjee2009fixed}
A.~Mukherjee and A.~L. Swindlehurst, ``Fixed-rate power allocation strategies
  for enhanced secrecy in {MIMO} wiretap channels,'' in \emph{10th Workshop on
  Signal Processing Advances in Wireless Communications (SPAWC)}.\hskip 1em
  plus 0.5em minus 0.4em\relax IEEE, 2009, pp. 344--348.

\bibitem{swindlehurst2009fixed}
A.~L. Swindlehurst, ``Fixed {SINR} solutions for the {MIMO} wiretap channel,''
  in \emph{International Conference on Acoustics, Speech and Signal Processing
  (ICASSP)}.\hskip 1em plus 0.5em minus 0.4em\relax IEEE, 2009, pp. 2437--2440.

\bibitem{goel2008guaranteeing}
S.~Goel and R.~Negi, ``Guaranteeing secrecy using artificial noise,''
  \emph{IEEE Trans. Wireless Commun.}, vol.~7, no.~6, pp. 2180--2189, 2008.

\bibitem{zhou2010secure}
X.~Zhou and M.~R. McKay, ``Secure transmission with artificial noise over
  fading channels: achievable rate and optimal power allocation,'' \emph{IEEE
  Trans. Veh. Technol.}, vol.~59, no.~8, pp. 3831--3842, 2010.

\bibitem{li2013transmit}
Q.~Li, M.~Hong, H.~T. Wai, Y.~F. Liu, W.~K. Ma, and Z.~Q. Luo, ``Transmit
  solutions for {MIMO} wiretap channels using alternating optimization,''
  \emph{IEEE J. Sel. Areas Commun.}, vol.~31, no.~9, pp. 1714--1727, 2013.

\bibitem{di2019smart}
M.~Di~Renzo, M.~Debbah, D.~T. Phan~Huy, A.~Zappone, M.~S. Alouini, C.~Yuen,
  V.~Sciancalepore, G.~C. Alexandropoulos, J.~Hoydis, H.~Gacanin \emph{et~al.},
  ``Smart radio environments empowered by reconfigurable {AI} meta-surfaces: an
  idea whose time has come,'' \emph{EURASIP J Wirel. Comm.}, vol. 2019, no.
  129, pp. 1--20, 2019.

\bibitem{qingqing2019towards}
\BIBentryALTinterwordspacing
Q.~Wu and R.~Zhang, ``Towards smart and reconfigurable environment:
  {I}ntelligent reflecting surface aided wireless network.'' [Online].
  Available: \url{https://arxiv.org/abs/1905.00152}
\BIBentrySTDinterwordspacing

\bibitem{huang2019holographic}
\BIBentryALTinterwordspacing
C.~Huang, S.~Hu, G.~C. Alexandropoulos, A.~Zappone, C.~Yuen, R.~Zhang,
  M.~Di~Renzo, and M.~Debbah, ``Holographic {MIMO} surfaces for 6{G} wireless
  networks: Opportunities, challenges, and trends.'' [Online]. Available:
  \url{https://arxiv.org/abs/1911.12296}
\BIBentrySTDinterwordspacing

\bibitem{wu2019beamforming}
Q.~Wu and R.~Zhang, ``Beamforming optimization for wireless network aided by
  intelligent reflecting surface with discrete phase shifts,'' \emph{IEEE
  Transactions on Communications}, vol.~68, no.~3, pp. 1838--1851, 2020.

\bibitem{huang2019reconfigurable}
C.~Huang, A.~Zappone, G.~C. Alexandropoulos, M.~Debbah, and C.~Yuen,
  ``Reconfigurable intelligent surfaces for energy efficiency in wireless
  communication,'' \emph{IEEE Trans. Wireless Commun.}, vol.~18, no.~8, pp.
  4157--4170, 2019.

\bibitem{huang2020reconfigurable}
\BIBentryALTinterwordspacing
C.~Huang, R.~Mo, C.~Yuen \emph{et~al.}, ``Reconfigurable intelligent surface
  assisted multiuser {MISO} systems exploiting deep reinforcement learning.''
  [Online]. Available: \url{https://arxiv.org/abs/2002.10072}
\BIBentrySTDinterwordspacing

\bibitem{wu2019towards}
Q.~Wu and R.~Zhang, ``Towards smart and reconfigurable environment: Intelligent
  reflecting surface aided wireless network,'' \emph{IEEE Communications
  Magazine}, vol.~58, no.~1, pp. 106--112, 2020.

\bibitem{hu2018beyond}
S.~Hu, F.~Rusek, and O.~Edfors, ``Beyond massive {MIMO}: The potential of data
  transmission with large intelligent surfaces,'' \emph{IEEE Trans. Signal
  Process.}, vol.~66, no.~10, pp. 2746--2758, 2018.

\bibitem{ford1984electromagnetic}
G.~W. Ford and W.~H. Weber, ``Electromagnetic interactions of molecules with
  metal surfaces,'' \emph{Phys. Rep.}, vol. 113, no.~4, pp. 195--287, 1984.

\bibitem{yang2017modulation}
G.~Yang, Y.~C. Liang, R.~Zhang, and Y.~Pei, ``Modulation in the air:
  Backscatter communication over ambient {OFDM} carrier,'' \emph{IEEE Trans.
  Commun.}, vol.~66, no.~3, pp. 1219--1233, 2017.

\bibitem{zhang2009optimal}
R.~Zhang, Y.~C. Liang, C.~C. Chai, and S.~Cui, ``Optimal beamforming for
  two-way multi-antenna relay channel with analogue network coding,''
  \emph{IEEE J. Sel. Areas in Commun.}, vol.~27, no.~5, pp. 699--712, 2009.

\bibitem{pan2019multicell}
\BIBentryALTinterwordspacing
C.~Pan, H.~Ren, K.~Wang, W.~Xu, M.~Elkashlan, A.~Nallanathan, and L.~Hanzo,
  ``Multicell {MIMO} communications relying on intelligent reflecting
  surface.'' [Online]. Available: \url{https://arxiv.org/abs/1907.10864}
\BIBentrySTDinterwordspacing

\bibitem{yu2019miso}
\BIBentryALTinterwordspacing
X.~Yu, D.~Xu, and R.~Schober, ``{MISO} wireless communication systems via
  intelligent reflecting surfaces.'' [Online]. Available:
  \url{https://arxiv.org/abs/1904.12199}
\BIBentrySTDinterwordspacing

\bibitem{yang2019intelligent}
\BIBentryALTinterwordspacing
Y.~Yang, B.~Zheng, S.~Zhang, and R.~Zhang, ``Intelligent reflecting surface
  meets {OFDM}: Protocol design and rate maximization.'' [Online]. Available:
  \url{https://arxiv.org/abs/1906.09956}
\BIBentrySTDinterwordspacing

\bibitem{wu2019intelligent}
Q.~Wu and R.~Zhang, ``Intelligent reflecting surface enhanced wireless network
  via joint active and passive beamforming,'' \emph{IEEE Trans. Wireless
  Commun.}, vol.~18, no.~11, pp. 5394--5409, 2019.

\bibitem{guo2019weighted}
\BIBentryALTinterwordspacing
H.~Guo, Y.~C. Liang, J.~Chen, and E.~G. Larsson, ``Weighted sum-rate
  optimization for intelligent reflecting surface enhanced wireless networks.''
  [Online]. Available: \url{https://arxiv.org/abs/1905.07920}
\BIBentrySTDinterwordspacing

\bibitem{nadeem2019large}
\BIBentryALTinterwordspacing
Q.~U.~A. Nadeem, A.~Kammoun, A.~Chaaban, M.~Debbah, and M.~S. Alouini, ``Large
  intelligent surface assisted {MIMO} communications.'' [Online]. Available:
  \url{https://arxiv.org/abs/1903.08127}
\BIBentrySTDinterwordspacing

\bibitem{zhou2019intelligent}
\BIBentryALTinterwordspacing
G.~Zhou, C.~Pan, H.~Ren, K.~Wang, W.~Xu, and A.~Nallanathan, ``Intelligent
  reflecting surface aided multigroup multicast {MISO} communication systems.''
  [Online]. Available: \url{https://arxiv.org/abs/1909.04606}
\BIBentrySTDinterwordspacing

\bibitem{bai2019latency}
\BIBentryALTinterwordspacing
T.~Bai, C.~Pan, Y.~Deng, M.~Elkashlan, and A.~Nallanathan, ``Latency
  minimization for intelligent reflecting surface aided mobile edge
  computing.'' [Online]. Available: \url{https://arxiv.org/abs/1910.07990}
\BIBentrySTDinterwordspacing

\bibitem{pan2019intelligent}
\BIBentryALTinterwordspacing
C.~Pan, H.~Ren, K.~Wang, M.~Elkashlan, A.~Nallanathan, J.~Wang, and L.~Hanzo,
  ``Intelligent reflecting surface enhanced {MIMO} broadcasting for
  simultaneous wireless information and power transfer.'' [Online]. Available:
  \url{https://arxiv.org/abs/1908.04863}
\BIBentrySTDinterwordspacing

\bibitem{yu2019enabling}
\BIBentryALTinterwordspacing
X.~Yu, D.~Xu, and R.~Schober, ``Enabling secure wireless communications via
  intelligent reflecting surfaces.'' [Online]. Available:
  \url{https://arxiv.org/abs/1904.09573}
\BIBentrySTDinterwordspacing

\bibitem{cui2019secure}
M.~Cui, G.~Zhang, and R.~Zhang, ``Secure wireless communication via intelligent
  reflecting surface,'' \emph{IEEE Wireless Commun. Lett.}, 2019.

\bibitem{shen2019secrecy}
\BIBentryALTinterwordspacing
H.~Shen, W.~Xu, S.~Gong, Z.~He, and C.~Zhao, ``Secrecy rate maximization for
  intelligent reflecting surface assisted multi-antenna communications.''
  [Online]. Available: \url{https://arxiv.org/abs/1905.10075}
\BIBentrySTDinterwordspacing

\bibitem{chen2019intelligent}
\BIBentryALTinterwordspacing
J.~Chen, Y.~C. Liang, Y.~Pei, and H.~Guo, ``Intelligent reflecting surface: A
  programmable wireless environment for physical layer security.'' [Online].
  Available: \url{https://arxiv.org/abs/1905.03689}
\BIBentrySTDinterwordspacing

\bibitem{feng2019physical}
\BIBentryALTinterwordspacing
K.~Feng and X.~Li, ``Physical layer security enhancement exploiting intelligent
  reflecting surface.'' [Online]. Available:
  \url{https://arxiv.org/abs/1911.02766}
\BIBentrySTDinterwordspacing

\bibitem{ma2010semidefinite}
Z.~Luo, W.~Ma, A.~M. So, Y.~Ye, and S.~Zhang, ``Semidefinite relaxation of
  quadratic optimization problems and applications,'' \emph{IEEE Signal
  Process. Mag.}, vol.~27, no.~3, pp. 20--34, 2010.

\bibitem{sun2016majorization}
Y.~Sun, P.~Babu, and D.~P. Palomar, ``Majorization-minimization algorithms in
  signal processing, communications, and machine learning,'' \emph{IEEE Trans.
  Signal Process.}, vol.~65, no.~3, pp. 794--816, 2016.

\bibitem{absil2009optimization}
P.~A. Absil, R.~Mahony, and R.~Sepulchre, \emph{Optimization algorithms on
  matrix manifolds}.\hskip 1em plus 0.5em minus 0.4em\relax Princeton
  University Press, 2009.

\bibitem{feng2019secure}
B.~Feng, Y.~Wu, and M.~Zheng, ``Secure transmission strategy for intelligent
  reflecting surface enhanced wireless system,'' in \emph{2019 11th
  International Conference on Wireless Communications and Signal Processing
  (WCSP)}.\hskip 1em plus 0.5em minus 0.4em\relax IEEE, 2019, pp. 1--6.

\bibitem{shi2019enhanced}
\BIBentryALTinterwordspacing
W.~Shi, X.~Zhou, L.~Jia, Y.~Wu, F.~Shu, and J.~Wang, ``Enhanced secure wireless
  information and power transfer via intelligent reflecting surface.''
  [Online]. Available: \url{https://arxiv.org/abs/1911.01001}
\BIBentrySTDinterwordspacing

\bibitem{guan2019intelligent}
\BIBentryALTinterwordspacing
X.~Guan, Q.~Wu, and R.~Zhang, ``Intelligent reflecting surface assisted secrecy
  communication via joint beamforming and jamming.'' [Online]. Available:
  \url{https://arxiv.org/abs/1907.12839}
\BIBentrySTDinterwordspacing

\bibitem{xu2019resource}
\BIBentryALTinterwordspacing
D.~Xu, X.~Yu, Y.~Sun, D.~W.~K. Ng, and R.~Schober, ``Resource allocation for
  secure {IRS}-assisted multiuser {MISO} systems.'' [Online]. Available:
  \url{https://arxiv.org/abs/1907.03085}
\BIBentrySTDinterwordspacing

\bibitem{tan2016increasing}
X.~Tan, Z.~Sun, J.~M. Jornet, and D.~Pados, ``Increasing indoor spectrum
  sharing capacity using smart reflect-array,'' in \emph{2016 IEEE
  International Conference on Communications (ICC)}.\hskip 1em plus 0.5em minus
  0.4em\relax IEEE, 2016, pp. 1--6.

\bibitem{hong2020robust}
\BIBentryALTinterwordspacing
S.~Hong, C.~Pan, H.~Ren, K.~Wang, K.~K. Chai, and A.~Nallanathan, ``Robust
  transmission design for intelligent reflecting surface aided secure
  communication systems with imperfect cascaded {CSI}.'' [Online]. Available:
  \url{https://arxiv.org/abs/2004.11580}
\BIBentrySTDinterwordspacing

\bibitem{shi2015secure}
Q.~Shi, W.~Xu, J.~Wu, E.~Song, and Y.~Wang, ``Secure beamforming for {MIMO}
  broadcasting with wireless information and power transfer,'' \emph{IEEE
  Trans. Wireless Commun.}, vol.~14, no.~5, pp. 2841--2853, 2015.

\bibitem{grant2014cvx}
M.~Grant and S.~Boyd, ``{CVX}: Matlab software for disciplined convex
  programming, version 2.1,'' 2014.

\bibitem{zhang2017matrix}
X.~D. Zhang, \emph{Matrix analysis and applications}.\hskip 1em plus 0.5em
  minus 0.4em\relax Cambridge University Press, 2017.

\bibitem{liang2009information}
Y.~Liang, H.~V. Poor \emph{et~al.}, ``Information theoretic security,''
  \emph{Foundations and Trends in communications and Information Theory},
  vol.~5, no. 4--5, pp. 355--580, 2009.

\bibitem{zhou2020framework}
\BIBentryALTinterwordspacing
G.~Zhou, C.~Pan, H.~Ren, K.~Wang, and A.~Nallanathan, ``A framework of robust
  transmission design for {IRS}-aided {MISO} communications with imperfect
  cascaded channels.'' [Online]. Available:
  \url{https://arxiv.org/abs/2001.07054}
\BIBentrySTDinterwordspacing

\end{thebibliography}


\end{document}